\documentclass[longauth]{aa}

\usepackage{euclid}
\usepackage{graphicx}
\usepackage{natbib}
\usepackage{scalerel}
\usepackage{multirow}   
\usepackage{mathrsfs}
\usepackage[table]{xcolor}

\newcommand{\pkd}{PKDGRAV-3}	
\newcommand{\ramses}{RAMSES}
\newcommand{\og}{Open-GADGET}
\newcommand{\gfour}{GADGET-4}
\newcommand{\concept}{CO\textit{N}CEPT}
\newcommand{\ahf}{AHF}
\newcommand{\subfind}{SUBFIND}
\newcommand{\velociraptor}{VELOCIraptor}
\newcommand{\denhf}{DENHF}
\newcommand{\aetiology}{AETIOLOGY}
\newcommand{\piccolo}{PICCOLO}
\newcommand{\tease}{TEASE}
\newcommand{\rockstar}{ROCKSTAR}
\newcommand{\consistent}{CONSISTENT}
\newcommand{\monofonic}{monofonIC}
\newcommand{\music}{MUSIC}
\newcommand{\msun}{\text{\ensuremath{h^{-1}\,{M}_{\odot}}}}
\newcommand{\mpc}{\text{\ensuremath{h^{-1}\,\textup{Mpc}}}}
\newcommand{\de}{\text{d}}

\usepackage{txfonts}
\usepackage[pdfencoding=auto,psdextra]{hyperref}
\hypersetup{
    colorlinks=true,
    linkcolor=blue,
    filecolor=magenta,      
    urlcolor=blue,
    citecolor=blue
}
\makeatletter
\renewcommand*\aa@pageof{, page \thepage{} of \pageref*{LastPage}}
\makeatother

\usepackage[utf8]{inputenc}

\usepackage[switch, modulo]{lineno}

\begin{document}
%
%
\title{\Euclid preparation. XXIV. Calibration of the halo mass function in $\Lambda(\nu)$CDM cosmologies}

\newcommand{\orcid}[1]{} 
\author{Euclid Collaboration: T.~Castro\orcid{0000-0002-6292-3228}$^{1,2,3}$\thanks{\email{tiago.batalha@inaf.it}}, A.~Fumagalli$^{4,1,2,3}$, R.~E.~Angulo$^{5,6}$, S.~Bocquet\orcid{0000-0002-4900-805X}$^{7}$, S.~Borgani\orcid{0000-0001-6151-6439}$^{4,3,1,2}$, C.~Carbone$^{8}$, J.~Dakin$^{9}$, K.~Dolag$^{10,11}$, C.~Giocoli\orcid{0000-0002-9590-7961}$^{12,13}$, P.~Monaco\orcid{0000-0003-2083-7564}$^{4,3,1,2}$
% ...
}

\institute{$^{1}$ INAF-Osservatorio Astronomico di Trieste, Via G. B. Tiepolo 11, I-34143 Trieste, Italy\\
$^{2}$ INFN, Sezione di Trieste, Via Valerio 2, I-34127 Trieste TS, Italy\\
$^{3}$ IFPU, Institute for Fundamental Physics of the Universe, via Beirut 2, 34151 Trieste, Italy\\
$^{4}$ Dipartimento di Fisica - Sezione di Astronomia, Universit\'a di Trieste, Via Tiepolo 11, I-34131 Trieste, Italy\\
$^{5}$ Donostia International Physics Center (DIPC), Paseo Manuel de Lardizabal, 4, 20018, Donostia-San Sebasti\'{a}n, Guipuzkoa, Spain\\
$^{6}$ IKERBASQUE, Basque Foundation for Science, 48013, Bilbao, Spain\\
$^{7}$ University Observatory, Faculty of Physics, Ludwig-Maximilians-Universit{\"a}t, Scheinerstr. 1, 81679 Munich, Germany\\
$^{8}$ INAF-IASF Milano, Via Alfonso Corti 12, I-20133 Milano, Italy\\
$^{9}$ Department of Physics and Astronomy, University of Aarhus, Ny Munkegade 120, DK-8000 Aarhus C, Denmark\\
$^{10}$ Universit\"ats-Sternwarte M\"unchen, Fakult\"at f\"ur Physik, Ludwig-Maximilians-Universit\"at M\"unchen, Scheinerstrasse 1, 81679 M\"unchen, Germany\\
$^{11}$ Max-Planck-Institut f\"ur Astrophysik, Karl-Schwarzschild Str. 1, 85741 Garching, Germany\\
$^{12}$ Istituto Nazionale di Astrofisica (INAF) - Osservatorio di Astrofisica e Scienza dello Spazio (OAS), Via Gobetti 93/3, I-40127 Bologna, Italy\\
$^{13}$ Istituto Nazionale di Fisica Nucleare, Sezione di Bologna, Via Irnerio 46, I-40126 Bologna, Italy\\
$^{14}$ Dipartimento di Fisica e Astronomia, Universit\'a di Bologna, Via Gobetti 93/2, I-40129 Bologna, Italy\\
$^{15}$ Laboratoire Univers et Th\'{e}orie, Observatoire de Paris, Universit\'{e} PSL, Universit\'{e} Paris Cit\'{e}, CNRS, F-92190 Meudon, France\\
$^{16}$ Institute of Cosmology and Gravitation, University of Portsmouth, Portsmouth PO1 3FX, UK\\
$^{17}$ Institut f\"ur Theoretische Physik, University of Heidelberg, Philosophenweg 16, 69120 Heidelberg, Germany\\
$^{18}$ INAF-Osservatorio di Astrofisica e Scienza dello Spazio di Bologna, Via Piero Gobetti 93/3, I-40129 Bologna, Italy\\
$^{19}$ INFN-Sezione di Bologna, Viale Berti Pichat 6/2, I-40127 Bologna, Italy\\
$^{20}$ Max Planck Institute for Extraterrestrial Physics, Giessenbachstr. 1, D-85748 Garching, Germany\\
$^{21}$ Dipartimento di Fisica, Universit\`{a} di Genova, Via Dodecaneso 33, I-16146, Genova, Italy\\
$^{22}$ INFN-Sezione di Roma Tre, Via della Vasca Navale 84, I-00146, Roma, Italy\\
$^{23}$ Department of Physics "E. Pancini", University Federico II, Via Cinthia 6, I-80126, Napoli, Italy\\
$^{24}$ INAF-Osservatorio Astronomico di Capodimonte, Via Moiariello 16, I-80131 Napoli, Italy\\
$^{25}$ Dipartimento di Fisica, Universit\'a degli Studi di Torino, Via P. Giuria 1, I-10125 Torino, Italy\\
$^{26}$ INFN-Sezione di Torino, Via P. Giuria 1, I-10125 Torino, Italy\\
$^{27}$ INAF-Osservatorio Astrofisico di Torino, Via Osservatorio 20, I-10025 Pino Torinese (TO), Italy\\
$^{28}$ Institut de F\'{i}sica d'Altes Energies (IFAE), The Barcelona Institute of Science and Technology, Campus UAB, 08193 Bellaterra (Barcelona), Spain\\
$^{29}$ Port d'Informaci\'{o} Cient\'{i}fica, Campus UAB, C. Albareda s/n, 08193 Bellaterra (Barcelona), Spain\\
$^{30}$ INAF-Osservatorio Astronomico di Roma, Via Frascati 33, I-00078 Monteporzio Catone, Italy\\
$^{31}$ INFN section of Naples, Via Cinthia 6, I-80126, Napoli, Italy\\
$^{32}$ Dipartimento di Fisica e Astronomia "Augusto Righi" - Alma Mater Studiorum Universit\'a di Bologna, Viale Berti Pichat 6/2, I-40127 Bologna, Italy\\
$^{33}$ INAF-Osservatorio Astrofisico di Arcetri, Largo E. Fermi 5, I-50125, Firenze, Italy\\
$^{34}$ Centre National d'Etudes Spatiales, Toulouse, France\\
$^{35}$ Institut national de physique nucl\'eaire et de physique des particules, 3 rue Michel-Ange, 75794 Paris C\'edex 16, France\\
$^{36}$ Institute for Astronomy, University of Edinburgh, Royal Observatory, Blackford Hill, Edinburgh EH9 3HJ, UK\\
$^{37}$ ESAC/ESA, Camino Bajo del Castillo, s/n., Urb. Villafranca del Castillo, 28692 Villanueva de la Ca\~nada, Madrid, Spain\\
$^{38}$ European Space Agency/ESRIN, Largo Galileo Galilei 1, 00044 Frascati, Roma, Italy\\
$^{39}$ Univ Lyon, Univ Claude Bernard Lyon 1, CNRS/IN2P3, IP2I Lyon, UMR 5822, F-69622, Villeurbanne, France\\
$^{40}$ Institute of Physics, Laboratory of Astrophysics, Ecole Polytechnique F\'{e}d\'{e}rale de Lausanne (EPFL), Observatoire de Sauverny, 1290 Versoix, Switzerland\\
$^{41}$ Departamento de F\'isica, Faculdade de Ci\^encias, Universidade de Lisboa, Edif\'icio C8, Campo Grande, PT1749-016 Lisboa, Portugal\\
$^{42}$ Instituto de Astrof\'isica e Ci\^encias do Espa\c{c}o, Faculdade de Ci\^encias, Universidade de Lisboa, Campo Grande, PT-1749-016 Lisboa, Portugal\\
$^{43}$ Department of Astronomy, University of Geneva, ch. d\'Ecogia 16, CH-1290 Versoix, Switzerland\\
$^{44}$ Universit\'e Paris-Saclay, CNRS, Institut d'astrophysique spatiale, 91405, Orsay, France\\
$^{45}$ Department of Physics, Oxford University, Keble Road, Oxford OX1 3RH, UK\\
$^{46}$ Jodrell Bank Centre for Astrophysics, Department of Physics and Astronomy, University of Manchester, Oxford Road, Manchester M13 9PL, UK\\
$^{47}$ Universit\'e Paris-Saclay, Universit\'e Paris Cit\'e, CEA, CNRS, Astrophysique, Instrumentation et Mod\'elisation Paris-Saclay, 91191 Gif-sur-Yvette, France\\
$^{48}$ Institute of Space Sciences (ICE, CSIC), Campus UAB, Carrer de Can Magrans, s/n, 08193 Barcelona, Spain\\
$^{49}$ Institut d'Estudis Espacials de Catalunya (IEEC), Carrer Gran Capit\'a 2-4, 08034 Barcelona, Spain\\
$^{50}$ INAF-Osservatorio Astronomico di Padova, Via dell'Osservatorio 5, I-35122 Padova, Italy\\
$^{51}$ Institute of Theoretical Astrophysics, University of Oslo, P.O. Box 1029 Blindern, N-0315 Oslo, Norway\\
$^{52}$ von Hoerner \& Sulger GmbH, Schlo{\ss}Platz 8, D-68723 Schwetzingen, Germany\\
$^{53}$ Technical University of Denmark, Elektrovej 327, 2800 Kgs. Lyngby, Denmark\\
$^{54}$ Institut d'Astrophysique de Paris, 98bis Boulevard Arago, F-75014, Paris, France\\
$^{55}$ Max-Planck-Institut f\"ur Astronomie, K\"onigstuhl 17, D-69117 Heidelberg, Germany\\
$^{56}$ Aix-Marseille Univ, CNRS/IN2P3, CPPM, Marseille, France\\
$^{57}$ Mullard Space Science Laboratory, University College London, Holmbury St Mary, Dorking, Surrey RH5 6NT, UK\\
$^{58}$ Universit\'e de Gen\`eve, D\'epartement de Physique Th\'eorique and Centre for Astroparticle Physics, 24 quai Ernest-Ansermet, CH-1211 Gen\`eve 4, Switzerland\\
$^{59}$ Department of Physics and Helsinki Institute of Physics, Gustaf H\"allstr\"omin katu 2, 00014 University of Helsinki, Finland\\
$^{60}$ NOVA optical infrared instrumentation group at ASTRON, Oude Hoogeveensedijk 4, 7991PD, Dwingeloo, The Netherlands\\
$^{61}$ Argelander-Institut f\"ur Astronomie, Universit\"at Bonn, Auf dem H\"ugel 71, 53121 Bonn, Germany\\
$^{62}$ Dipartimento di Fisica e Astronomia "Augusto Righi" - Alma Mater Studiorum Universit\`{a} di Bologna, via Piero Gobetti 93/2, I-40129 Bologna, Italy\\
$^{63}$ European Space Agency/ESTEC, Keplerlaan 1, 2201 AZ Noordwijk, The Netherlands\\
$^{64}$ AIM, CEA, CNRS, Universit\'{e} Paris-Saclay, Universit\'{e} de Paris, F-91191 Gif-sur-Yvette, France\\
$^{65}$ Space Science Data Center, Italian Space Agency, via del Politecnico snc, 00133 Roma, Italy\\
$^{66}$ Institute of Space Science, Bucharest, Ro-077125, Romania\\
$^{67}$ Instituto de Astrof\'isica de Canarias, Calle V\'ia L\'actea s/n, E-38204, San Crist\'obal de La Laguna, Tenerife, Spain\\
$^{68}$ Departamento de Astrof\'{i}sica, Universidad de La Laguna, E-38206, La Laguna, Tenerife, Spain\\
$^{69}$ Dipartimento di Fisica e Astronomia "G.Galilei", Universit\'a di Padova, Via Marzolo 8, I-35131 Padova, Italy\\
$^{70}$ INFN-Padova, Via Marzolo 8, I-35131 Padova, Italy\\
$^{71}$ Jet Propulsion Laboratory, California Institute of Technology, 4800 Oak Grove Drive, Pasadena, CA, 91109, USA\\
$^{72}$ Departamento de F\'isica, FCFM, Universidad de Chile, Blanco Encalada 2008, Santiago, Chile\\
$^{73}$ Centro de Investigaciones Energ\'eticas, Medioambientales y Tecnol\'ogicas (CIEMAT), Avenida Complutense 40, 28040 Madrid, Spain\\
$^{74}$ Universidad Polit\'ecnica de Cartagena, Departamento de Electr\'onica y Tecnolog\'ia de Computadoras, 30202 Cartagena, Spain\\
$^{75}$ Kapteyn Astronomical Institute, University of Groningen, PO Box 800, 9700 AV Groningen, The Netherlands\\
$^{76}$ Infrared Processing and Analysis Center, California Institute of Technology, Pasadena, CA 91125, USA\\
$^{77}$ INAF-Osservatorio Astronomico di Brera, Via Brera 28, I-20122 Milano, Italy\\
$^{78}$ INAF-Istituto di Astrofisica e Planetologia Spaziali, via del Fosso del Cavaliere, 100, I-00100 Roma, Italy\\
$^{79}$ Junia, EPA department, F 59000 Lille, France\\
$^{80}$ SISSA, International School for Advanced Studies, Via Bonomea 265, I-34136 Trieste TS, Italy\\
$^{81}$ Dipartimento di Fisica e Scienze della Terra, Universit\'a degli Studi di Ferrara, Via Giuseppe Saragat 1, I-44122 Ferrara, Italy\\
$^{82}$ Istituto Nazionale di Fisica Nucleare, Sezione di Ferrara, Via Giuseppe Saragat 1, I-44122 Ferrara, Italy\\
$^{83}$ Institut de Physique Th\'eorique, CEA, CNRS, Universit\'e Paris-Saclay F-91191 Gif-sur-Yvette Cedex, France\\
$^{84}$ Institut de Recherche en Astrophysique et Plan\'etologie (IRAP), Universit\'e de Toulouse, CNRS, UPS, CNES, 14 Av. Edouard Belin, F-31400 Toulouse, France\\
$^{85}$ NASA Ames Research Center, Moffett Field, CA 94035, USA\\
$^{86}$ INAF, Istituto di Radioastronomia, Via Piero Gobetti 101, I-40129 Bologna, Italy\\
$^{87}$ INFN-Bologna, Via Irnerio 46, I-40126 Bologna, Italy\\
$^{88}$ Universit\'e C\^{o}te d'Azur, Observatoire de la C\^{o}te d'Azur, CNRS, Laboratoire Lagrange, Bd de l'Observatoire, CS 34229, 06304 Nice cedex 4, France\\
$^{89}$ Instituto de Astrof\'isica e Ci\^encias do Espa\c{c}o, Faculdade de Ci\^encias, Universidade de Lisboa, Tapada da Ajuda, PT-1349-018 Lisboa, Portugal\\
$^{90}$ Institute for Theoretical Particle Physics and Cosmology (TTK), RWTH Aachen University, D-52056 Aachen, Germany\\
$^{91}$ Department of Physics \& Astronomy, University of California Irvine, Irvine CA 92697, USA\\
$^{92}$ University of Lyon, UCB Lyon 1, CNRS/IN2P3, IUF, IP2I Lyon, France\\
$^{93}$ INFN-Sezione di Genova, Via Dodecaneso 33, I-16146, Genova, Italy\\
$^{94}$  Universit\'e Paris Cit\'e, CNRS, Astroparticule et Cosmologie, F-75013 Paris, France\\
$^{95}$ Instituto de F\'isica Te\'orica UAM-CSIC, Campus de Cantoblanco, E-28049 Madrid, Spain\\
$^{96}$ Department of Physics, P.O. Box 64, 00014 University of Helsinki, Finland\\
$^{97}$ Ruhr University Bochum, Faculty of Physics and Astronomy, Astronomical Institute (AIRUB), German Centre for Cosmological Lensing (GCCL), 44780 Bochum, Germany\\
$^{98}$ Department of Physics, Lancaster University, Lancaster, LA1 4YB, UK\\
$^{99}$ Universit\'{e} Paris-Saclay, CNRS/IN2P3, IJCLab, 91405 Orsay, France\\
$^{100}$ Department of Physics and Astronomy, University College London, Gower Street, London WC1E 6BT, UK\\
$^{101}$ Astrophysics Group, Blackett Laboratory, Imperial College London, London SW7 2AZ, UK\\
$^{102}$ Univ. Grenoble Alpes, CNRS, Grenoble INP, LPSC-IN2P3, 53, Avenue des Martyrs, 38000, Grenoble, France\\
$^{103}$ Dipartimento di Fisica, Sapienza Universit\`a di Roma, Piazzale Aldo Moro 2, I-00185 Roma, Italy\\
$^{104}$ University of Applied Sciences and Arts of Northwestern Switzerland, School of Engineering, 5210 Windisch, Switzerland\\
$^{105}$ Aix-Marseille Univ, CNRS, CNES, LAM, Marseille, France\\
$^{106}$ Zentrum f\"ur Astronomie, Universit\"at Heidelberg, Philosophenweg 12, D- 69120 Heidelberg, Germany\\
$^{107}$ Department of Mathematics and Physics E. De Giorgi, University of Salento, Via per Arnesano, CP-I93, I-73100, Lecce, Italy\\
$^{108}$ INAF-Sezione di Lecce, c/o Dipartimento Matematica e Fisica, Via per Arnesano, I-73100, Lecce, Italy\\
$^{109}$ INFN, Sezione di Lecce, Via per Arnesano, CP-193, I-73100, Lecce, Italy\\
$^{110}$ Observatoire de Sauverny, Ecole Polytechnique F\'ed\'erale de Lau- sanne, CH-1290 Versoix, Switzerland\\
$^{111}$ Institute for Computational Science, University of Zurich, Winterthurerstrasse 190, 8057 Zurich, Switzerland\\
$^{112}$ Higgs Centre for Theoretical Physics, School of Physics and Astronomy, The University of Edinburgh, Edinburgh EH9 3FD, UK\\
$^{113}$ Universit\'e St Joseph; Faculty of Sciences, Beirut, Lebanon\\
$^{114}$ Department of Astrophysical Sciences, Peyton Hall, Princeton University, Princeton, NJ 08544, USA\\
$^{115}$ Helsinki Institute of Physics, Gustaf H{\"a}llstr{\"o}min katu 2, University of Helsinki, Helsinki, Finland\\
$^{116}$ Department of Mathematics and Physics, Roma Tre University, Via della Vasca Navale 84, I-00146 Rome, Italy}

%
%
\abstract{
\textit{Euclid}'s photometric galaxy cluster survey has the potential to be a very competitive cosmological probe. The main cosmological probe with observations of clusters is their number count, within which the halo mass function (HMF) is a key theoretical quantity. We present a new calibration of the analytic HMF, at the level of accuracy and precision required for the uncertainty in this quantity to be subdominant with respect to other sources of uncertainty in recovering cosmological parameters from \textit{Euclid} cluster counts. Our model is calibrated against a suite of \textit{N}-body simulations using a Bayesian approach taking into account systematic errors arising from numerical effects in the simulation. First, we test the convergence of HMF predictions from different \textit{N}-body codes, by using initial conditions generated with different orders of Lagrangian Perturbation theory, and adopting different simulation box sizes and mass resolution. Then, we quantify the effect of using different halo finder algorithms, and how the resulting differences propagate to the cosmological constraints. In order to trace the violation of universality in the HMF, we also analyse simulations based on initial conditions characterised by scale-free power spectra with different spectral indexes, assuming both Einstein--de Sitter and standard $\Lambda$CDM expansion histories. Based on these results, we construct a fitting function for the HMF that we demonstrate to be sub-percent accurate in reproducing results from 9 different variants of the $\Lambda$CDM model including massive neutrinos cosmologies. The calibration systematic uncertainty is largely sub-dominant with respect to the expected precision of future mass--observation relations; with the only notable exception of the effect due to the halo finder, that could lead to biased cosmological inference.
}
%
%
%
%
   \titlerunning{\textit{Euclid} preparation. XXIV. Calibration of the halo mass function in $\Lambda(\nu)$CDM cosmologies}
   \authorrunning{Castro et al.}
   
   \maketitle
%
%
%
%
   
\section{\label{sc:Intro}Introduction}
Structure formation in the Universe follows a hierarchical process, with larger structures forming from the collapse of smaller ones. Galaxy clusters sit at the top of this hierarchy as the most massive virialized objects in the Universe. Their cosmological abundance and evolution make them competitive cosmological probes of both the geometry of our Universe and the growth of density perturbations~\citep[][for reviews]{Allen:2011zs, KravtsovBorgani:2012}.

Number-counts experiments represent the main cosmological probe from cluster surveys~\citep{Holder:2001db,DSDD:2009php,Hasselfield:2013wf,Planck:2013lkt, Bocquet:2014lmj,Mantz:2014paa, Planck:2015lwi, SPT:2018njh, DES:2020ahh, DES:2020cbm, Lesci:2020qpk}. The idea behind number counts is remarkable in its simplicity; assuming the halo mass function (HMF) is precisely predicted in a given cosmological model, one can confront the number of observed clusters in a survey with this theoretical prediction and, from this, constrain cosmological parameters.  

\citet{Press:1973iz} presented the first theoretical model for the HMF based on mapping the peaks of a primordial, Gaussian density field to the number of collapsed structures at low redshift. Later, \citet{Bond:1990iw} introduced the excursion-set formalism that embedded the Press--Schechter derivation in a more rigorous and rich mathematical framework. Despite the success of these models in providing an insightful, qualitative description of the abundance of halos, their agreement with increasingly accurate \textit{N}-body simulations is quantitatively poor~\citep[see, e.g.][]{Sheth:1999mn}. Later, \citet{Maggiore:2009rv, Maggiore:2009rw, Maggiore:2009rx} extended the excursion-set solution to more realist collapse barriers leading to better agreement with \textit{N}-body simulations~\citep[see also][]{Corasaniti:2011dr,Corasaniti:2010zt}.

Several works presented semi-analytical extensions of the Press--Schechter formalism calibrated to reproduce the results of numerical simulations~\citep[see, e.g.][]{Sheth:1999mn, Sheth:1999su,Jenkins:2000bv,Warren:2005ey,Tinker:2008ff,Watson:2012mt,Despali:2015yla}. These extensions rely on the feature of the HMF being predominantly universal, that is, being independent of the underlying cosmological model, if expressed as a function of the Gaussian density field statistics.

The departure of the HMF from universality depends on several aspects, and modelling it is a daunting task. Recent works have started using emulators to circumvent the complex analytical modelling to predict the HMF~\citep{McClintock:2018uyf, Nishimichi:2018etk, Bocquet:2020tes}. Although practical and straightforward, this strategy has limitations as emulators are not rigorously supported by a robust underlying model and are known to perform poorly outside the regime in which they have been built. Furthermore, building an emulator does not necessarily lead to a better understanding of nature, lacking a theoretical legacy. Still, it is essential to note that emulators have pushed the precision and accuracy of the HMF predictions to a few percent, outperforming fitting functions, which range in accuracy from $10$ to $30$ percent.

\textit{N}-body simulations are central to both approaches as they are the only theoretical tool to with which to stringently assess the non-linear regime where galaxy clusters are deep-seated. The ever-increasing requirements for larger and high-resolution simulations pushed the development of fast and efficient gravity solvers, each of them relying on different algorithms with different approximations. Keeping the validity, accuracy, and precision of those approximations under control is utterly necessary in order to guarantee the robustness of the final HMF model~\citep[see e.g.][for a detailed discussion of the methodology]{Angulo:2021kes}.

The luminous matter composition of our Universe, although subdominant, is known to impact the formation of structure in a significant way~\citep{Cui:2014aga,Velliscig:2014,Bocquet:2015pva,Castro:2020yes}. The modelling of the baryonic feedback in hydrodynamical simulations is a controversial subject in vogue. However, at the scales of galaxy clusters, it is well established that baryonic feedback cannot disrupt structures; instead, its net effect is to rearrange the halo composition, changing its mass with respect to the same object simulated using a collisionless scheme. As hydrodynamical simulations are thousands of times more expensive than purely gravitational \textit{N}-body simulations, the standard approach is to characterise the HMF using the latter and model how baryonic physics changes the mass of halos in post-processing~\citep[see, e.g.][]{Schneider:2015wta, Arico:2020lhq}. This method will not be addressed in this paper.

The impact of systematic effects in determining the HMF has already been addressed. For instance, \citet{Knebe:2011rx} and \citet{Garcia:2019xel} showed that differences in the halo finder algorithm result in changes in the HMF that can be as large as several percent. \citet{Salvati:2020exw} and \citet{Artis:2021tjj} claimed that such discrepancies on the HMF model in future surveys could lead to severe biases in the cosmological constraints, raising awareness of the degree of accuracy required by the quality of next generation data.

In the present work, we aim to calibrate an HMF fitting function with a target accuracy of 1 percent for objects more massive than a few times~$10^{13}\,\msun$, as demanded by the large number of galaxy clusters expected to be observed by the ESA \Euclid satellite~\citep[see][]{Sartoris:2016}. In particular, the wide, \textit{Euclid} survey will cover 15\,000 ${\rm deg}^2$ of the extra-galactic sky \citep{EUCLID:2011zbd} in optical and near-infrared (NIR) bands. By relying on suitable algorithms to identify galaxy clusters as galaxy concentrations in photometric redshift space \citep{Adam:2019}, \Euclid will provide a survey of optically identified clusters that is unique in terms of the depth and area covered, and that has the potential to provide tight constraints on cosmological models \citep{Sartoris:2016}.

To fully exploit the cosmological potential of the \textit{Euclid} Cluster Survey, several observational and theoretical systematic effects must be controlled. As for the latter, we concentrate in this paper on an accurate calibration of the HMF. To this purpose, we follow the approach of~\citet{Ondaro-Mallea:2021yfv} and explicitly model the non-universality of the HMF. To guarantee the robustness of our results, we start by presenting a detailed study of the numerical systematic errors on \textit{N}-body simulations and how they impact the HMF. We then employ a suite of \textit{N}-body simulations and calibrate a model to predict the abundance of halos as a function of mass, cosmology, and redshift. Lastly, we forecast the impact of numerical and systematic uncertainties on \textit{Euclid}'s number counts analysis.

This paper is organised as follows: in Sect.~\ref{sec:methodology} we present our methodology. The study of the convergence of the simulation setup used in this work is presented in detail in Sect.~\ref{sec:mastering}. In Sect.~\ref{sec:modeling}, we present our modelling of the HMF. Our results are presented in Sect.~\ref{sec:results}. Finally, concluding remarks are provided in Sect.~\ref{sec:conclusions}. Additionally, in Appendixes~\ref{sec:App-convergence}, \ref{sec:App-hf}, and \ref{sec:App-quality}, we present further numerical convergence tests for our adopted setup, a comparison between the HMF calibrated in different halo finders, and the robustness of our calibration as a function of redshift, respectively.

\section{\label{sec:methodology}Methodology}

In this section, we present a short overview of the main concepts of the HMF (Sect.~\ref{sec:theory}), a description of the sets of simulations carried out for our analysis (Sect.~\ref{sec:sims}), the halo finders adopted (Sect.~\ref{sec:halocats}), the Bayesian framework used for the HMF calibration (Sect.~\ref{sec:likelihood}), and the pipeline to forecast the effects of uncertainties on the HMF calibration on the cosmological constraints from the \textit{Euclid} photometric survey of galaxy clusters (Sect.~\ref{sec:forecast}).

\subsection{\label{sec:theory}The halo mass function}

The comoving number density of halos with mass in the range $[M, M+\de M] $ is given by:
\begin{equation}
	\frac{\de n}{\de M} \de M = \frac{\rho_{\rm m}}{M} \, \nu f(\nu) \, \de \ln \nu\,.
	\label{eq:hmf}
\end{equation}
Equation~\eqref{eq:hmf} is known as the differential halo mass function, where $\rho_{\rm m}$ is the comoving cosmic mean matter density, $\nu$ is the peak height, and the function $\nu f(\nu)$ is known as the multiplicity function. The peak height $\nu$ is a measure of how rare a halo is and is defined as $\nu=\delta_{\rm c}/\sigma(M, z)$, where $\delta_{\rm c}$ is the critical density for spherical halo collapse extrapolated to $z=0$~\citep{peebles2020large} and $\sigma^2(M, z)$ is the mass variance at redshift $z$, which can be expressed in terms of the linear matter power spectrum $P_{\rm m}(k,z)$ as:
\begin{equation}
	\sigma^2(M,z) = \frac{1}{2\pi^2} \int_0^\infty \de k\,k^2\,P_{\rm m}(k,z)\,W^2 \left(k\,R \right)\,,
\label{eq:massvariance}
\end{equation}
where $R(M)= \left[3\,M\,/\,(4\,\pi\,\rho_{\rm m})\right]^{1/3} $ is the Lagrangian radius of a sphere containing the mass $M$, and $W(k\,R)$ is the top-hat filter in $k$-space.

The multiplicity function is said to be universal if its cosmological dependence comes only through its dependence on the peak height $\nu$, with the functional form of this dependence being independent of cosmology. While this assumption of HMF universality holds to first approximation, a number of independent analyses based on \textit{N}-body simulations have demonstrated that the HMF systematically deviates from universality at late times \citep{Crocce:2009mg, Courtin:2010gx, Watson:2012mt, Diemer:2020rgd, Ondaro-Mallea:2021yfv}.

The non-universality of the HMF has been observed to depend on both the halo definition \citep{Watson:2012mt,Despali:2015yla, Diemer:2020rgd, Ondaro-Mallea:2021yfv} and the residual dependence of $\delta_{\rm c}$ on cosmology. For instance, although several works have used the mass variance $\sigma^2(M, z)$ as a characteristic parameter \citep{ Jenkins:2000bv, Reed:2003sq, Warren:2005ey, Tinker:2008ff, Crocce:2009mg,  Watson:2012mt,  Bocquet:2015pva, Castro:2020yes}, it has been shown by \citet{Courtin:2010gx} that correctly using the peak height $\nu(M, z)=\delta_{\rm c}/\sigma(M, z)$, thus including the cosmological dependence of the linear collapse barrier $\delta_{\rm c}$, provides more universal results. The effect is more prominent at high masses, which is the regime relevant for cluster cosmology and  where we dedicate particular attention in this paper.

In our analysis, halos are defined as spherical overdensities with average enclosed density equal to $\Delta_{\rm vir}(z)$ times the background density, where $\Delta_{\rm vir}(z)$ is the non-linear density contrast of virialized structures as predicted by spherical collapse~\citep[see, e.g.][]{Eke:1996ds,Bryan_1998}. In the following, we parameterize the HMF for a given cosmology at a given redshift according to the fitting function introduced by~\citet{Bhattacharya:2010wy},
\begin{equation}
	\label{eq:mult}
	\nu\,f(\nu)=A(p,q) \sqrt{\frac{2a\nu^2}{\pi}} \mathrm{e}^{-a\nu^2/2} \left(1+ \frac{1}{(a\nu^2)^p} \right) (\nu\sqrt{a})^{q-1} \,.
\end{equation}
The dependence of the fitting parameters $\{a, p, q\}$ on cosmology is presented in Sect.~\ref{sec:modeling}. As for $\delta_{\rm c}$, we use the fitting formula presented by~\cite{Kitayama:1996ne}.

According to the Halo Model~\citep[see e.g.][and references therein]{Cooray:2002dia}, all matter in the Universe is contained in halos. Thus, the integral of the HMF should be normalised to unity. In the case where $a$, $p$, and $q$ do not depend on $\nu$, this condition is satisfied by:
\begin{equation}
	A(p,q) = \Bigg\{ \frac{2^{-1/2-p+q/2}}{\sqrt{\pi}} \, \left[ 2^p \, \Gamma\left(\frac{q}{2}\right)+ \Gamma\left(-p+\frac{q}{2}\right)  \right] \Bigg\}^{-1}\, ,
\label{eq:Anorm}
\end{equation}
where $\Gamma$ denotes the Gamma function. However, we note that not all HMF models presented in the literature obey this condition. Furthermore, this normalisation is not likely to hold in our Universe ~\citep[see e.g.][]{Angulo:2009hf}. As discussed below, we write the parameters $\{a,p,q\}$ explicitly as functions of the matter power spectrum shape and background evolution; thus, our model violates the condition of normalisation to unity even if it formally uses Eq.~\eqref{eq:Anorm}. In this sense, Eq.~\eqref{eq:Anorm} is used here only to guarantee an asymptotic behaviour to the HMF when extrapolated to lower masses. 

\subsection{\label{sec:sims}Simulations}

Due to its intrinsically non-linear dynamics, the formation of virialized halos can only be studied in detail by resorting to  cosmological simulations. The need for an accurate \textit{N}-body simulation to numerically solve the evolution of billions of particles during many dynamical time scales entails the design and development of high performance and scalable algorithms and codes. Each gravity solver algorithm comes with different technical aspects characterising its validity, accuracy, and precision. A comprehensive comparison of the publicly available codes is beyond the scope of this paper~\citep[see e.g.][and references therein]{Angulo:2021kes}. Still, understanding and quantifying differences in the HMFs predicted by different, and widely used,  \textit{N}-body codes is key to establishing confidence in the convergence and robustness of the \textit{N}-body solution of the HMF.  

The HMF analysis that we present in this paper is based on three sets of simulations specifically designed for this purpose and which are summarised in Table~\ref{tab:sims}. The \emph{TEsting simulAtion SEt} (\tease) is used in Sects.~\ref{sec:mastering} and~\ref{sec:modeling} to quantify the aforementioned impact of numerical systematic effects on the HMF. The set consists of simulations of $500\,\mpc$ cosmological boxes run with different setups and different codes. We used the codes: \og, \gfour,\footnote{\url{https://wwwmpa.mpa-garching.mpg.de/gadget4/}} \pkd,\footnote{\url{https://bitbucket.org/dpotter/pkdgrav3}} \concept,\footnote{\url{https://jmd-dk.github.io/concept}} and \ramses,\footnote{\url{https://www.ics.uzh.ch/~teyssier/ramses}} which respectively deploy the following gravity solver algorithms: tree-particle mesh~\citep[tree-PM;][]{Xu:1994fk,Bagla:1999tx,Springel:2005mi}, fast multipole method-particle mesh~\citep[FMM-PM;][]{Springel:2020plp}, ~\citep[FMM;][]{Potter:2016ttn}, particle particle - particle mesh ~\citep[P$^3$M;][]{Dakin:2021ivb}, and adaptive mesh refinement ~\citep[AMR;][]{Teyssier:2001cp}. Initial conditions have been generated using either the Zeldovich approximation~\citep{Zeldovich:1969sb} at $z=99$ with \music~\citep{Hahn:2011uy}\footnote{\url{https://bitbucket.org/ohahn/music}} or third-order Lagrangian perturbation theory (3LPT) at $z=24$ with \monofonic~\citep{Michaux:2020yis}.\footnote{\url{https://bitbucket.org/ohahn/monofonic}} For this simulation set, the background expansion history and linear matter power spectrum are both kept fixed, while resolution is varied (see Table~\ref{tab:sims}).

The \emph{scAle frEe simulaTIons fOr cLuster cOsmoloGY} (\aetiology) set of simulations is used in Sect.~\ref{sec:modeling} to model the departure of the HMF from universality. The set consists of simulations with $1024^3$ particles in boxes of $1000\,\mpc$ on a side. The simulations and the initial conditions (2LPT at $z=99$) were run with \gfour. This set is specifically designed to discriminate between the non-universality arising from the presence of characteristic timescales in the background expansion history, and from the presence of characteristic length scales in the matter power spectrum. For this purpose, the set combines simulations with either power-law or $\Lambda$CDM linear power spectrum on a background that is either Einstein--de Sitter -- that is $(\Omega_{\rm m}, \Omega_{\Lambda})=(1,0)$; EdS hereafter -- or $\Lambda$CDM. The self-similarity of scale-free simulations has been extensively used by~\cite{Leroy:2020fzc}, ~\cite{Garrison:2021zpr},~\cite{Garrison:2021hus},~\cite{Joyce:2020qxv}, and~\cite{Maleubre:2021guf} to carry out controlled tests of numerical convergence for several cosmological algorithms. Here, instead, we focus on self-similarity to isolate the effect of the power spectrum shape from the effect of the background evolution on the HMF universality. We note that the idea of carrying out simulations with independent background and power spectrum parameters was also used by~\citet{Ondaro-Mallea:2021yfv}.

Lastly, the \emph{ PathfInder Cluster COsmoLOgy} (\piccolo) set is used in Sect.~\ref{sec:results} to calibrate our HMF model. The set comprises $26$ cosmological boxes of $2000\,\mpc$ in size and $4\times 1280^3$ particles. These simulations were carried out with \og. Initial conditions were generated using \monofonic\ according to 3LPT at $z=24$. The set of cosmological parameters for this set are varied by uniformly drawing them from the $95$ percent confidence level hyper-volume of the joint SPT and DES cluster abundance constraints presented by~\citet{DES:2020cbm} (see Table~\ref{tab:cosm}). Two realisations were simulated for each cosmology, with the exception of the reference $C0$ model, for which ten realisations were carried out.

In this paper, we deliberately ignored the effect of massive neutrinos in the simulations. \citet{Castorina:2014} showed that the effects of massive neutrinos on the HMF are mostly absorbed into the computation of the mass variance once the cold-dark-matter power spectrum is considered instead of the total matter power \citep[see also][]{Costanzi:2013bha}. We will rely on this approach to extend our modelling to scenarios with massive neutrinos; however, as we recognise the importance of validating this approach ensuring it provides the accuracy required for the application to the \textit{Euclid} cluster survey, in Sect.~\ref{sec:neutrinos} we compare our model to two external sets of simulations that include neutrinos at the particle level.

\begin{table*}
	\centering
	\caption{The suite of simulations. The corresponding cosmology is given in Table~\ref{tab:cosm}.}
	\label{tab:sims}
	\resizebox{0.995\textwidth}{!}{%
	{\renewcommand{\arraystretch}{1.2}
	\begin{tabular}{c|c|c|c|c|c|c|c|c}
	\hline
	Set & $L_{\rm box}$ & $N_{\rm p}$ & Background & $P_{\rm lin}(k)$ & \multicolumn{3}{c|}{Initial Conditions} & Grav. Solver \\\cline{6-8}
	& [\mpc] &&&& Code & LPT Order & $z$ & \\\hline
	\multirow{7}{*}{\tease} & \multirow{7}{*}{$500$} & $256^3$ & \multirow{7}{*}{$C0$} & \multirow{7}{*}{$\Lambda$CDM} & \multirow{4}{*}{\music} & \multirow{4}{*}{Zel.} & \multirow{4}{*}{$99$} & tree-PM, FMM-PM\\
		& & $512^3$ & & & & & & FMM, P$^3$M, AMR$\color{blue}^\dagger$ \\
		& & $1024^3$ & & & & & & \\
		&  & $4\times160^3$ &  &  & & & \\\cline{6-9}
		&  & $4\times320^3$ &  &  & \multirow{3}{*}{\monofonic} & \multirow{3}{*}{3LPT} &  \multirow{3}{*}{24} & tree-PM, FMM-PM \\
		&  & $4\times640^3$ &  &  &  &  & & FMM, P$^3$M \\
		&  & $4\times1280^3$ &  &  &  & & &  \\\hline
	\multirow{4}{*}{\aetiology} & \multirow{4}{*}{$1000$} & \multirow{4}{*}{$1024^3$} & \multirow{2}{*}{EdS} & Power-law & \multirow{4}{*}{\gfour} & \multirow{4}{*}{2LPT}& \multirow{4}{*}{99} & \multirow{4}{*}{FMM-PM} \\
        &  & & & $\Lambda$CDM $(C0)$ &  &  &  \\\cline{4-5}
		&  & & \multirow{2}{*}{$C0$} & Power-law &  &  &  \\
		&  & & & $\Lambda$CDM &  &  &  \\\hline
		
		\piccolo\ & $2000$ & $4\times1280^3$ & $C0-C8$ & $\Lambda$CDM & \monofonic & 3LPT & $24$ & tree-PM \\\hline
		\hline
	\end{tabular}}}
	\begin{flushleft}
		$\color{blue}^\dagger$\footnotesize{Corresponding to the gravity solver algorithms deployed in \og, \gfour, \pkd, \concept, and \ramses, respectively.}
	\end{flushleft}
\end{table*}
\begin{table}
	\centering
	\caption{The cosmological parameters of the simulations presented in Table~\ref{tab:sims}. The parameters have been uniformly drawn from the $95\%$ confidence hyper-volume of the cluster abundances constraints presented in~\citet{DES:2020cbm}.}
	\label{tab:cosm}
	\begin{tabular}{c|c|c|c|c|c}
		\hline
		Name & $\Omega_{\rm m}$ & $h$ & $\Omega_{\rm b}$ & $n_{\rm s}$ & $\sigma_8$ \\\hline
		$C0$ & $0.3158$ & $0.6732$ & $0.0494$ & $0.9661$ & $0.8102$ \\
		$C1$ & $0.1986$ & $0.7267$ & $0.0389$ & $0.9775$ & $0.8590$ \\
		$C2$ & $0.1665$ & $0.7066$ & $0.0417$ & $0.9461$ & $0.8341$ \\
		$C3$ & $0.3750$ & $0.6177$ & $0.0625$ & $0.9778$ & $0.7136$ \\
		$C4$ & $0.3673$ & $0.6353$ & $0.0519$ & $0.9998$ & $0.7121$ \\
		$C5$ & $0.1908$ & $0.6507$ & $0.0527$ & $0.9908$ & $0.8971$ \\
		$C6$ & $0.2401$ & $0.8087$ & $0.0357$ & $0.9475$ & $0.8036$ \\
		$C7$ & $0.3020$ & $0.5514$ & $0.0674$ & $0.9545$ & $0.8163$ \\
		$C8$ & $0.4093$ & $0.7080$ & $0.0446$ & $0.9791$ & $0.7253$ \\
        \hline
		\hline
	\end{tabular}
\end{table}

We emphasise that, in order to guarantee a fair comparison between the different gravity solvers, we proceeded as follows. 
\begin{itemize}
	\item We carried out convergence tests for the adopted setup of \og\ and \gfour\ (see Sect.~\ref{sec:mastering}).
	\item We used the default setup of the \concept\ code. The default setup of \concept\ was extensively tested and compared with other codes by~\citet{Dakin:2021ivb}.
	\item We used the conservative \pkd\ example setup (D. Potter, 2021, priv. comm.).
	\item We used the recommended \ramses\ setup suggested by the \music\ and \monofonic\ initial conditions generators.
\end{itemize}

\subsection{\label{sec:halocats}Halo finders}

For the purpose of testing the HMF calibration against the choice of the halo finder, we used \ahf,\footnote{\url{http://popia.ft.uam.es/AHF/Download.html}} \subfind,\footnote{See n. 1.} \velociraptor,\footnote{\url{https://github.com/pelahi/VELOCIraptor-STF}} \denhf, and \rockstar\footnote{\url{https://bitbucket.org/gfcstanford/rockstar}} to extract halo catalogues from the different simulation sets. While all these algorithms are based on a spherical overdensity (SO) method to define halo boundaries, they differ in the method applied to identify the centre from which spheres are grown. For each of these halo finders, we provide a very short description of their main characteristics below, while we refer to the original papers where they have been presented. We further refer to \citet{Knebe:2011rx} for an extensive and detailed comparison of the different halo finders. 
\begin{itemize}
    \item \ahf~\citep{Gill:2004km,Knollmann:2009pb} deploys an AMR algorithm to identify prospective halos centres as density peaks on the end-leaves of the refined tree. 
    \item \denhf~\citep{Despali:2015yla} determines prospective halo centres as peaks of the density field estimated using the inverse of the cubic distance to the $n$-th closest neighbours. 
    \item \subfind~\citep{Springel:2000qu,Springel:2020plp} determines halo centres by first executing a parallel implementation of the 3D friends-of-friends~\citep[FOF, see e.g.][]{Davis:1985rj} algorithm and then by assigning it to the position of the particle with the lowest potential. 
    \item \velociraptor~\citep{Elahi:2019wap} can operate very similarly to \subfind\ if required. Additionally, the user can choose three group finders: 3DFOF, 6DFOF~\citep[see e.g.][]{Diemand:2006ey}, and a 6DFOF with adaptive velocity scale. Lastly, halo centres are defined as the group's centre of mass.
    \item \rockstar~\citep{Behroozi:2011ju} starts by partitioning the simulation volume into 3DFOF groups for straightforward parallelization. An adaptive and recursive 6DFOF algorithm is then run for each group, creating a hierarchical set of FOF subgroups. Halo centres are computed by averaging the positions of the particles belonging to the innermost subgroup in the hierarchy. We run the \consistent\footnote{\url{https://bitbucket.org/pbehroozi/consistent-trees}} algorithm on top of the previously extracted \rockstar\ catalogues. \consistent\ was shown in~\citet{Behroozi:2011js} to improve the consistency of the halo catalogues using a dynamical tracking of halo progenitors.
\end{itemize}
It is important to note that we employ the same halo mass definition; thus, the algorithms above will differ only in the centring definition and, more importantly, in their hierarchical conditions, that is, the requirements to discriminate between structures and substructures. Unless stated otherwise, we only consider inclusive SO masses, that is, including all particles inside the virial radius regardless of whether they are gravitationally bound.

All halo catalogues have been logarithmically binned according to the number of halo particles to minimise the effect of mass discretization. This effect and the adopted binning are further discussed in Sect.~\ref{sec:ics}.

\subsection{\label{sec:likelihood}HMF calibration: maximum likelihood approach}

In the Press--Schechter formalism, halos are assumed to form due to the gravitational collapse of matter over-densities, filtered over a given mass scale $M$, above the linear collapse threshold~$\delta_{\rm c}$. The unconditional distribution of the number of excursions above a given threshold on a Gaussian field follows a Poisson distribution. This motivates us to write the likelihood $\mathcal{L}(N_i|\pmb{\theta}, z)$ for the number of halos $N_i^{\rm sim}$ with masses $M_{\rm halo} \in [M_i, M_{i+1})$ in the snapshot at redshift $z$ as:
\begin{equation}
	\ln \mathcal{L}(N_i|\pmb{\theta}, z) \,=  N_{i}^{\rm sim} \ln \left(\frac{N_{i}(\pmb{\theta},z)}{N_{i}^{\rm sim}}\right) - \, N_{i}(\pmb{\theta}, z) + N_{i}^{\rm sim},
	\label{eq:lkpoisson}
\end{equation}
where $N_i(\pmb{\theta}, z)$ is the theoretical expectation of halos computed by integrating Eq.~\eqref{eq:hmf} and multiplying it by the volume of the simulated cosmological box. Lastly, $\pmb{\theta}$ is the vector of parameters of Eq.~\eqref{eq:mult} including their dependency on cosmology.

However, numerical systematic effects, such as round-off errors, can affect the distribution of halos introducing further scatter between the predicted HMFs. This problem is more prominent for low-mass halos as the abundance of halos quickly grows with decreasing mass and Poisson errors under-predict the total error budget. In this paper, we use the composite likelihood instead:
\begin{equation}
	\ln \mathcal{L}(N_i|\pmb{\theta}, z) =
	\begin{cases} 
		N_{i}^{\rm sim} \ln \left(\frac{N_{i}(\pmb{\theta},z)}{N_{i}^{\rm sim}}\right) -  N_{i}(\pmb{\theta}, z) + N_{i}^{\rm sim}  & N_{i}^{\rm sim}\leq 25 \\ \\
		\frac{1}{2}\ln\left(2\pi \sigma^2\right) + \frac{\left( N_{i}(\pmb{\theta}, z) -  N_{i}^{\rm sim} \right)^2}{2\,\sigma^2} & N_{i}^{\rm sim} > 25\,,
	\end{cases}
	\label{eq:lkhmf}
\end{equation}
where $\sigma=\sqrt{ N_{i}^{\rm sim}+\sigma_{\rm sys}^2 }$ is the standard deviation resulting from the convolution of the Gaussian approximation to the Poisson distribution with a noise distribution assumed to be Gaussian with zero mean and variance $\sigma_{\rm sys}^2$. We note that our likelihood presents a discontinuity at $N_i^{\rm sim}=25$. The impact of the discontinuity is two-fold. Firstly, the discontinuity amplitude depends on the chosen value for $\sigma_{\rm sys}$. As we discuss here below, we assume $\sigma_{\rm sys}\lesssim1$ percent, thus minimising the impact on the transition regime. Secondly, there is the transition from the purely Poisson error to a symmetric Gaussian approximation. The chosen transition value for $N_{i}^{\rm sim}$ guarantees that the transition is again smooth as the Poisson contribution largely dominates the total error budget while negative values for $N_i(\pmb\theta, z)$ are very rare ($\gtrsim 5\,\sigma$) fluctuations of a Gaussian distribution. We also checked a more complex composite likelihood function with a two-sided $\sigma$ following~\citet[][]{Watson:2012mt}:
\begin{equation}
\sigma^2=\left(\sqrt{N_{i}^{\rm sim} + 0.25} \pm 0.5 \right)^2 +\sigma_{\rm sys}^2\,.
\end{equation}
The relative difference in the log-likelihood between the two functional forms is below $10^{-5}$ around the best-fit values and provides statistically indistinguishable results. We stick to the functional form presented in Eq.~\eqref{eq:lkhmf} for simplicity.

The final log-likelihood is computed by summing Eq.~\eqref{eq:lkhmf} over all redshifts, mass bins, and simulations. This amounts to assuming that different mass bins at fixed redshift and simulation outputs at different redshifts are independent of each other. However, we note that when using the output of the same simulation at different redshifts, the results are clearly not independent, as the binning in a given redshift will contain the progenitors or descendants of the object in another redshift. To minimise the impact of the correlation, we follow~\citet{Bocquet:2015pva} and use a time-spacing of roughly $1.7$ Gyr; such spacing is larger than the characteristic dynamical time of galaxy clusters. 

\subsection{\label{sec:forecast}Forecasting \textit{Euclid}'s cluster counts observations}

To understand the impact of the HMF systematic errors on cosmological constraints it is important to realistically forecast the cosmological information to be extracted from the \textit{Euclid} photometric cluster survey. Synthetic cluster abundance data are generated in the following way: we consider a \textit{Euclid}-like light cone covering $15\,000$~deg$^2$, with redshift range $z=[0, 2]$ ~\citep{EUCLID:2011zbd}. As discussed in detail by \cite{Adam:2019}, clusters from the Euclid Wide Survey \citep{Scaramella:2021} will be identified as overdensities in photometric redshift space by applying two cluster finders, which have been demonstrated to be the most accurate in terms of completeness and purity among those considered, namely AMICO \citep{Bellagamba:2018} and PZWav \citep{Gonzalez:2014}. Once clusters are identified an optical richness is assigned to them.

The abundance of halos is sampled assuming our primary calibration for the HMF presented in Sect.~\ref{sec:results}. Optical richness $\lambda$ is assigned to the halos according to the richness--mass relation $\langle \lambda | M_{\rm vir}, z  \rangle$~\citep[see e.g.][]{DES:2015mqu}:
\begin{align}
	\langle \ln \lambda | M_{\rm vir}, z  \rangle =& \ln A_\lambda + B_\lambda \ln{\left(\frac{M_{\rm vir}}{3\times10^{14}\msun}\right)} \nonumber\\ 
	& + C_\lambda \ln{\left(\frac{E(z)}{E(z=0.6)}\right)}\,,
	\label{eq:lambda}
\end{align}
where $E(z)$ is the ratio of the Hubble parameter at redshift $z$ and 0. We assume a richness range $\lambda = [20,2000]$ and a log-normal scatter given by: 
\begin{equation}
	\sigma_{\ln \lambda | M_{\rm vir}, z}^2 = D_\lambda^2.
	\label{eq:lambda-var}
\end{equation}
We use the following fiducial values for the parameters of Eqs.~\eqref{eq:lambda} and~\eqref{eq:lambda-var} $A_\lambda=37.8,\ B_\lambda = 1.16,\ C_\lambda = 0.91,\ D_\lambda = 0.15$. These parameter values were determined by converting the richness--mass relation presented by~\citet{DES:2015mqu} for $M_{\rm 500c}$ (presented in their Table 2) to the virial mass definition, assuming that halos have a Navarro–Frenk–White (NFW) profile~\citep{Navarro:1996gj} and follow the mass--concentration relation given by~\citet{Diemer:2018vmz}. The adopted values are in agreement with the results presented by~\citet{Castignani:2016lvp}.

Lastly, Poisson and sample variance fluctuations are added through a multivariate Gaussian distribution with amplitude given by the covariance model of~\citet{Hu:2002we}, which was validated by~\citet{Euclid:2021api}.

\section{\label{sec:mastering}Setup of \textit{N}-body simulations}

In this section, we define an accurate and precise numerical setup for our primary \textit{N}-body code, \og. We present convergence tests for the adopted configuration and parameter values (Sect.~\ref{sec:parconf}), and the simulation resolution (Sect.~\ref{sec:res}). Lastly, we compare the convergence of our configuration setup to other \textit{N}-body solvers (Sect.~\ref{sec:codecomp}).

\subsection{\label{sec:parconf}Parameter and configuration setting}

The chosen values for the internal key  simulation parameters used within \og\ are presented in Table~\ref{tab:og3-set}. Parameter file variables are set at run time, controlling time integration accuracy, the maximum time step allowed, the tree opening criterion, the value of the critical angle for tree opening, and the force accuracy. The configuration file parameters are set at compilation time. These latter control the maximum distance used to compute short-range forces ({\tt RCUT}) and the matching region between short-range and long-range forces ({\tt ASMTH}). See details of the implementation in \cite{Springel:2005mi} and in the official user guide.\footnote{ \url{https://wwwmpa.mpa-garching.mpg.de/gadget/users-guide.pdf}}
\begin{table}
\caption{Chosen values for the \og\ runs. Parameter file parameters corresponds to the variables set at run time while configuration parameters are set at compilation time. From top to bottom they control: the time integration accuracy, the maximum time step allowed, the tree opening criterion, the angle threshold for tree opening, and the force accuracy, the maximum distance used to compute short-range forces, and the matching region between short-range and long-range forces.}
\label{tab:og3-set}
\begin{center}
	\begin{tabular}{|l|c|}
		\hline
		Variable & Value \\
		\hline
		\multicolumn{2}{c}{Parameter File} \\ \hline
		{\tt ErrTolIntAccuracy} & 0.05  \\ 
		{\tt MaxSizeTimeStep}   & 0.05  \\ 
		{\tt TypeOfOpeningCriterion} & 1 \\
		{\tt ErrTolTheta}   & 0.4  \\ 
		{\tt ErrTolForceAcc}   & 0.0025  \\  \hline
		\multicolumn{2}{c}{Configuration File} \\ \hline
		{\tt RCUT}   & 6.0  \\ 
		{\tt ASMTH}   & 1.25  \\ 
		\hline
		\hline
	\end{tabular}
\end{center}
\end{table}

We checked that our parameter set provides sub-Poisson differences relative to higher precision sets (see Appendix~\ref{sec:App-convergence} for details), deviating by less than a fraction of a percent in the abundance of halos more massive than $3\times10^{13}\,\msun$. In this test and the following ones, we use the cumulative HMF as a test for the convergence instead of the differential HMF for three reasons: firstly, the cumulative version presents less noise than the differential HMF; secondly, we do not want to assume a binning to compute the differential HMF at this stage before discussing the impact of binning (see Sect.~\ref{sec:ics} below); lastly, for cluster cosmology, most of the constraining power comes from measuring the total number of objects above a given mass threshold; the cumulative mass function is the limit where the mass--observable relation tends to a Dirac delta function. We assessed the numerical convergence of the results by directly comparing it to more precise setups where each parameter was set to to half the value presented in Table~\ref{tab:og3-set}.  

Testing the convergence of the simulation results to the configuration settings of parameters presented in Table~\ref{tab:sims} is slightly more delicate than the parameter settings as the configuration variables control the raw structure of the gravity solver algorithm. For instance, if one selects very large values for {\tt RCUT} in \og, the tree-PM algorithm results will converge to the tree. The tree algorithm struggles to provide accurate force calculations if the particle distribution is close to a regular uniform grid, as is commonly the case for initial conditions generated with low-order LPT. In this latter case, tweaking the parameters in order to search for a stable point does not guarantee convergence to accurate results~\citep[see e.g.][]{Springel:2020plp}.

Instead, we test the accuracy of our configuration set through a comparison with \gfour. The rationale for this approach is that the fifth order FMM-PM algorithm deployed in \gfour\ delivers more accurate force calculations than the tree-PM algorithm deployed in \og\ at fixed accuracy parameters and has smaller error correlations due to the possibility of randomising the box centre at every domain decomposition step. We assessed the convergence of the results for three different initial conditions: $512^3$ particles displaced from a uniform regular grid according to 3LPT at $z=24$, the same number of particles using Zeldovich at $z=99$, and $4\times 320^3$ particles displaced from a face centred cubic (FCC) grid according to 3LPT at $z=24$. In all configurations, we verified that the agreement between \og\ and \gfour\ is better than a fraction of a percent for all masses of interest. However, due to the higher degree of symmetry, the FCC configuration shows even better agreement between the two codes. In this configuration, the standard tree algorithm delivers accurate forces from the simulation start, suppressing the spatial correlations of the force calculation errors.

\subsection{\label{sec:res}Resolution convergence}

 Previous works \citep[see e.g.][]{Joyce:2005nr,Marcos:2008rb,Garrison:2016vvp} showed that the mass element discretization and their departure from the fluid limit at initial conditions introduces transient systematic effects on \textit{N}-body simulations. \citet{Michaux:2020yis} showed that the impact of these transient effects is significantly suppressed if simulations start from initial conditions generated with higher order Lagrangian perturbation theory using a grid of elements with more planes of symmetry at the closest redshift prior to shell-crossing. 
 
\begin{figure*}
	\includegraphics[width=0.99\textwidth]{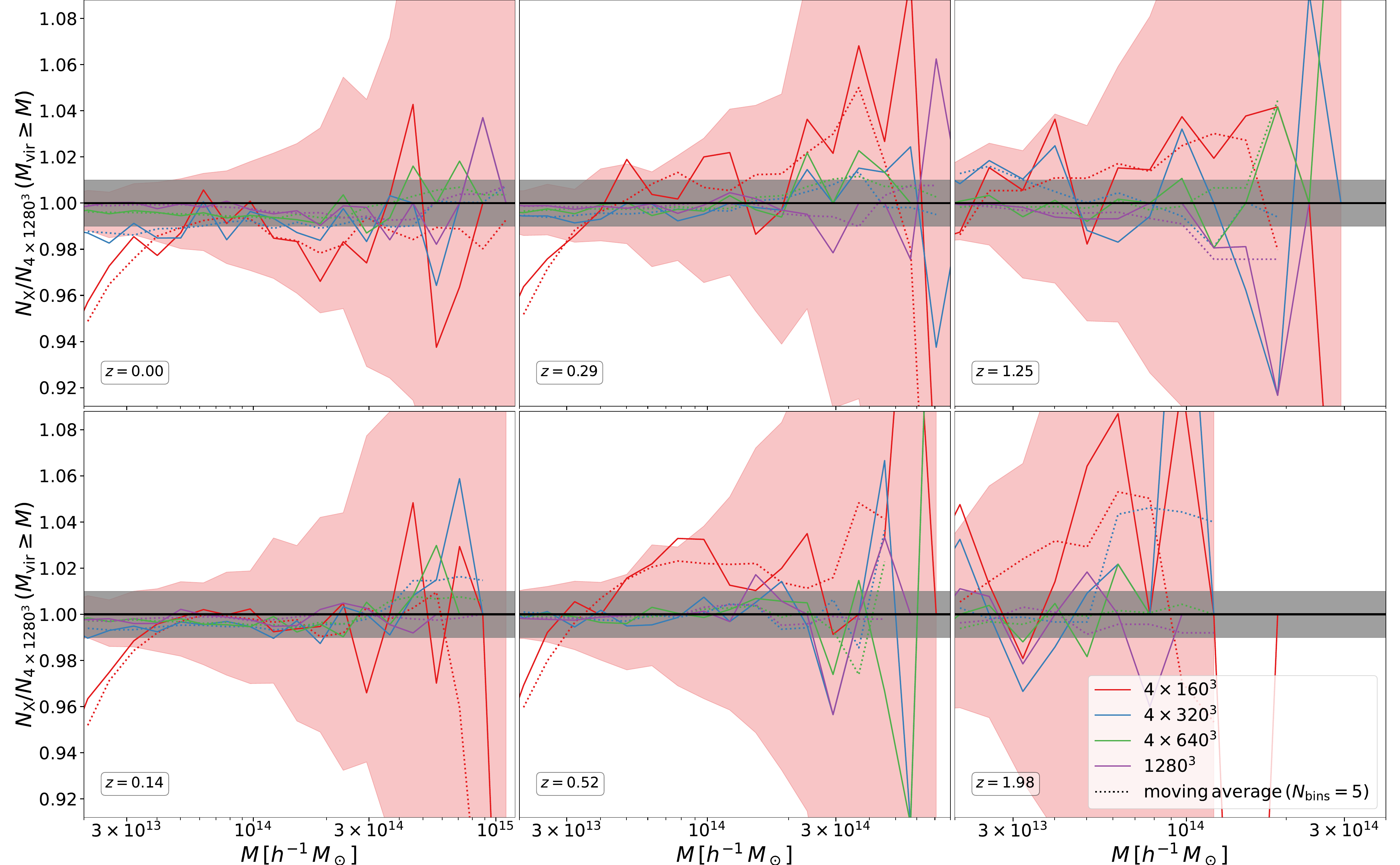}
	\caption{Convergence test for the cumulative mass function with respect to the effects of both transients and modes sampled in the initial conditions, at fixed box size. In each panel, we report results for  five particle masses: $\{6.50\times 10^{11}, 8.13\times 10^{10}, 1.02\times 10^{10}, 5.08\times 10^{9}, 1.27\times 10^{9}\}\,\msun$; respectively $\{4\times160^3, 4\times320^3, 4\times640^3, 1280^3, 4\times1280^3\}$ particles in a box of $500\,\mpc$. The rationale for the chosen number of particles is that $4\times 160^3$ differs by $\sim 2$ percent from $256^3$, and therefore one can directly compare the results presented here with Fig.~\ref{fig:convergence} at similar computational cost. The filled region in grey represents the 1 percent region around unity and the filled region in red marks the 68 percent confidence level assuming Poisson distribution for the abundance of halos in both simulations. Each panel shows the comparison for a different redshift, as labelled.}
	\label{fig:convergence-2}
\end{figure*}

 In Fig.~\ref{fig:convergence-2}, we present the convergence test for the cumulative mass function concerning both transients and modes sampled in the initial conditions. We consider five resolutions, corresponding to particle masses: $\{6.50\times 10^{11}, 8.13\times 10^{10}, 1.02\times 10^{10}, 5.08\times 10^{9}, 1.27\times 10^{9}\}\,\msun$; these values for the particle masses correspond to $\{4\times160^3, 4\times320^3, 4\times640^3, 1280^3, 4\times1280^3\}$ particles in a box of $500\,\mpc$, respectively. The rationale behind such choices for the number of particles is that $4\times 160^3$ differs by $\sim 2$ percent from $256^3$, thus one can directly compare with other commonly used particle numbers at similar computational cost. The gravitational softening is set to one-fortieth of the mean inter-particle distance in all simulations. Initial conditions are created by \monofonic\ using 3LPT at $z=24$ and FCC grid for all simulations but the simulation with $1280^3$ particles, which uses a standard equally spaced grid. The reference in all redshift panels is the simulation run at the highest resolution. For easier inspection, we add the moving average over five bins for each case. We note that the lower resolution simulation tends to suppress the formation of halos at all masses at $z=0$, with a more severe effect at low masses. At $z=0.14$, we observe good agreement with higher resolution simulations for objects more massive than $5\times10^{13}\,\msun$. For the other redshifts, the lower resolution simulation presents either deviations or fluctuations comparable to the Poisson noise (shown in red). However, we note that the Poisson noise was computed assuming no correlation between the simulations, which is not true as the simulations share the same initial conditions. However, as assessing the correlation between the simulations would require many more realisations, we still decided to present the Poisson estimates, considering that they represent a very conservative estimation of the true scatter.
 
 From the next simulation in resolution order, we see a subpercent agreement in the number of objects more massive than $5\times10^{13}\,\msun$. At the same computational cost, the simulation with $4\times 320^3$ (Fig.~\ref{fig:convergence-2}) particles presents a slightly better convergence than $512^3$ (Fig.~\ref{fig:convergence}) at $z=0$. Lastly, comparing the two most costly simulations at $z=1.98$, we observe that the simulation $4\times640^3$ has a more stable agreement with the higher resolution simulation than the simulation with $1280^3$ particles, despite the factor of two increase in the total number of particles; this illustrates one of the advantages of using FCC grids instead of the standard ones for creating initial conditions.

\subsection{\label{sec:codecomp}Comparison of different \textit{N}-body solvers}

\begin{figure*}
\begin{center}
  \includegraphics[height=0.75  \paperheight]{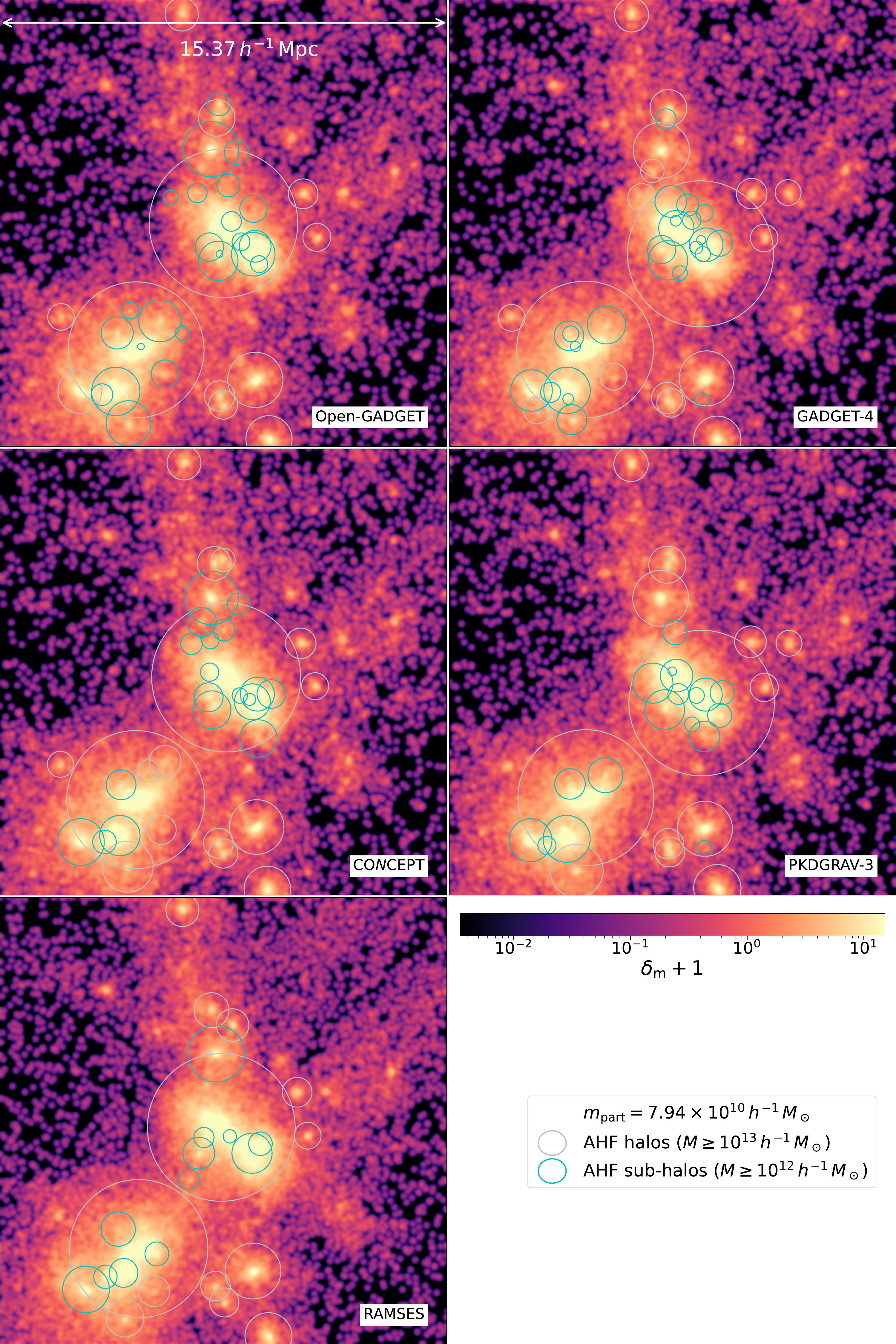}
  \caption{Matter density contrast in a 2D projection of a zoomed-in volume of $(15.37 \mpc)^3$ around the position of the most massive halo found by \ahf\ in the corresponding \og\ simulation. The box size of this volume corresponds to six times the virial radius of the \og\ central halo. Silver and cyan circles denote the virial radius of halos and subhalos identified in this region, respectively. For the sake of clarity, a mass threshold of ($10^{12}$) $10^{13} \msun$ was imposed to select the (sub)halos presented.}
  \label{fig:zoom}
\end{center}
\end{figure*}

To gain insight into the impact of the different gravity solvers on the halo statistics, in Fig.~\ref{fig:zoom} we present the matter density contrast of the 2D projection of a zoomed-in volume of $(15.37\,\mpc)^3$ around the position of the most massive halo found by \ahf. The box size of this volume corresponds to six times the virial radius of the central halo. The corresponding mass of the central halo in each simulation is $\{2.007, 1.904, 2.029, 1.887, 1.975\}\,\times 10^{15}\,\msun$ for \og, \gfour, \pkd, \concept, and \ramses. Silver and cyan circles denote the virial radius of halos and subhalos identified in this region, respectively. For the sake of clarity, a mass threshold of ($10^{12}$) $10^{13}\,\msun$ was imposed to select the (sub)halos presented in Fig.~\ref{fig:zoom}. We observe that all \textit{N}-body codes produce similar distributions of the most massive objects, however, due to slight differences in the evolution, the relation between a large halo and its surrounding depends on the \textit{N}-body code as some structures are detected as an isolated halo in a simulation but as a subhalo in others. 

Figure~\ref{fig:zoom} also shows a stronger code signature on the distribution of substructures identified by AHF. While in this paper we are not interested in any detailed analysis of substructures, we trace them here for the sole purpose of comparing different \textit{N}-body solvers in detail. \og, \gfour, \pkd, and \concept\ produce similar numbers of substructures in large objects, but a very heterogeneous spatial distribution of them. On the other hand, we observe that \ramses\ produces a smoother mass distribution than the other codes, significantly reducing the number of detected substructures. Similar results were previously reported by~\citet{Pascal:2016}. The tendency of \ramses\ to give a smoother mass distribution is also confirmed by measuring the NFW concentration parameter $c$ of the central object: $c=\{5.587, 5.283, 6.170, 5.007, 4.610\}$ for  \og, \gfour, \pkd, \concept, and \ramses. 

\begin{figure}
	\begin{center}
		\includegraphics[width=0.99 \columnwidth]{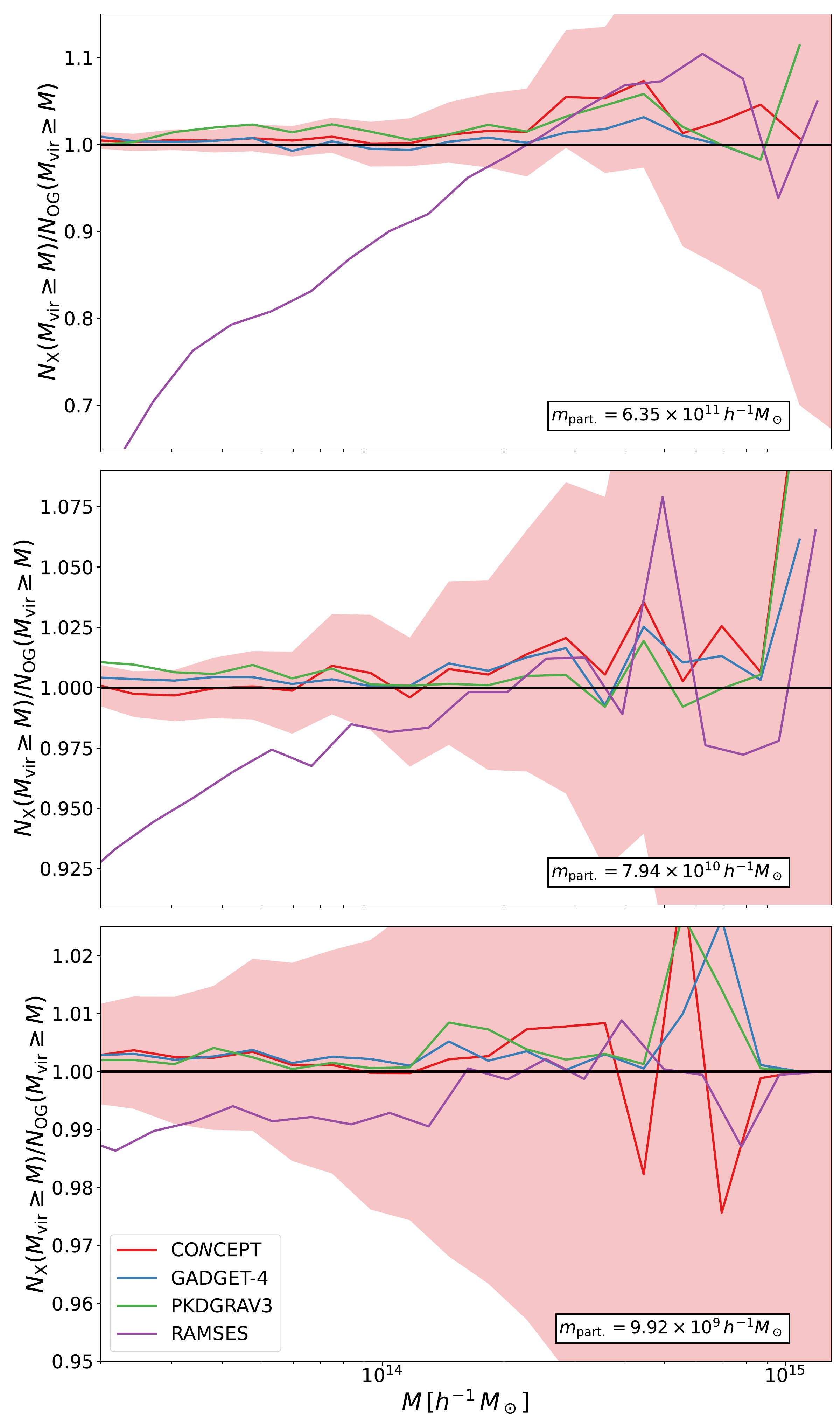}
		\caption{Ratio of the cumulative mass function at $z=0$ to that measured from the \og\ simulation. From top to bottom, the panels correspond to a particle mass resolution of $\{6.35\times 10^{11}, 7.94\times 10^{10}, 9.92\times 10^{9}\}\,\msun$; respectively, $\{256^3, 512^3, 1024^3\}$ particles in a box of $500\,\mpc$. The red filled area represents the 68 percent confidence level for the \concept\ case assuming that the number of halos observed in the two simulations follows an uncorrelated Poisson distribution.}
		\label{fig:comp-nbody}
	\end{center}
\end{figure}

The impact of different \textit{N}-body solvers on the final results is expected to depend on the simulation resolution as critical parameters are usually set as a function of the number of particles; for example the softening length, the maximum refinement level, and the refinement strategy. Figure~\ref{fig:comp-nbody} presents the ratio of the cumulative mass function at $z=0$ to that observed in \og. The three panels correspond to three different mass resolutions of $\{6.35\times 10^{11}, 7.94\times 10^{10}, 9.92\times 10^{9}\}\,\msun$; respectively $\{256^3, 512^3, 1024^3\}$ particles in a box of $500\,\mpc$. For reference, we present the 68 percent confidence level for the \concept\ case assuming that the number of halos observed in the two simulations is described by a Poisson realization of the mass function. 

While the differences of the other codes with respect to \og\ are stable and at the few-percent level for all resolutions for halos more massive than $3 \times 10^{13}\,\msun$, the difference between \ramses\ and \og\ is quite sensitive to resolution. At the lowest resolution considered in Fig.~\ref{fig:comp-nbody}, \ramses\ agrees with the other codes for halos more massive than $\approx 3 \times 10^{14}\,\msun$ while producing a significantly smaller number of less massive objects. At the highest resolution considered, this difference is partially reduced and \ramses\ agrees to better than 1 percent with the other codes for halos more massive than $\approx 4 \times 10^{13}\,\msun$. The suppression in the  halo abundance and the production of smoother density fields are known signatures of the \ramses\ AMR gravity solver. As the adaptive refinement cannot be repeated indefinitely as it is bound to stop before producing empty voxels, the total number of particles in the simulation also sets the maximum force resolution achievable in \ramses. Lastly, from Figs.~\ref{fig:convergence-2} and~\ref{fig:comp-nbody}, we conclude that the HMF convergence with respect to particle mass is achieved first for the other codes before \ramses.

\section{\label{sec:modeling}Modelling}

In this section, we present our modelling for the HMF and the numerical and theoretical systematic effects that influence its assessment, including their dependence on initial conditions~(Sect.~\ref{sec:ics}) and the impact of the simulated volume~(Sect.~\ref{sec:samplevariance}). We revisit the implications of assuming different halo definitions~(Sect.~\ref{sec:halodef}); globally, comparing different halo finders~(Sect.~\ref{sec:hfs}) and internally to a single halo finder (AHF), using different centring~(Sect.~\ref{sec:socentres}) and different halo mass assignment~(Sect.~\ref{sec:somasses}). Lastly, we present our modelling for the non-universality of the HMF~(Sect.~\ref{sec:universality}), modelling it as a function of the shape of the power spectrum~(Sect.~\ref{sec:shape}) and the background evolution~Sect.~\ref{sec:background}).

\subsection{\label{sec:ics}Sensitivity of the HMF to initial conditions}

\begin{figure}
	\includegraphics[width=0.99\columnwidth]{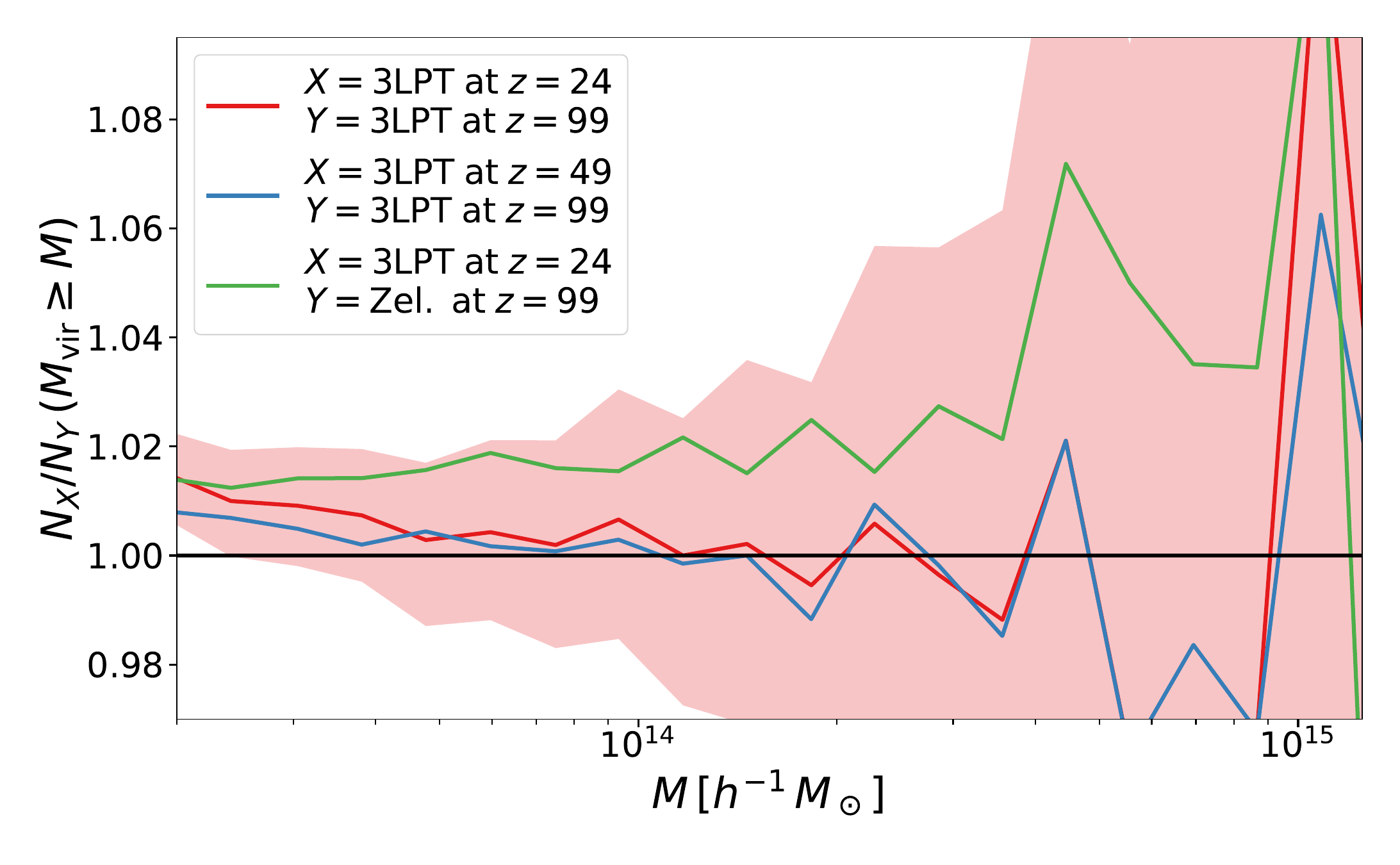}
	\caption{Ratio of the cumulative mass function measured from simulations started with same seed, but different LPT order (3LPT vs. Zeldovich approximation) and starting redshifts ($z=24$, $z=49$, and $z=99$), as indicated by the different coloured lines in the legend.}
	\label{fig:lpt3-vs-zel}
\end{figure}

In Fig.~\ref{fig:lpt3-vs-zel}, we present the ratio of the cumulative mass function computed from simulations started with the same seed, but different LPT orders and redshifts. We consider the 3LPT and Zeldovich approximation and the following starting redshifts: $z=24$, $z=49$, and $z=99$. The rationale for the chosen starting redshifts is two-fold: $z=99$ and $z=49$ have been extensively used in the literature to start simulations using Zeldovich and high-order LPT, respectively. Furthermore,~\citet{Michaux:2020yis} showed that starting the simulation at $z=24$ using 3LPT is a good compromise between the convergence of the LPT (see, for instance, their Fig. 4) and the effect of particle discreteness on several summary statistics. While third-order LPT gives percent-level accuracy on the cumulative mass function for objects more massive than $5\times 10^{13}\,\msun$ independent of the starting redshift considered, setting initial conditions at $z=99$ using the Zeldovich approximation suppresses the formation of structures by $\gtrsim 1$ percent with respect to 3LPT. Our results are in agreement with previous studies~\citep{Crocce:2006ve,Reed:2012ih,Michaux:2020yis}. We also note that, for objects less massive than $5\times 10^{13}\,\msun$, the 3LPT initial conditions set at $z=24$ slightly boosts the formation of structures. The reason for this is two-fold: firstly, as discussed in Sect.~\ref{sec:mastering}, there is a difference in the tree force accuracy calculations when the particle distribution is close to the initial grid. Secondly, shell crossing is known to artificially boost the formation of smaller objects~\citep{Power:2016usj}. In summary, although we have not run tests for 2LPT, from Fig.~\ref{fig:lpt3-vs-zel} we infer that the configuration used for the \aetiology\ set, that is, 2LPT at $z=99$, provides results that agree to better than $\lesssim 2\%$ with 3LPT at $z=24$ as it should range between the green and red lines.

\begin{figure}
	\includegraphics[width=0.99\columnwidth]{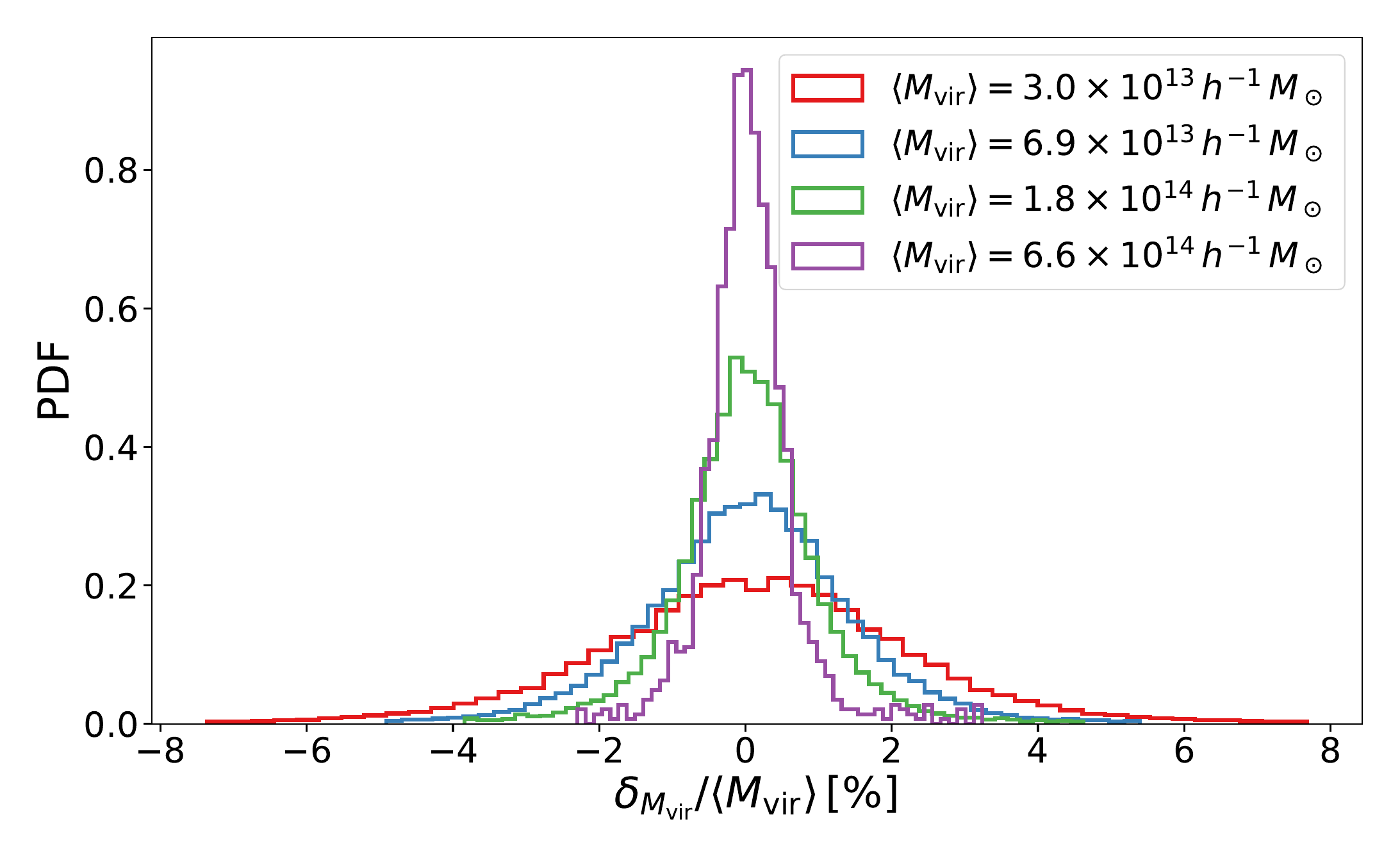}
	\includegraphics[width=0.99\columnwidth]{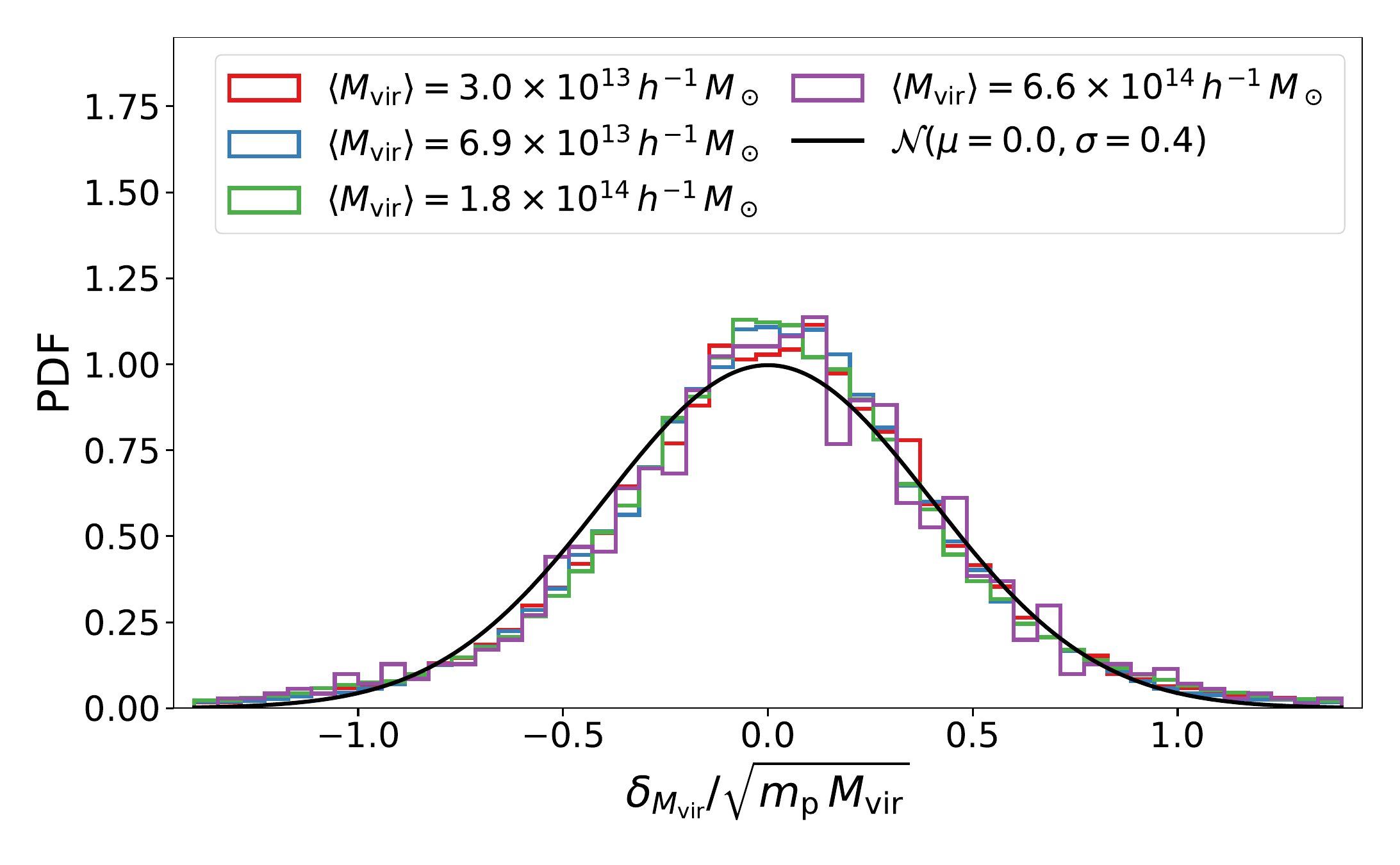}
	\caption{Distribution of the relative difference between masses of halos  at $z=0$ matched between different simulations with initial conditions perturbed by $\lesssim  1$ pc. \emph{Top:} Distribution of the relative mass difference. \emph{Bottom:} Same distribution scaled by the number of particles corresponding to the object mass.}
	\label{fig:roundoff-single}
\end{figure}

Besides the sensitivity to the LPT order used to generate initial conditions, structure formation is also sensitive to small perturbations in the initial positions of particles, such as those caused by round-off errors. This is due to the intrinsically chaotic dynamics obeyed by the several thousands of particles whose orbits are integrated by an \textit{N}-body solver during many dynamical times as they follow the collapse of a halo. The variation of a simulation result on small perturbations in the initial conditions is dubbed the butterfly effect. ~\cite{Genel:2018dfb} thoroughly discuss this effect and how it is amplified in hydrodynamic simulations by thermal processes. Correspondingly, to assess the dependence of the HMF on small perturbations to the initial conditions in purely collisionless simulations, we ran ten simulations for which the initial positions of the particles were randomly displaced by a single unit in the last significant single-precision digit. For the box size of $500~\mpc$, this perturbation corresponds to a random displacement of $\lesssim 1\,{\rm pc}$. We note that, for isolating the effect of perturbing initial conditions from round-off errors due to the use of single precision, those simulations were run using \gfour\ with long integer (i.e., 64 bit) positions. The fluctuations in the HMF caused by the perturbation in initial positions are due to the increasing sensitivity of the non-linear structure formation to initial conditions. However, such small fluctuations cannot disrupt large groups. Instead, they can cause differences in the history of these objects that grow in time resulting in particles accreted by a given group in one simulation ending in a different group in another. In Fig.~\ref{fig:roundoff-single}, we present the distribution of the mass of halos with similar mass matched by their position between different simulations at $z=0$. In the top panel, we present the distribution of the relative mass difference for halos with masses residing in six intervals, each of $0.3$ dex width, between $10^{13}$ and $10^{15}~\msun$. We observe that the impact on halo masses depends strongly on the object mass; whereas halos with masses of a few times $10^{13}~\msun$ can have their masses affected by several percent, the impact reduces to $\approx 1$ percent for the most massive objects. In the bottom panel, we repeat the former exercise scaling the distribution by the number of particles $N_{\rm p}$, corresponding to the object mass: $N_{\rm p}=M_{\rm vir}/m_{\rm p}$, where $m_{\rm p}$ is the particle mass. We note that the effect is rather universal if presented in these terms, and agrees with a Gaussian distribution of zero mean and standard deviation $\sigma \simeq 0.4\,\sqrt{m_{\rm p} \, M_{\rm vir}} \equiv 0.4 \, m_{\rm p} \sqrt{ N_{\rm p}}$. Thus, the limited precision in the initial conditions propagates to a sub-Poisson fluctuation in the number of particles belonging to a given halo at low redshift. In relative terms, the effect is larger for objects with fewer particles and represents only a subpercent effect for objects with more than $1500$ particles.

\begin{figure}
	\includegraphics[width=0.99\columnwidth]{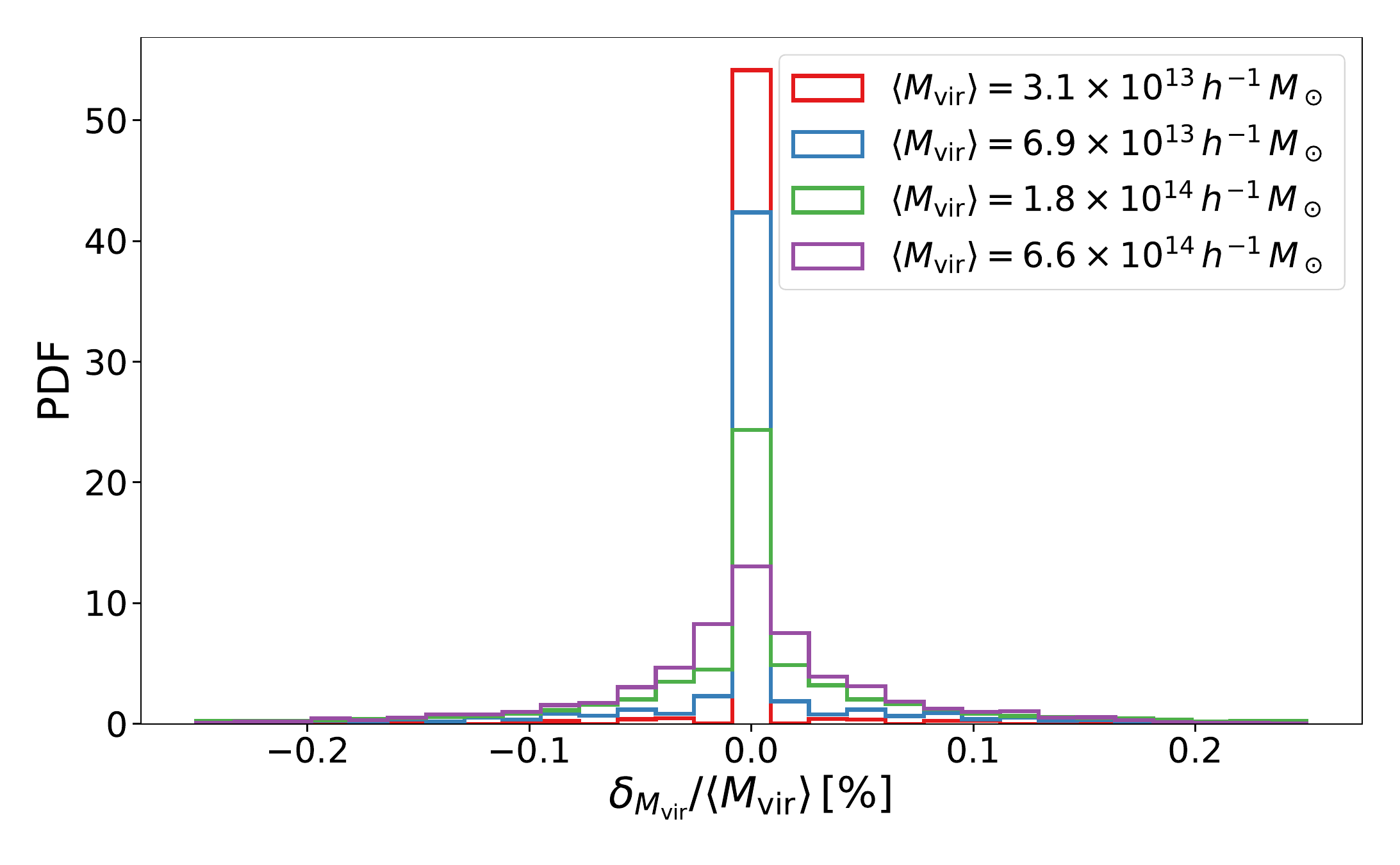}
	\includegraphics[width=0.99\columnwidth]{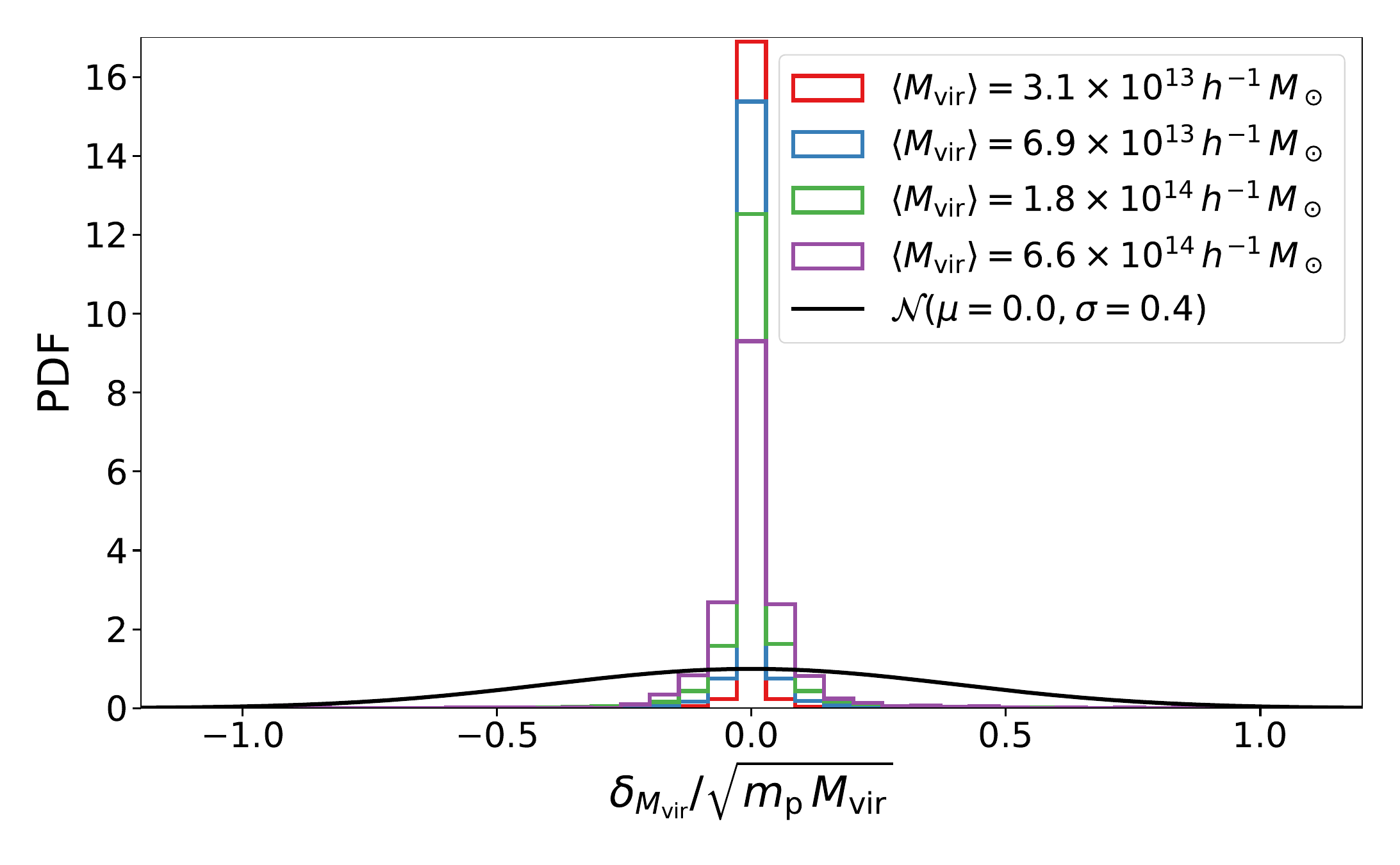}
	\caption{Same as Fig.~\ref{fig:roundoff-single} but with double precision. }
	\label{fig:roundoff-double}
\end{figure}

Figure~\ref{fig:roundoff-double} presents the results of the same analysis, when double precision (i.e., 64 bit floating-point) is used. For double precision the effect is not only strongly suppressed, but is also no longer universal when scaled by the number of particles. The significant suppression in the scatter of the mass of the objects indicates that double precision should be used to setup initial conditions of cosmological simulations and internally by the gravity solver, assuming one can afford the factor two increase in the memory and storage. As storing double-precision outputs is not the default option for several codes, the propagation of double-precision round-off errors will not be further discussed in this work.

\begin{figure}
	\includegraphics[width=0.99\columnwidth]{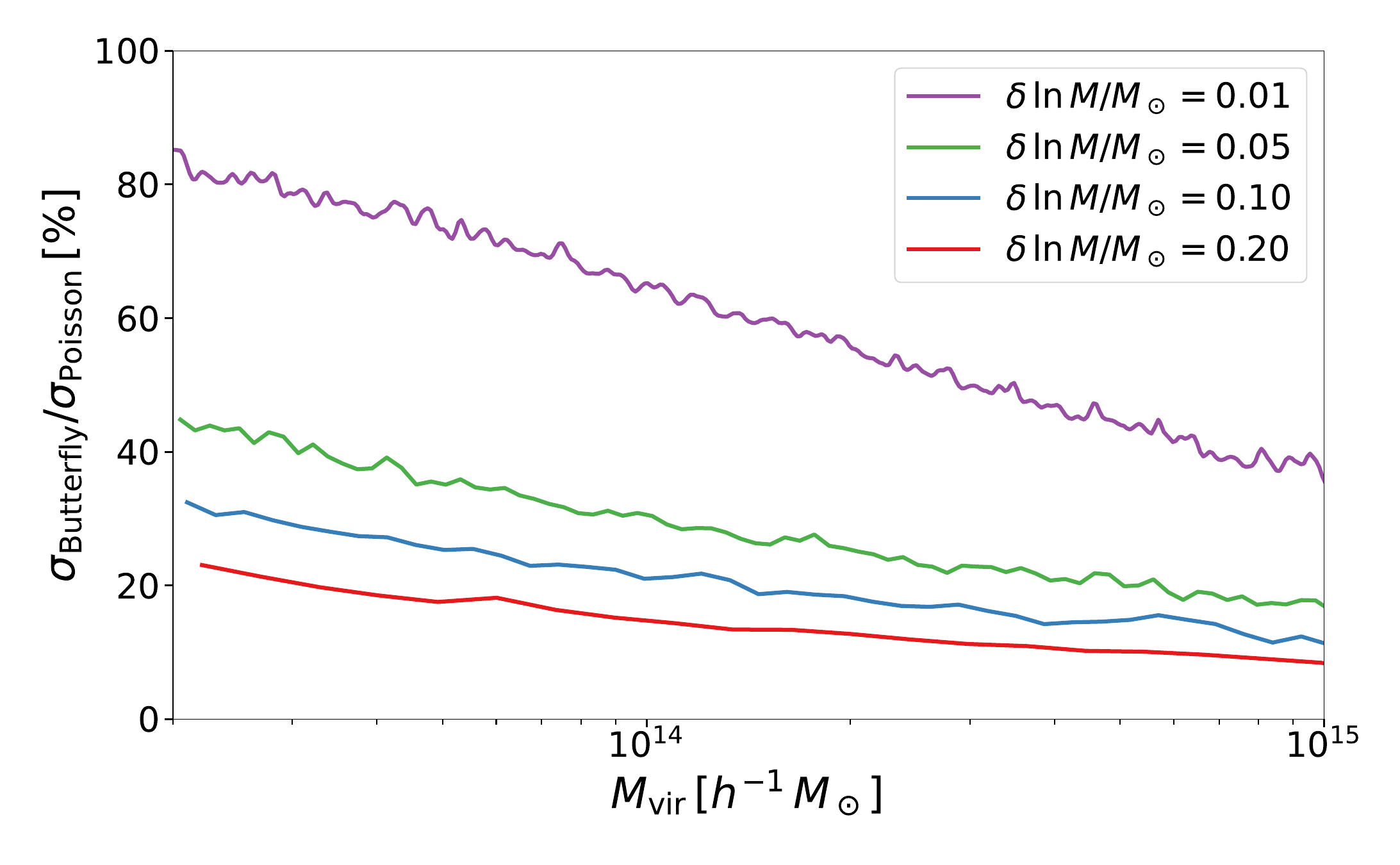}
	\caption{The r.m.s. variation in the HMF induced by the noise in the initial conditions, normalised to the expected Poisson noise, as a function of halo mass. Curves with different colours correspond to different binnings of halo masses, as indicated in the legend. }
	\label{fig:buttefly-2}
\end{figure}

The mass fluctuations presented in Fig.~\ref{fig:roundoff-single} can be strongly amplified in binned statistics depending on the bin width. In Fig.~\ref{fig:buttefly-2}, we show the r.m.s. variation in the HMF induced by the noise in the initial conditions, normalised to the expected Poisson noise, as a function of halo mass. Different curves correspond to different binnings of halo masses.  As is commonly done for the calibration of the HMF, we considered logarithmically spaced mass bins. The results shown in Fig.~\ref{fig:buttefly-2} were obtained by creating a synthetic halo catalogue with masses distributed according to the HMF presented by~\citet{Tinker:2008ff}. After that, several halo catalogues were created by perturbing the halo masses according to the distribution presented in Fig.~\ref{fig:roundoff-single}. Lastly, the HMF was extracted from these catalogues binning the halos in mass and dividing by the volume times the mass bin width. Although we have to tacitly assume a volume in the previous exercise, we have observed negligible impact of the nominal volume assumed by considering two different volumes: $(2000\,\mpc)^3$ and $(500\,\mpc)^3$.

From Fig.~\ref{fig:buttefly-2} it is clear that the binning has to be carefully chosen to reduce the butterfly impact. Ideally, the bin width should be much larger than the scatter caused by the round-off errors. This leads to the condition:
\begin{equation}
	\delta \ln M \gg 1/\sqrt{N_{\rm min}}\,,
\end{equation}
where $N_{\rm min}$ is the number of particles of the smallest object of interest.

\subsection{\label{sec:samplevariance}Impact of the simulated volume}

The effect of the simulation volume on the HMF is two-fold: firstly, the computation of the mass variance $\sigma^2(M)$ presented in Eq.~\eqref{eq:massvariance} has to be truncated to the fundamental mode of the cosmological box; secondly, by construction, only a few independent modes are contained in the simulated volume for the first multiples of the fundamental mode, thus introducing an effect of sample variance into the computation of the mass variance at large halo masses.

Quantifying the effect of the simulated volume in the HMF directly with simulations would require a much more extensive set of simulations than the ones used in this paper. Instead, we assess the impact of the simulated volume through its impact on the mass variance calculation and propagate this effect to the HMF, assuming the analytical prescription.

\begin{figure}
    \centering
    \includegraphics[width=\columnwidth]{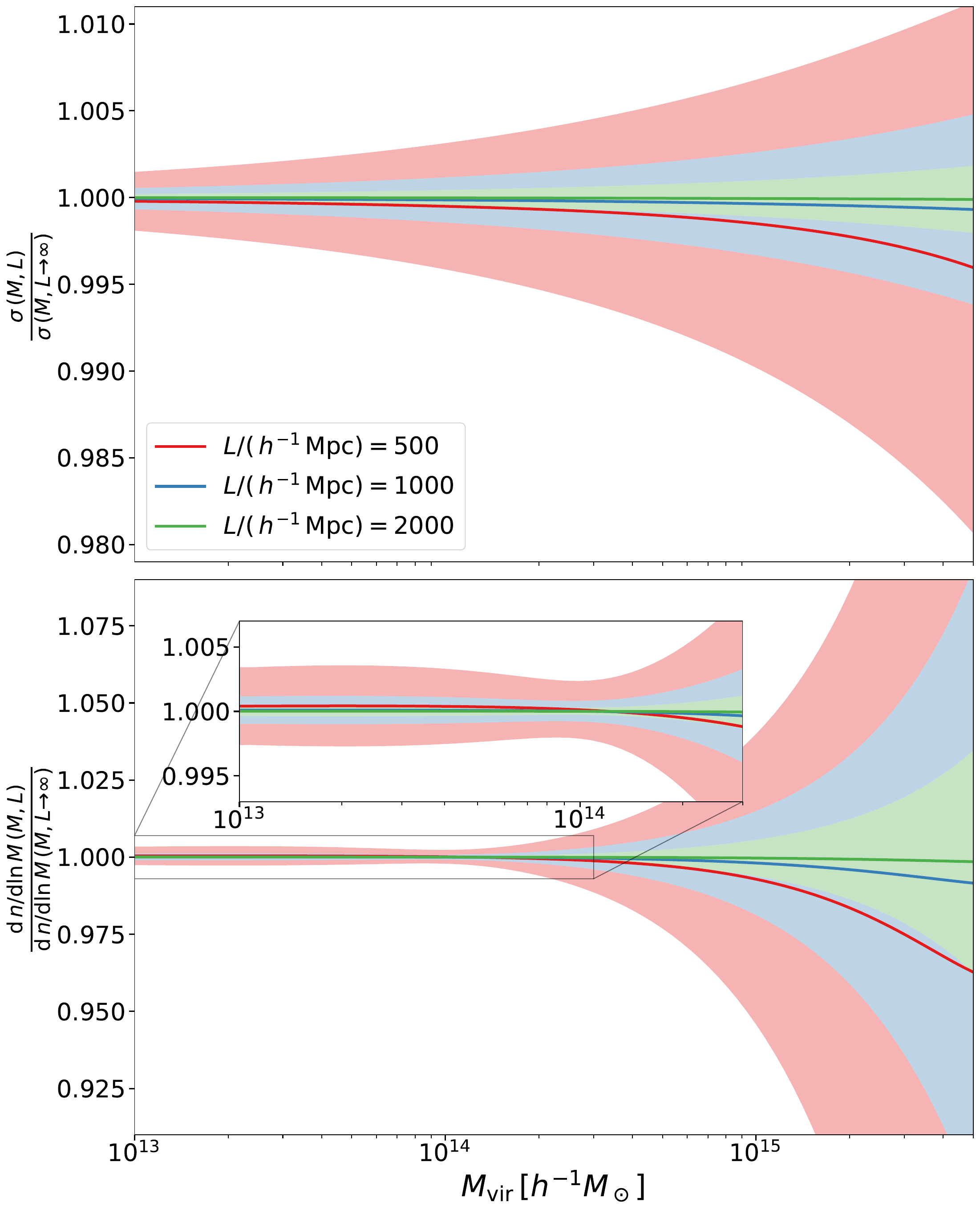}
    \caption{Impact of the simulation box size $L$ on the mass variance (\emph{top panel}) and the differential HMF (\emph{bottom panel}) for three different cases $L=\{500, 1000, 2000\}\,\mpc$ depicted in red, blue, and green, respectively. Solid lines represent the mean effect due to the truncation of the mass variance integration to the fundamental mode. The corresponding shaded regions correspond to the 68 percent interval due to the sample variance.}
    \label{fig:samplevariance}
\end{figure}

In Fig.~\ref{fig:samplevariance}, we present the impact of the simulation box size $L$ on the HMF for three different cases $L=\{500, 1000, 2000\}\,\mpc$ shown in red, blue, and green, respectively. Solid lines represent the mean effect due to the truncation of the mass variance integration to the fundamental mode. The corresponding shaded regions correspond to the 68 percent interval due to the sample variance. In the top panel, we present the effect on the calculation of the mass variance itself, while in the bottom we propagate the impact on the mass variance to the differential HMF. The 68 percent regions were determined creating 1000 synthetic realisations of a matter power spectrum assuming the $C0$ cosmology for each box. The matter power spectrum was sampled between the fundamental and the 4096th modes. Sample variance was added to the power spectrum perturbing it with a Gaussian fluctuation of amplitude $\sigma_P(k)$ given by:
\begin{equation}
\sigma_P(k)=\sqrt{\frac{1}{N(k)}}\, P_{\rm m}(k)\,,
\end{equation}
where $N(k)=2\pi\,(\delta_k/k)^2$ is the number of independent modes inside the bin with width $\delta_k$ in $k$-space. Finally, the differential HMF in the bottom panel of Fig.~\ref{fig:samplevariance} has been calculated using the model of~\citet{Tinker:2008ff}.

In Fig.~\ref{fig:samplevariance}, we observe that the exponential dependence of the HMF on the mass variance at high-masses significantly amplify the impact of both the truncation and sample-variance impact on the mass variance. The absence of modes larger than $500\,\mpc$ causes the suppression of the formation of objects more massive than $8\times10^{14}\,\msun$. No significant suppression on the formation of objects is observed for the two larger box sizes. On the other hand, the impact of sample variance on the HMF is below one percent for all boxes considered for halos lighter than $3\times10^{14}\,\msun$. For the $1000\,\mpc$ and $2000\,\mpc$ box sizes, the $1$ percent threshold is exceeded for halos more massive than $8\times10^{14}\,\msun$ and $2\times10^{15}\,\msun$, respectively. 

One way of taking into account the effect of the box size on the calibration of the HMF is to calculate the mass variance, using the matter power spectrum computed from the initial conditions~\citep[see e.g.][]{Despali:2015yla}. This approach presents a few challenges: for instance, the computation of the power spectrum at initial conditions might be affected by the mesh used to create the initial displacement field -- assuming one has not used a `glass'\footnote{Glass pre-initial conditions are random samples that are dynamically evolved assuming a gravitational interaction with reversed sign until they settle in a quasi-equilibrium state.} unstructured mesh from which to displace particles. Also, one has to choose the $k$-binning for the computation of the matter power spectrum wisely, as an overly thin shell would produce a noisy measurement, while an excessively broad shell would affect the mass variance integral accuracy. We instead advocate that to keep the impact of sample variance to a minimum, one should use boxes larger than $1000\,\mpc$ and produce initial conditions with fixed amplitudes of the initial conditions random field to the desired theoretical power spectrum, as presented in~\citet{Angulo:2016hjd}. Following this approach one circumvents the above-mentioned challenges as the realised and theoretical power spectrum match exactly.

\subsection{\label{sec:halodef}Impact of the halo definition}

\subsubsection{\label{sec:hfs}Sensitivity to different halo finders}

\begin{figure*}
	\centering
	\includegraphics[width=0.99\textwidth]{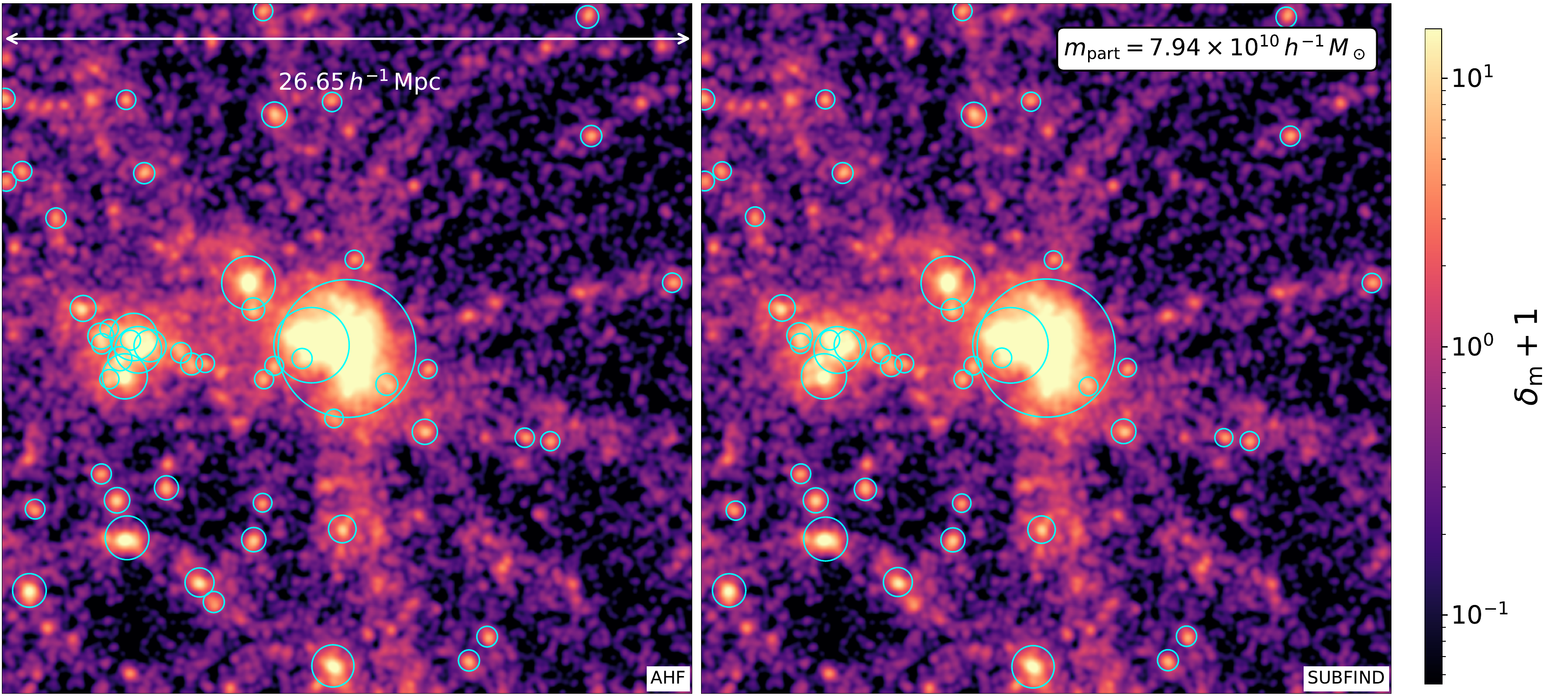}
	\caption{Comparison between \ahf\ and \subfind\ halos extracted from the same $(26.65\,\mpc)^3$ volume. The largest and smallest halo masses depicted are $2\times 10^{15}\,\msun$ and $10^{13}\,\msun$, respectively. The cyan circles denote the virial radius of halos identified in this region.}
	\label{fig:delta-hf}
\end{figure*}

For a visual inspection of the effect of different halo finders on the HMF, Fig.~\ref{fig:delta-hf} shows a comparison between \ahf\ and \subfind\ halos identified within a $(26.65\,\mpc)^3$ volume extracted from a $500~\mpc$ box with $512^3$ particles started at $z=99$ using the Zeldovich approximation. The mass of the largest halo located at the centre is $2\times 10^{15}\,\msun$ while the smallest halo marked in each panel is $10^{13}\,\msun$. Whereas \ahf\ and \subfind\ both find the same large groups, for smaller groups, we  notice a non-negligible suppression on the number of objects, with \subfind\ tending to group together smaller objects into a larger one. The effect is more evident along the stream on the centre left of the larger object presented in Fig.~\ref{fig:delta-hf}. The tendency of pure FOF-based methods to merger smaller, dynamically distinct objects along tidal streams, building  `particle bridges' between structures, is well known~\citep[see e.g.][and references therein]{Knebe:2011rx}.

\begin{figure}
	\includegraphics[width=0.99\columnwidth]{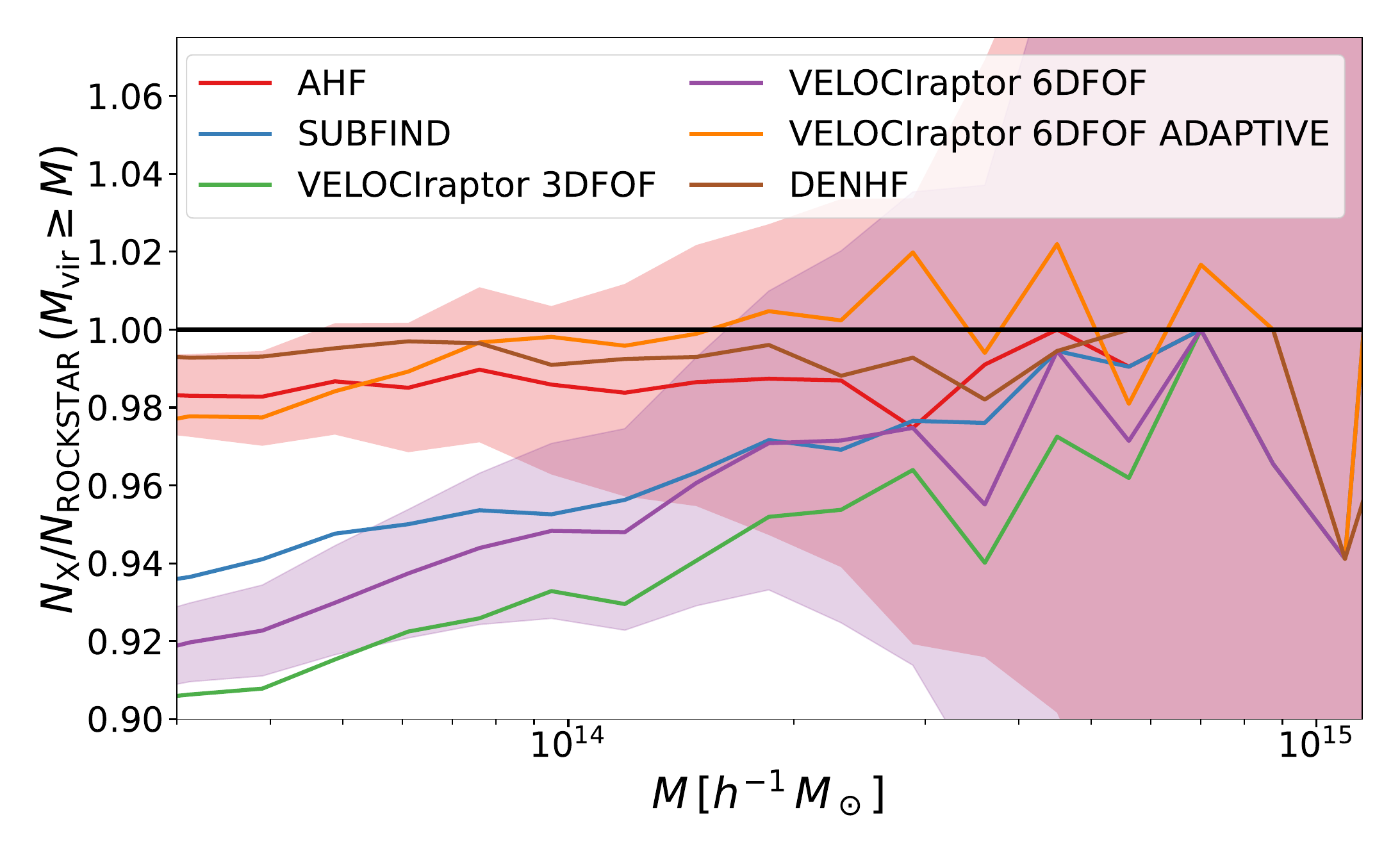}
	\caption{Ratio of the cumulative mass function extracted from \ahf, \subfind, \velociraptor, and \denhf\ to that extracted from \rockstar. The filled regions in red and purple represent the $68\%$ confidence regions for the \ahf\ and \velociraptor\ (6DFOF Adaptive), respectively, assuming that the number of objects in each catalogue follows a Poisson distribution. }
	\label{fig:comp-hf}
\end{figure}

Figure~\ref{fig:comp-hf} presents the ratio of the cumulative mass functions extracted from \ahf, \subfind, \velociraptor, and \denhf\ to that extracted from \rockstar. The filled regions in red and purple represent the $68$ percent confidence level region, for \ahf\ and \velociraptor\ (6DFOF Adaptive), respectively, assuming that the number of objects in each catalogue follows a Poisson distribution.
Figure~\ref{fig:comp-hf} clearly shows a separation of the algorithms considered here into two groups, with the 3DFOF and non-adaptive 6DFOF suppressing the number of halos less massive than $10^{14}~\msun$. 

We caution that it is not our goal here to model the specifics of each halo finder. The reason is two-fold; firstly, this is a complex task, as many parameters control the different algorithms. Secondly, investigating how the results from different finders compare to each other when changing the respective parameters, even in great detail, would not address the fundamental question of how does the definition of a halo adopted in simulations compares with the observed clusters. The analysis presented is simply designed to deliver a flexible model for the HMF that can accommodate different halo definitions. With such a model calibrated against different halo catalogues, we will assess the impact of different definitions of halos in the HMF calibration on \textit{Euclid}'s cluster counts analysis.

\subsubsection{Impact of the halo centre}
\label{sec:socentres}

We assessed the impact of the choice for the halo centre on the cumulative mass function using \ahf\ on one of the \tease\ simulations. \ahf\ allows the user to choose the  prospective halo centre alternatively as the geometrical centre of the refinement patch, the cell with the lowest potential, the cell with the highest density, or the centre of mass of the particles inside the refinement patch. The latter is our default choice, following the official \ahf\ documentation.

We verified that the \ahf\ cumulative mass function has a percent-level robustness to the choice of the halo centre. We remind that \velociraptor\ (based on a 3DFOF) and \subfind\ differ only for the choice of the halo centre. Nevertheless, they differ from each other (see Fig.~\ref{fig:comp-hf}) by an amount that is larger than the differences between different halo centres in the AHF. This is, again, due to particle bridges that connect two dynamically distinct objects, which causes a stronger impact on the halo centre choice.

\subsubsection{\label{sec:somasses}Total halo mass versus bound mass}

\begin{figure}
	\includegraphics[width=0.99\columnwidth]{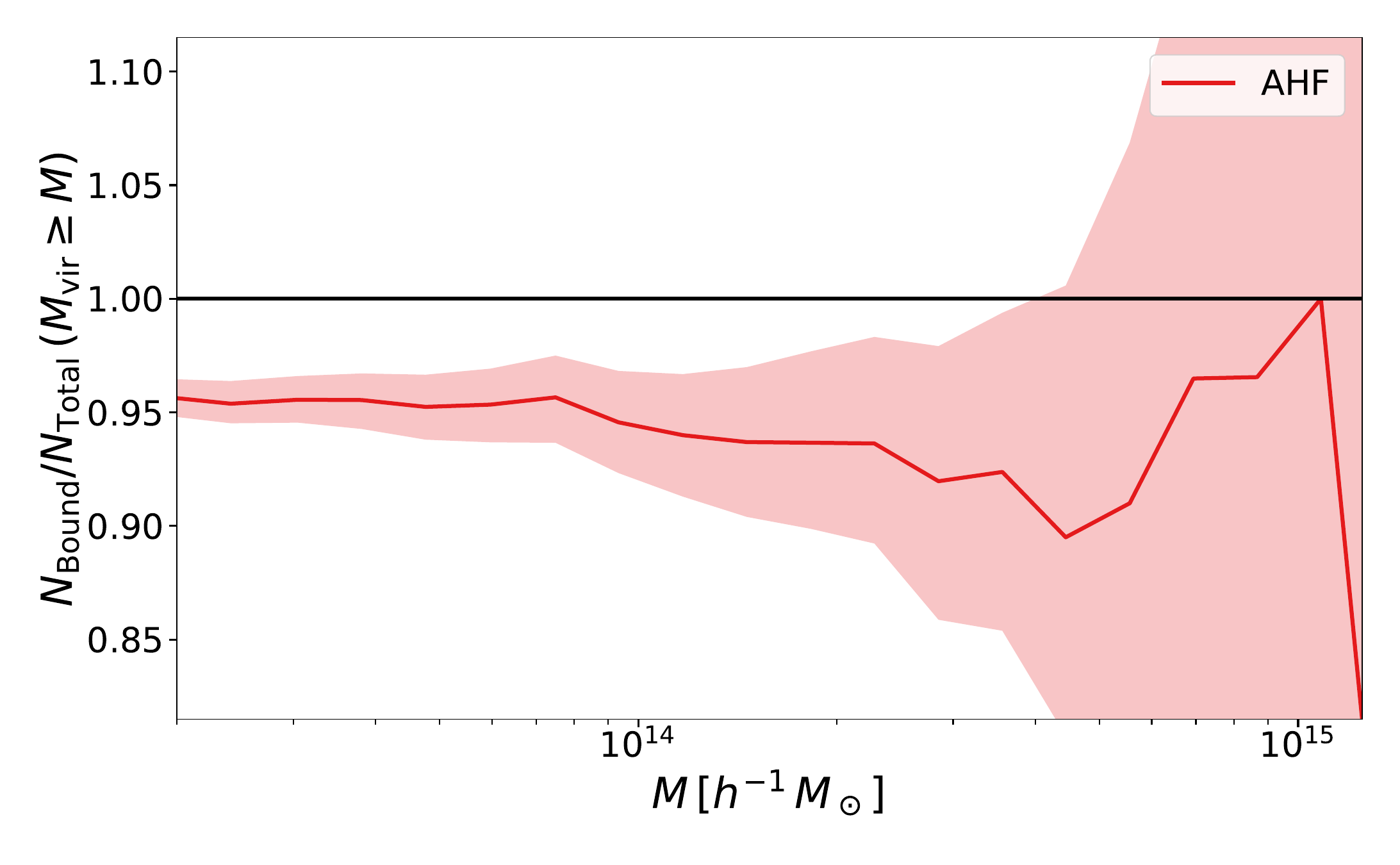}
	\caption{Difference in the cumulative mass function when all particles inside the spherical overdensity are considered versus just the bound particles. }
	\label{fig:masses}
\end{figure}

All halo finders considered in this paper include all particles within a given radius, when computing the spherical overdensity masses, regardless of whether the particles are bound to a halo or not. In order to test the effect of this assumption, in Fig.~\ref{fig:masses}, we present the difference in the cumulative HMF between including all particles inside the spherical overdensity and only the contribution from bound particles. \ahf\ determines as bound the particles with velocities smaller than the local escape velocity multiplied by a constant parameter \texttt{VescTune} which we set to unity. The simulation used for this test is the same as the one used in Sect.~\ref{sec:socentres}. Not surprisingly, assigning to halos only bound particles reduces the cumulative mass function by about 5 percent. That is due to the fact that the gravitationally bound mass within a fixed halo radius is by construction smaller than the total mass within the same radius. It is important to note that both are valid definitions to weigh a halo. The most adequate choice will depend on the method used to measure cluster masses from observations. Arguably, if one works with, for instance, masses obtained from the gravitational lensing, the total mass definition should be adopted, as this is the one that contributes to the lensing signal. In the remainder of this paper, we refer only to the total mass definition.

\subsection{\label{sec:universality}Non-universality of the virial mass function}

\citet{Despali:2015yla,Diemer:2020rgd,Ondaro-Mallea:2021yfv} showed that the HMF preserves most of its universality when described as a function of the virial mass, as predicted by the spherical collapse model \citep{Eke:1996ds,Bryan_1998}. Still, departure from  universality can reach up to $20$ percent for the high-end tail of the mass function~\citep[see e.g.][]{Diemer:2020rgd}. 

In the following sections, and for the specific purpose of tracing the origin of any departure from universality, we use the \aetiology\ set of simulations to model the dependence of the virial HMF both as a function of the shape of the matter power spectrum and of the background evolution. Unless stated otherwise, the simulations used here were run using \gfour\, with initial conditions generated at $z=99$ according to 2LPT. Halo catalogues were extracted on the fly with the \gfour \subfind\ implementation. Halos were binned according to their mass using $50$ log-spaced intervals in the number of particles. 

For the calibration of the HMF, we impose a mass cut that corresponds to a minimum number of 300 particles so as to minimise the noise in the identification of low-mass halos~\citep[see e.g.][]{Leroy:2020fzc}. We assume the likelihood presented in Sect.~\ref{sec:likelihood} with $\sigma_{\rm sys}=0.01\, N_i^{\rm sim}$. This choice for the systematic error means that the relative error has a floor of $1$ percent in the total error budget, thus avoiding over-weighting mass bins with many halos, for which the purely Poisson error under-predicts the true data variance, as discussed in Sects.~\ref{sec:ics} and~\ref{sec:samplevariance}. Its value was chosen so that the best-fit neither over- nor under-fits the simulations over the entire redshift range considered, that is, keeping the fit-quality constant. We use the snapshots at $z=\{2.00, 1.25, 0.90, 0.52, 0.29, 0.14, 0.0\}$.

\subsubsection{\label{sec:shape}Dependence on the power spectrum slope}

\begin{figure}
	\includegraphics[width=0.99\columnwidth]{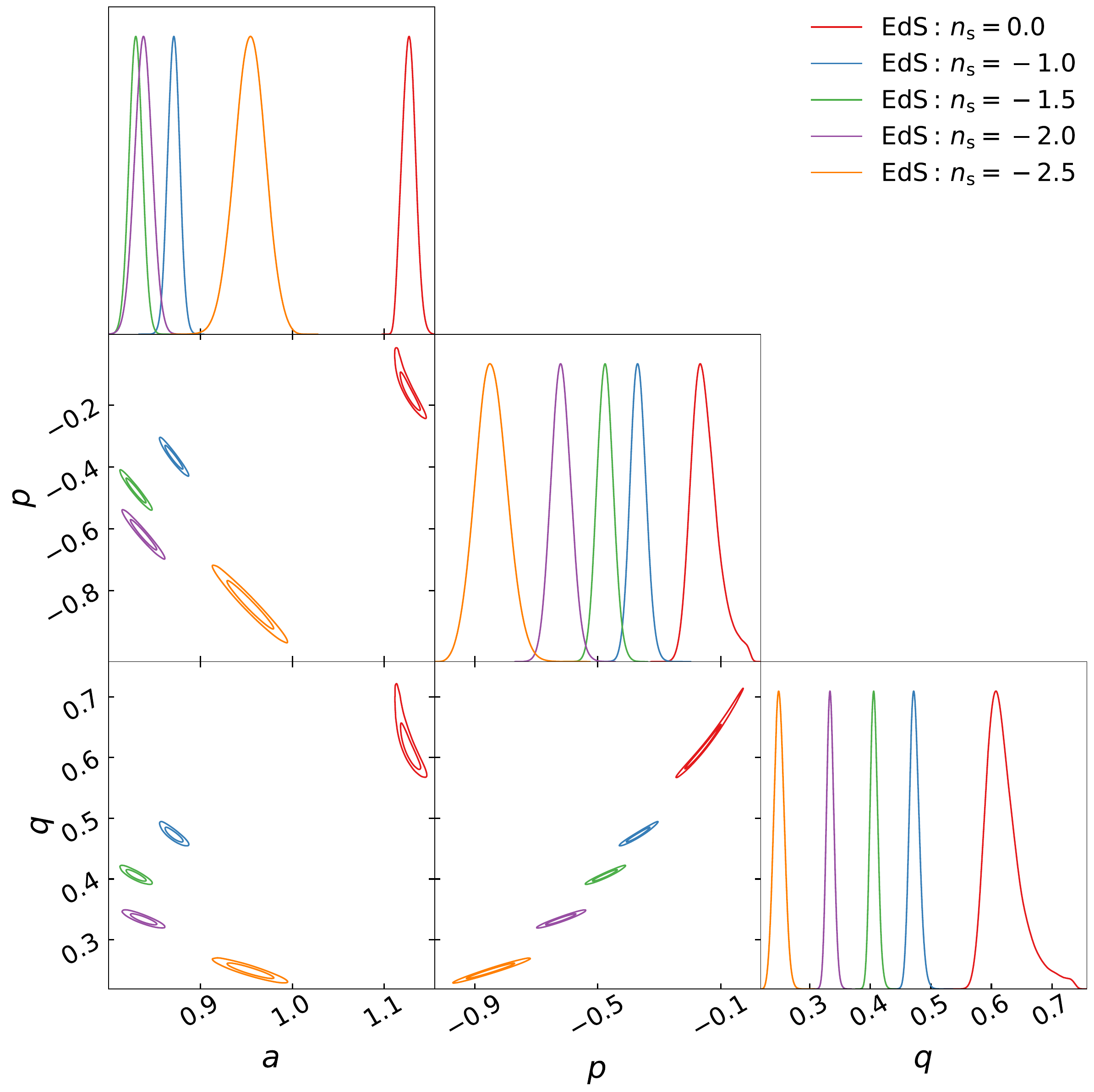}
	\caption{The $68\%$ and $95\%$ confidence level contours of the calibration carried out independently in several simulations assuming a power-law linear power spectrum in an EdS background, i.e., $P_{\rm m}(k)\approx k^{n_{\rm s}}$ and $\Omega_{\rm m}=1$.}
	\label{fig:model-shape}
\end{figure}

In order to understand the dependence of the HMF parameters on the power spectrum shape, in Fig.~\ref{fig:model-shape}, we present the $68$ and $95$ percent confidence level contours of the calibration carried out independently in several simulations assuming a power-law linear power spectrum in an EdS background, that is, $P_{\rm m}(k)\propto k^{n_{\rm s}}$ and $(\Omega_{\rm m}, \Omega_{\Lambda})=(1,0)$.

The first interesting result is that, even for self-similar cosmologies, the HMF is not universal against changing the spectral index. While the parameters $\{p, q\}$ of  Eq.~\eqref{eq:mult} seem to have a linear dependence on $n_{\rm s}$, the parameter $a$ exhibits a non-monotonic dependence with a local minimum around $n_{\rm s} = -1.75$. The overall simple and well-behaved dependence of the parameters of the HMF  on the spectral index over a range of values covering all the relevant regimes for structure formation motivates our approach to select a flexible fitting function that precisely accommodates the HMF shape on a self-similar cosmology. At the same time, the non-universality is modelled through the explicit dependence of the HMF parameters on cosmology.

\subsubsection{\label{sec:background}Dependence on the background evolution}

\begin{figure}
	\includegraphics[width=0.99\columnwidth]{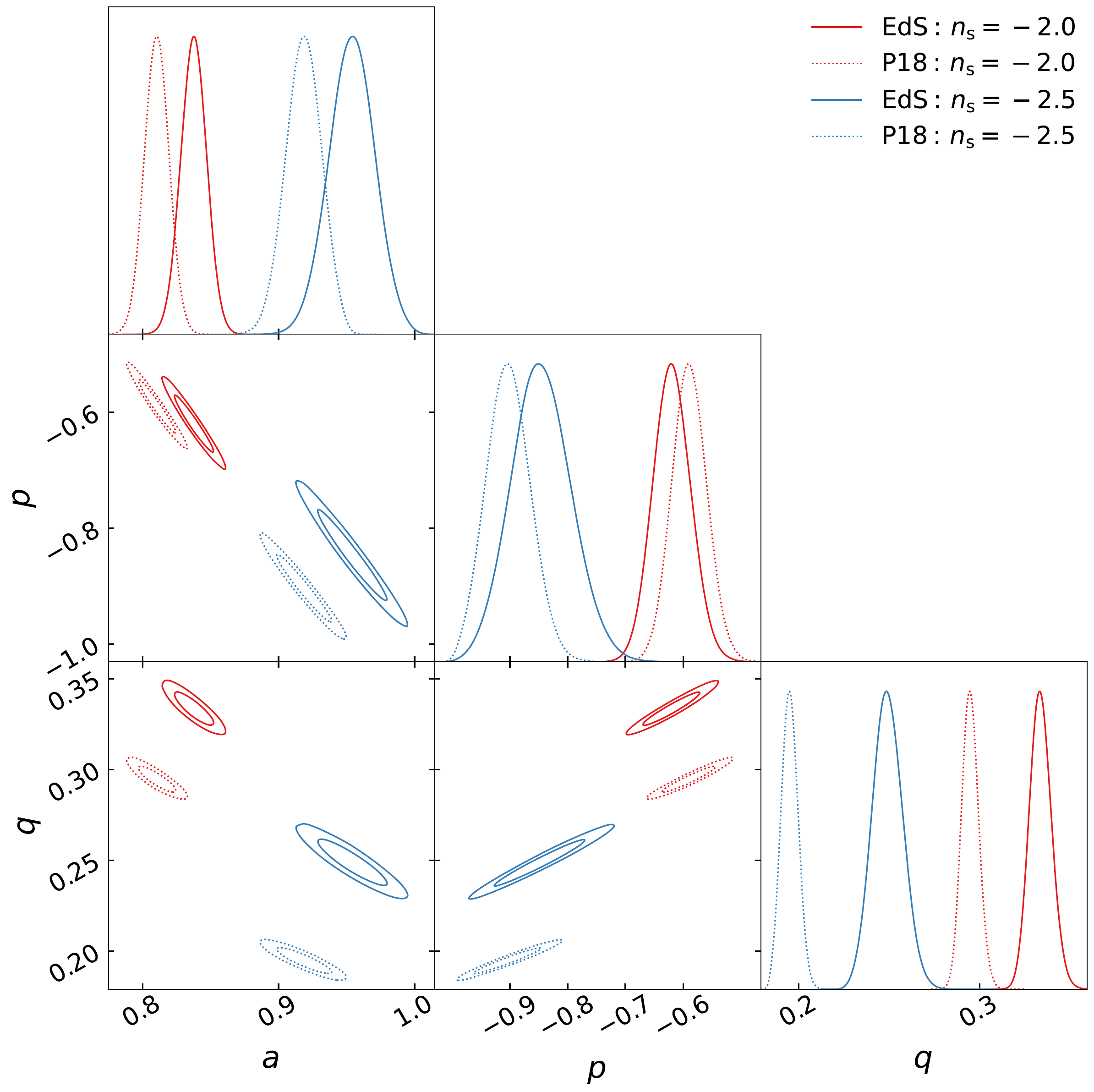}
	\caption{The $68\%$ and $95\%$ confidence level contours of the calibration carried out independently in simulations with $n_{\rm s}=\{-2.0, -2.5\}$ and background evolution given by either a cosmology in agreement with \citet{Planck:2018vyg} or EdS. }
	\label{fig:model-bkg}
\end{figure}

In order to understand the dependence of the HMF parameters on the background evolution we follow an approach similar to that presented in \citet{Ondaro-Mallea:2021yfv}. In Fig.~\ref{fig:model-bkg}, we present the $68$ and $95$ percent confidence level contours of the calibration carried out independently in simulations with $n_{\rm s}=\{-2.0, -2.5\}$ and background evolution given by either a cosmology in agreement with the Planck 2018 (P18) results for the cosmological parameters~\citep{Planck:2018vyg} or EdS. For the specific purpose of this exercise, the chosen values for the spectral index are not important, and other combinations produce similar trends; still, we choose $n_{\rm s}=\{-2.0, -2.5\}$ as this range corresponds roughly to the slope of the $\Lambda$CDM matter power spectrum on cluster scales.

At fixed $n_{\rm s}$, we note that $p$ is consistent with being independent of the background evolution. On the other hand, the parameters $\{a, q\}$ consistently show a dependence on the chosen background, with departures from the EdS scenario providing smaller values of such parameters.

\subsubsection{The HMF model}

From the above results, we define the following model to capture the dependence of the multiplicity function on redshift and power spectrum shape as:
\begin{align}
	\label{eq:parevol1}
	a &= a_R \,  \Omega_{\rm m}^{a_z}(z) \\
	p &= p_1 + p_2 \,  \left( \frac{\de \ln \sigma}{\de \ln R} + 0.5 \right) \\
	q &= q_R \, \Omega_{\rm m}^{q_z}(z) \,.
\end{align}
where:
\begin{align}
a_R &= a_1 + a_2 \,  \left( \frac{\de \ln \sigma}{\de \ln R} + 0.6125 \right)^2 \\
q_R &= q_1 + q_2 \,  \left( \frac{\de \ln \sigma}{\de \ln R} + 0.5 \right) \,.
\label{eq:parevol2}
\end{align}
The growth rate, that is, $\de \ln G/\de\ln (1+z)^{-1}$ where $G(z)$ is the linear growth of density perturbations, was used by \citet{Ondaro-Mallea:2021yfv} to characterise the non-universality of the HMF instead of $\Omega_{\rm m}(z)$. For the cosmological models considered here this is well approximated by~\citep[see e.g.][]{amendola2010dark}:
\begin{equation}
  \frac{\de \ln G}{\de \ln (1+z)^{-1}}=\Omega_{\rm m}^\gamma(z)\,,  
\end{equation}
with $\gamma=0.55$. Therefore,  Eqs.~\eqref{eq:parevol1} to~\eqref{eq:parevol2} produce exactly the same results for the cosmological models studied here once one substitutes:
\begin{equation}
\Omega_{\rm m}(z)\rightarrow \left(\frac{\de \ln G}{\de \ln (1+z)^{-1}}\right)^{1/\gamma}\,.
\end{equation}
Whether such a  substitution leads to universal extensions of the model to cosmologies with growth histories given by modified gravity theories is left for further investigation. We concentrate here on the description as a function of $\Omega_{\rm m}$, as it is more straightforward to compute and produces the same results.

Regarding the characterisation of the dependence of the multiplicity function on the shape of the power spectrum, the first obvious choice is to use its logarithmic slope $\de \ln P_{\rm m}(k)/\de \ln k$ which reduces to the spectral index $n_{\rm s}$ for power-law cosmologies. However, in more realistic cosmologies, the logarithmic slope of the power spectrum contains fluctuations due to the baryonic acoustic oscillations (BAOs) for the characteristic length of the most massive halos. One can circumvent this problem by smoothing the BAO before the computation of the slope, as was done in~\citet{Diemer:2014gba}. Conversely, we propose the logarithmic slope of the mass variance $\de\ln \sigma(R)/\de \ln\,R$. For the power-law cosmologies this reduces to:
\begin{equation}
\frac{\de \ln \sigma}{\de \ln R}=-\frac{(n_{\rm s}+3)}{2}\,.
\end{equation}

\section{\label{sec:results}Results}

In this section, we present the calibration of the HMF model (Sect.~\ref{sec:hmf}), compare it with previous works (Sect.~\ref{sec:comparing}), and forecast the impact of numerical and statistical uncertainties related to our calibration on \textit{Euclid}'s cluster counts analysis~(Sect.~\ref{sec:cc}).

\subsection{\label{sec:hmf}Calibration of the HMF}

First, we provide a short recap of the rationale behind the setup of the \piccolo\ simulations. Before presenting the calibration of the HMF model validated in the \aetiology\ suite, we present a summary of the numerical and theoretical systematic effects on the HMF and how the \piccolo\ suite was designed to address them. The \piccolo\ simulations were run with \og\ set according to the parameters presented in Table~\ref{tab:sims}. This choice of the parameters is shown in Sect.~\ref{sec:parconf} and~\ref{sec:res} to produce results of subpercent accuracy and the robustness of the results is assessed by code-comparison in Sect.~\ref{sec:codecomp}. This suite of simulations uses FCC grids as pre-initial conditions to mitigate the impact of transients due to fluid discretization~\citep[see][]{Michaux:2020yis}. The FCC pre-initial conditions and initial displacements computed with 3LPT at $z=24$ also reduce the impact of correlated errors on the force computations at early redshifts. We refer to Fig.~\ref{fig:og3-vs-g4}, where we compare \og\ and \gfour, and to the discussion of force accuracy presented by~\citet{Springel:2020plp}. To mitigate the effect of round-off errors on the HMF, we use the binning condition presented at the end of Sect.~\ref{sec:ics}. To minimise the impact of sample-variance, box sizes are $2000\,\mpc$  and initial conditions have been created with fixed amplitudes of initial density perturbations~\citep{Angulo:2016hjd} as discussed in Sect.~\ref{sec:samplevariance}. 

In order to calibrate the parameters in Eqs.~\eqref{eq:parevol1} to~\eqref{eq:parevol2}, we use the \piccolo\ set of simulations presented in Table~\ref{tab:sims}. The halo catalogues were binned in $50$  logarithmically spaced bins in number of particles, corresponding to roughly $\delta \ln M \approx 0.15$. We assumed the likelihood presented in Sect.~\ref{sec:likelihood} with $\sigma_{\rm sys}=0.005\, N_i^{\rm sim}$. As for the \aetiology\ fit, its value was chosen such that the best-fit neither over- nor under-fits the simulations and this value is in agreement with the expected non-Poisson scatter caused by round-off errors discussed in Sect.~\ref{sec:ics} for the most populated bin considered in the calibration. Its impact is rapidly diluted as the Poisson contribution grows quickly with mass and redshift. Lastly, we imposed a mass cut below which a halo is not considered as resolved, which corresponds to a minimum of $300$ particles. Finally, we analyzed the outputs of simulations at redshifts $z=\{2.00, 1.25, 0.90, 0.52, 0.29, 0.14, 0.0\}$.

In Table~\ref{tab:hmf_param_piccolo}, we present the best fitting parameters of our calibration for our primary halo finder (\rockstar) and three other auxiliary halo finders (\ahf, \subfind, and \velociraptor). Details of the difference between the calibrations for different halo finders are further discussed in Appendix~\ref{sec:App-hf}. The \ahf, \subfind, and \velociraptor\ fits were performed using one simulation for each cosmology presented in Table~\ref{tab:cosm}. For the \rockstar\ fit, we used two simulations for each cosmology including $C0$ with which we carried out ten realisations. We decided not use the remaining eight $C0$ simulations for calibration in order to avoid over-weighting this cosmology in our derivation of the HMF fit; we only use them to reduce the Poisson fluctuations of rare halos, which allows us to assess our uncertainty in the calibration at this regime.
\begingroup
\setlength{\tabcolsep}{2.0pt} 
\renewcommand{\arraystretch}{1.15} 
\begin{table}
	\centering
	\caption{Parameters of the multiplicity function $f(\nu,z)$ presented in Eq.~\ref{eq:mult} for the \piccolo\  set of simulations.}
	\label{tab:hmf_param_piccolo}
	\resizebox{0.995\columnwidth}{!}{%
	\begin{tabular}{cccccccc}		
		$a_1$ & $a_2$ & $a_z$ & $p_1$ & $p_2$ & $q_1$  & $q_2$ & $q_z$ \\ \hline
		\multicolumn{8}{c}{\rockstar}\\
		$0.7962$ & $0.1449$ & $-0.0658$ & $-0.5612$ & $-0.4743$ & $0.3688$ & $-0.2804$ & $0.0251$ \\ \hline
		\multicolumn{8}{c}{\ahf}\\
		$0.7937$ & $0.1119$ & $-0.0693$ & $-0.5689$ & $-0.4522$ & $0.3652$ & $-0.2628$ & $0.0376$ \\ \hline
		\multicolumn{8}{c}{\subfind}\\
		$0.7953$ & $0.1667$ & $-0.0642$ & $-0.6265$ & $-0.4907$ & $0.3215$ & $-0.2993$ & $0.0330$ \\ \hline
		\multicolumn{8}{c}{\velociraptor\ (adaptive 6DFOF)}\\
		$0.7987$ & $0.1227$ & $-0.0523$ & $-0.5912$ & $-0.4088$ & $0.3634$ & $-0.2732$ & $0.0715$ \\		
		\hline\hline
	\end{tabular}}
\end{table}
\endgroup

\begin{figure}
	\includegraphics[width=0.99 \columnwidth]{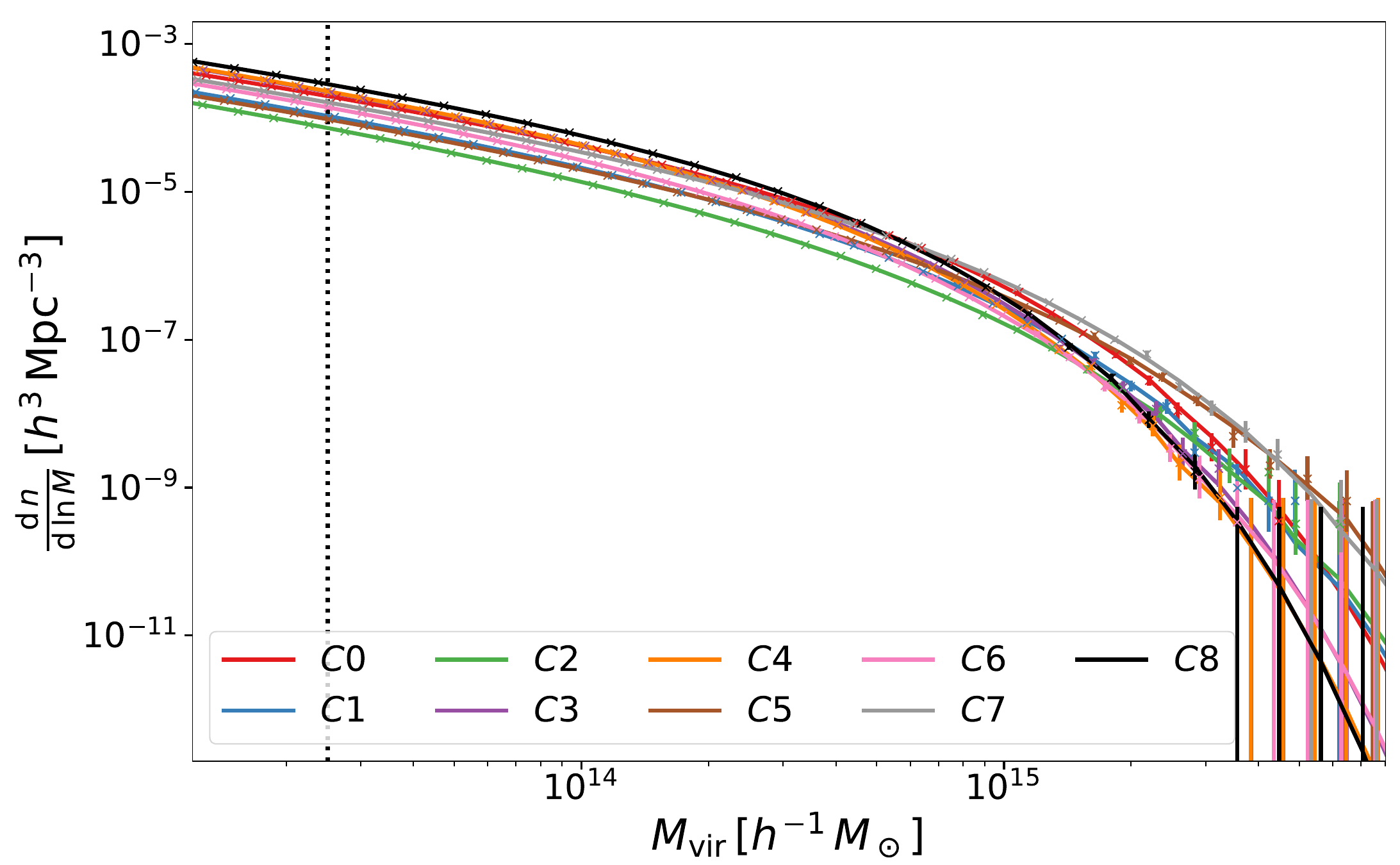}
	\caption{HMF at $z=0$ for the \piccolo\ cosmologies and the \rockstar best-fit prediction. The vertical dotted line represents the cut in mass corresponding to the minimum number of $300$ particles for the $C0$ runs. Error bars are shown assuming a Poisson distribution of the expected number of objects.}
	\label{fig:hmf-mc}
\end{figure}

\begin{figure}
	\includegraphics[width=0.99 \columnwidth]{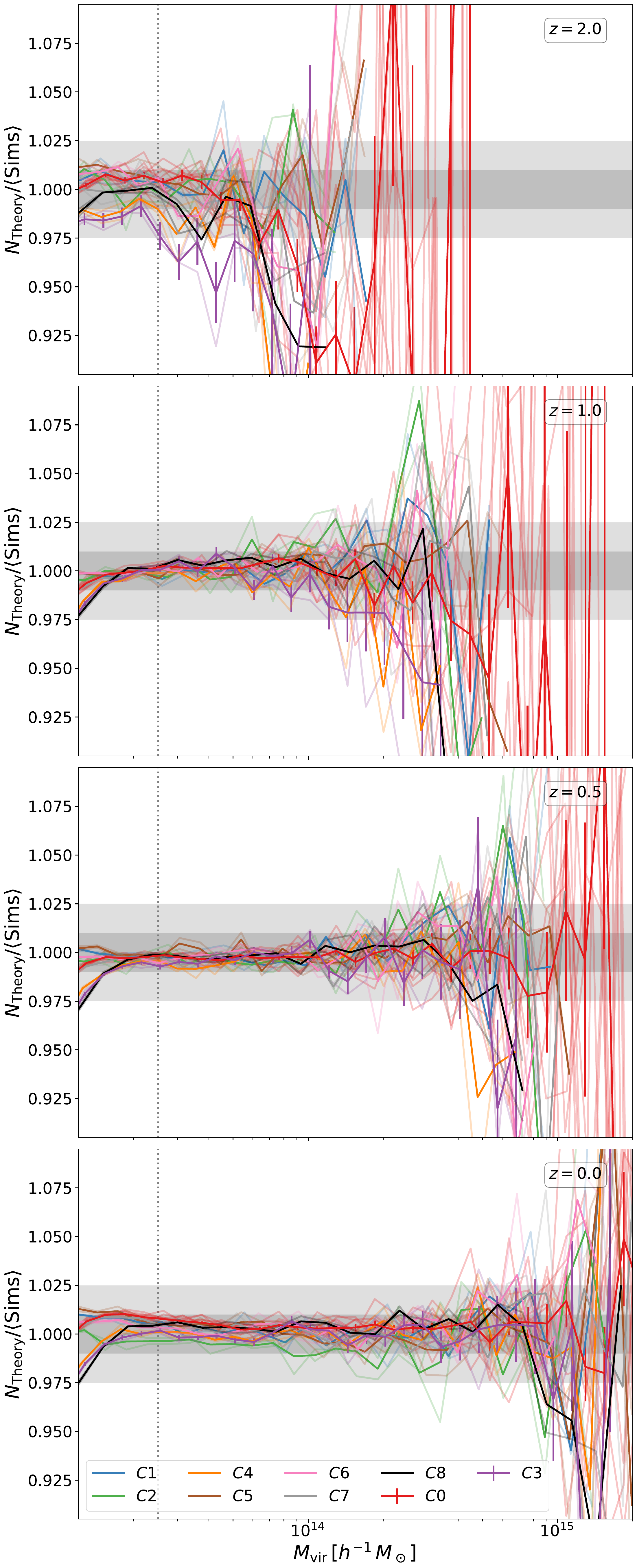}
	\caption{Ratio of \rockstar best-fit to the mean abundance of halos extracted from the simulations at $z=\{0, 0.5, 1.0, 2.0\}$. The vertical dotted line represents the cut in mass corresponding to the minimum number of $300$ particles for the $C0$ runs. The regions in grey represent the relative $1\%$ and $2.5\%$ regions. The Poisson error bars correspond to the $C0$ and $C3$ (our cosmology with fewer halos), which count with ten and two realisations, respectively. We only show mass bins with more than $50$ halos for improved readability.}
	\label{fig:cal-qual-1}
\end{figure}

In Fig.~\ref{fig:hmf-mc}, we present the HMF for \rockstar\ halos and the prediction of the respective best fit from the \piccolo\ simulations at $z=0$. The vertical dotted line represents the cut in mass corresponding to the cut in the minimum number of $300$ particles for the $C0$ runs. Figure~\ref{fig:cal-qual-1} presents the ratio of \rockstar best-fit to the mean abundance of halos extracted from the simulations at $z=\{0.0, 0.5, 1.0, 2.0\}$. As in Fig.~\ref{fig:hmf-mc}, the vertical dotted line represents the cut in mass corresponding to the cut in the minimum number of $300$ particles for the $C0$ runs. The regions in grey represent the relative 1 percent and 2.5 percent regions. For the sake of illustration, we present the Poisson error bars corresponding to the $C0$ and $C3$ (our cosmology with fewer halos) -- the former counts with ten realisations while the latter counts with two. 

Quite remarkably, our model shows a percent-level agreement with the simulations over the entire mass and redshift ranges considered in the calibration despite the much larger difference between the cosmological models presented in Fig.~\ref{fig:hmf-mc}. This result demonstrates that our model HMF is capable of accurately describing the cosmology dependence of the deviations from universality, for a given set of simulations and for a given choice of the halo finder. In Appendix~\ref{sec:App-quality}, we present further tests of the robustness of our calibration as a function of redshift.

\subsection{\label{sec:neutrinos}Cosmologies with massive neutrinos}

The response of the HMF to neutrinos $\mathcal{R}$ is defined as
\begin{equation}
    \mathcal{R}(x) \equiv \frac{\de n/\de m_{\sum M_\nu=x}}{\de n/\de m_{\sum M_\nu=0}}\,,
\label{eq:ratio}
\end{equation}
where $\sum M_\nu$ is the sum of the masses of the three neutrino species. 

\begin{figure}
    \includegraphics[width=0.99 \columnwidth]{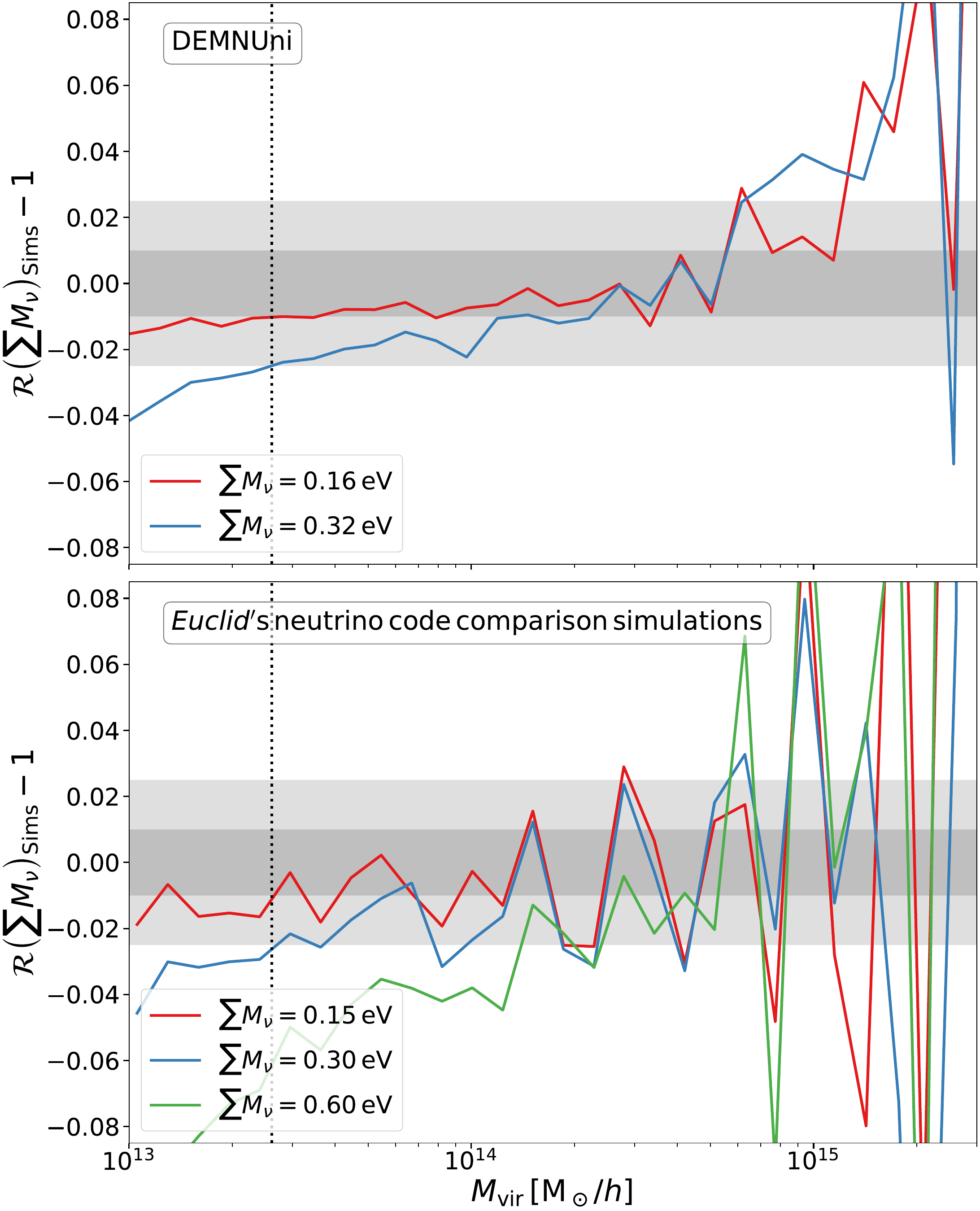}
    \caption{Impact of massive neutrinos on the HMF. We present the ratio of the HMF with massive neutrinos to the corresponding massless neutrinos counterpart as observed in two independent external simulations ${\rm DEMNUni}$ (\emph{top}) and the \og\ subset of \textit{Euclid}’s neutrino code comparison simulations (\emph{bottom}). See the definition of $\mathcal{R}$ in Eq.~\eqref{eq:ratio}.}
    \label{fig:neutrino-impact}
\end{figure}

\begin{figure}
    \includegraphics[width=0.99 \columnwidth]{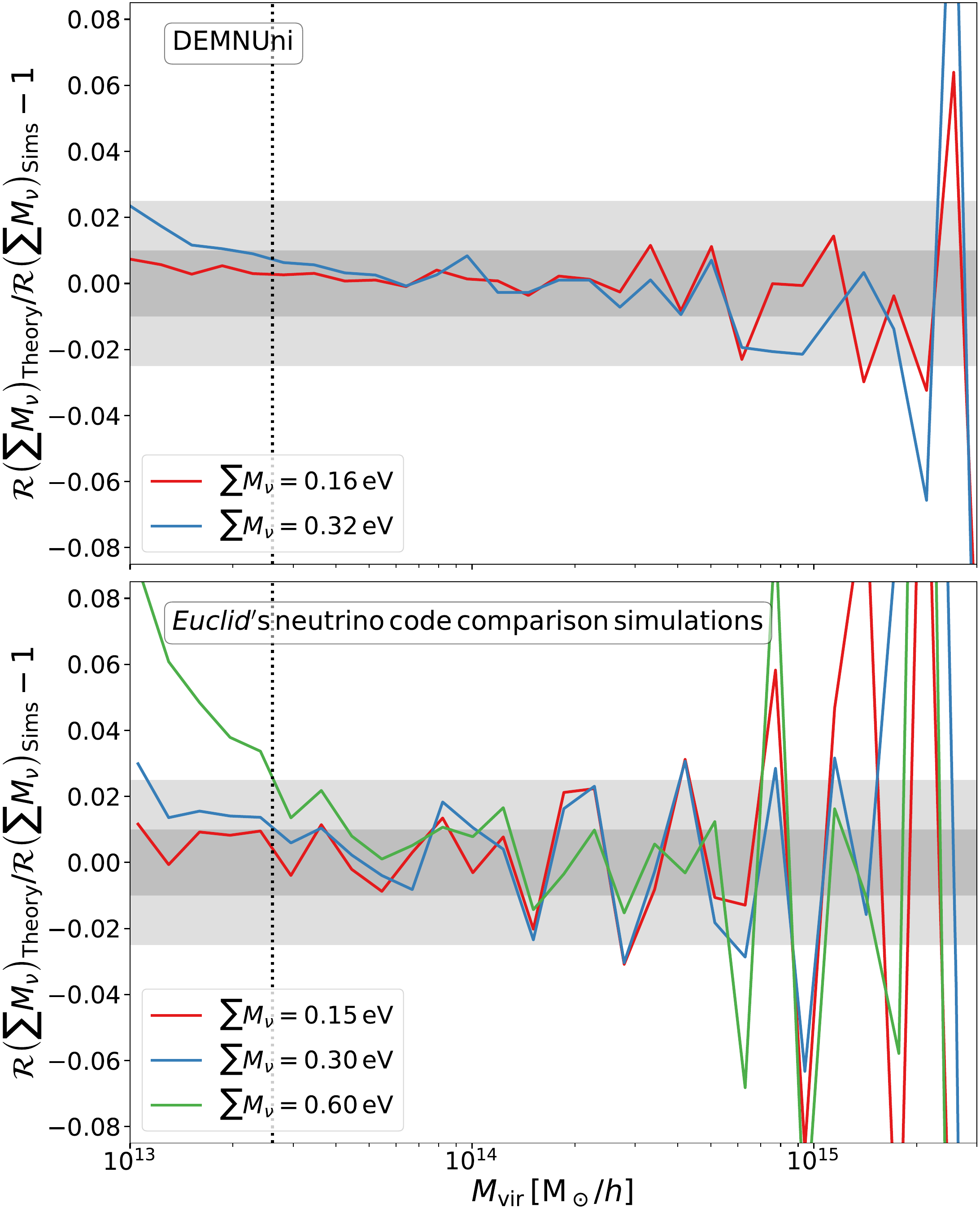}
    \caption{Accuracy of the model presented in~\citet{Castorina:2014} in accounting for the impact of massive neutrinos on the HMF. We compare the ratio of the HMF with massive neutrinos to the corresponding massless neutrinos counterpart, assuming the model of~\citet{Castorina:2014} divided by the same quantity as observed in two independent external simulations: ${\rm DEMNUni}$ (\emph{top}) and the \og\ subset of \textit{Euclid}’s neutrino code comparison simulations (\emph{bottom}). See the definition of $\mathcal{R}$ in Eq.~\eqref{eq:ratio}.}
    \label{fig:neutrino}
\end{figure}

In Figs.~\ref{fig:neutrino-impact} and~\ref{fig:neutrino}, we present the response of the HMF to massive neutrinos and the accuracy of the model presented in~\citet{Castorina:2014} in order to take this latter into account. Briefly, the model assumes that neutrinos impact the HMF passively, that is, through its effect on the background evolution. Operationally, implementation of the model only requires that we replace the total matter power spectrum $P_{\rm m}$ in the computation of the mass variance (Eq.~\ref{eq:massvariance}) with the cold dark matter plus baryons power spectrum and that we ignore the contribution of neutrinos to the mass of the Lagrangian patch corresponding to a given halo. To assess the accuracy of the model, we compare the response $\mathcal{R}$ predicted by the model and from simulations. The rationale for using $\mathcal{R}$ to assess the accuracy of the model is to mitigate the impact of systematic effects due to differences in the setup between the external simulations and the ones used for the calibration of the HMF in this work.
 
 We used two independent external simulations: the DEMNUni~\citep[see][for details of the simulation]{Carbone:2016nzj}\footnote{\url{https://www.researchgate.net/project/DEMN-Universe-DEMNUni}} and the \og\ subset of the \textit{Euclid}’s neutrino code comparison simulations~\citep[][]{Euclid:2022qde}. Halos were extracted in both sets using \subfind. Thus, the results presented in this subsection rely on our \subfind\ calibration presented in Table~\ref{tab:hmf_param_piccolo}.
 
 The DEMNUni set consists of three simulations of $2000$ \mpc\ boxes assuming a cosmology in agreement with Planck 2013 results for the cosmological parameters~\citep{Planck:2013pxb}.\footnote{\url{https://github.com/jmd-dk/nucodecomp-data}} The reference simulation considers massless neutrinos, while the other two simulations use the particle-based implementation of neutrinos, assuming a total neutrino masses of $0.16$ and $0.32$ eV, respectively. The reference simulations include $2048^3$ dark matter particles, while the latter includes the same number of dark matter particles and as many $2048^3$ neutrino particles. 
 
 The \og\ subset of the \textit{Euclid}’s neutrino code comparison shares the same implementation, a similar mass resolution, and a similar cosmological parameters as the DEMNUni but in $1000$ \mpc\ boxes. Instead, the total neutrino masses simulated are $0.0$, $0.15$, $0.30$, and $0.60$ eV. These simulations have been extensively compared with different neutrino implementations. The agreement for the power spectrum, bispectrum, and HMF was observed to be better than 1 percent for the range of interest.

In Fig.~\ref{fig:neutrino-impact}, we observe the well-established suppression in the cluster abundance caused by massive neutrinos and its dependence on the total neutrino mass. In Fig.~\ref{fig:neutrino}, we observe that, for total neutrino masses smaller than $0.32$ eV and for both simulations, the model of~\citet{Castorina:2014} agrees with simulations to better than $1$ percent (dark shaded area) over the entire mass range of validity of our calibration; as in Fig.~\ref{fig:cal-qual-1}, the dotted line represents the $300$-particle mass cut used for the calibration. We only note that at lower masses, below $\sim 3\times 10^{13} \msun$ the model HMF tends to underestimate the effect of massive neutrinos, an effect that increases with neutrino masses. While we are not interested in the HMF calibration in this mass range, we ascribe this difference to the effect of neutrino free-streaming on the growth of CDM perturbations, which cannot be captured by simply ignoring the neutrino component in the linear matter power spectrum. 

For the most massive neutrinos considered, we only rely on the simulations carried out for the code comparison. Due to its smaller volume, larger fluctuations are observed with respect to the DEMNUni, but at the mass threshold of validity of our calibration, we nevertheless observe that the model agrees with the simulation within a few percent, and for lower masses the model starts to over-predict the impact of neutrinos. The agreement for masses larger than $4\times10^{13}$ \msun\ is again consistent with 1 percent scatter.

\subsection{\label{sec:comparing}Comparison to other works}

\begin{figure*}[ht]
	\includegraphics[width=0.99 \textwidth]{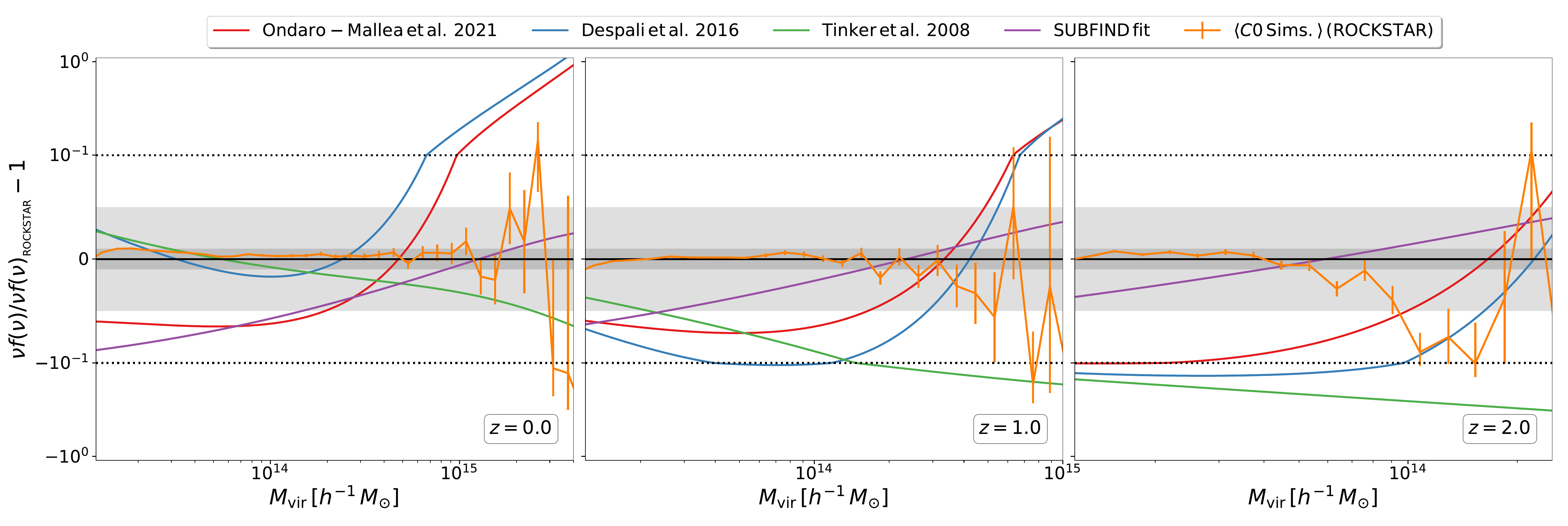}
	\caption{Comparison of our HMF model to the works of~\citet{Tinker:2008ff},~\citet{Despali:2015yla}, and~\citet{Ondaro-Mallea:2021yfv}. We use our \rockstar\ calibration as the reference and present our \subfind\ calibration for easy comparison with~\citet{Ondaro-Mallea:2021yfv}. We also present the mean of the multiplicity function $\nu f(\nu)$ measured from the ten \piccolo\ $C0$ runs. The grey regions depict the $1$ and $5$ percent regions. We adopted a symmetric-log scale on the $y$-axis, where the region between the dashed lines is presented in a linear scale.}
	\label{fig:other-works}
\end{figure*}

In Fig.~\ref{fig:other-works}, we compare our HMF model to the works of~\citet{Tinker:2008ff},~\citet{Despali:2015yla}, and~\citet{Ondaro-Mallea:2021yfv}. We use our \rockstar\ calibration as a reference and present our \subfind\ calibration as our \subfind\ setup matches those used by~\citet{Ondaro-Mallea:2021yfv}, making the comparison easier. WE note that, in Fig.~\ref{fig:comp-hf}, we present a comparison of the cumulative mass function produced with \rockstar\ with that produced by \denhf. The latter, used by~\citet{Despali:2015yla}, differs from the former by roughly 1 percent. Lastly, \citet{Behroozi:2011ju} compared \rockstar\ catalogues with the prediction of~\citet{Tinker:2008ff} and presented an agreement within 5 percent at $z=0$. We also present the mean of the multiplicity function $\nu f(\nu)$ measured from the ten \piccolo\ $C0$ runs. The grey areas depict the $1$ and $5$ percent regions. To embed the significant differences with respect to the other works, we adopted a symmetric log scale on the $y$-axis, where the region between the dotted lines is presented in a linear scale.

As already shown in Fig.~\ref{fig:cal-qual-1}, our model accurately reproduces results from simulations over fairly large ranges of masses and redshifts. Globally, the differences are larger at both large masses and high redshifts, where the statistics are poorer. At $z=0$, the model of~\citet{Tinker:2008ff} differs from ours by a maximum of 3 percent for halos less massive than $10^{15}\,\msun$; it crosses the $5$ percent threshold at $2\times10^{15}\,\msun$, and beyond that it deviates from our model, predicting significantly fewer halos. The model of~\citet{Despali:2015yla} crosses the $5$ percent threshold over a narrower mass range; it deviates from our model by more than $5$ percent beyond $4\times10^{14}\,\msun$, and over-predicts the number of halos more massive than $10^{15}\,\msun$ by more than $20$ percent. Comparing the model by~\citet{Ondaro-Mallea:2021yfv} to our HMF calibration based on \subfind\, we note that they start differing by more than $5$ percent for halos more massive than $7\times10^{14}\,\msun$, above which the model of these latter starts to follow the same trend as the model of~\citet{Despali:2015yla}. The picture at $z=1$ and $z=2$ are similar; at smaller masses, all models tend to predict fewer halos than observed in our simulations and predicted by our model. For~\citet{Despali:2015yla} and~\citet{Ondaro-Mallea:2021yfv}, the trend flips at intermediate masses and starts to over-predict the abundance of the most massive halos. Similar results were obtained by comparing our calibration directly with the simulations used and made available by~\citet{Ondaro-Mallea:2021yfv}, which reassures us of the robustness of our calibration.

\subsection{\label{sec:cc}Impact on cluster counts analysis}

In this subsection, we forecast the impact of the uncertainties on the HMF calibration on the cosmological constraints to be derived from the  \textit{Euclid} cluster counts.

\subsubsection{Impact of the halo finder}

In Table~\ref{tab:IOI&FOM}, we summarise the impact of the different halo finders on the inference of the marginalised cosmological parameters $\Omega_{\rm m}$ and $\sigma_8$. To do so, we compare the results obtained when one assumes \velociraptor, \subfind\ or \ahf\ calibrations to create the synthetic catalogue while the fiducial \rockstar\ calibration is used for the analysis. We perform this forecast following the methodology described in Sect.~\ref{sec:forecast}. For the likelihood analysis, we assume flat priors on the cosmological parameters and Gaussian priors with amplitudes on the mass--observable parameters of 1, 3, and 5 percent. The likelihood sampling is performed with ZEUS~\citep[see][]{karamanis2020ensemble,karamanis2021zeus}. The impact of the different halo finders' calibrations is quantified using the index of inconsistency \citep[IOI;][]{Lin:2017ikq}, which is calculated as
\begin{equation}
    {\rm IOI} = \frac{\boldsymbol{\delta}^{\rm t}\,\boldsymbol{\Sigma}^{-1}\,\boldsymbol{\delta}}{2}\, ,
\end{equation}
where $\boldsymbol{\delta}$ is the two-dimensional difference vector between the best-fit and the assumed cosmological parameters $\{\Omega_{\rm m}, \sigma_8\}=\{0.30711, 0.8288\}$. Additionally, $\boldsymbol{\Sigma}$ is the covariance matrix between these parameters, which we assume to be Gaussian distributed. In all cases, the tension in the $(\Omega_{\rm m},\sigma_8)$ plane increases monotonically as the priors on the richness--mass relation tightens. For both \velociraptor\ and \subfind\ the tension goes from $\lesssim 1\,\sigma$ for $5$ and $3$ percent priors to $\lesssim 2\,\sigma$ when the prior tightens to $1$ percent. The tension for the \ahf\ case is $\lesssim 1\,\sigma$ for all priors considered, which is a result of the similarity observed in Figs.~\ref{fig:comp-hf} and~\ref{fig:comp-hf-2}  between \ahf\ and \rockstar\ calibrations.

Therefore, we conclude that differences in the HMF calibration associated with different choices of the halo finder propagate into systematic effects in the measurements of cosmological parameters that are comparable to the formal uncertainties on such parameters. For instance, if the cluster richness--mass relation from \textit{Euclid} data could be calibrated at $<3$ percent precision, then a crucial factor in deriving cosmological constraints from cluster number counts would become the way in which a halo is defined and identified in simulations in the process of the HMF calibration. Lastly, if one increases the error budget for the HMF calibration until it comprises the different halo-finder prescriptions studied here, the IOI presented in Table~\ref{tab:IOI&FOM} also provides the expected reduction factor in the FOM of the cosmological constraining power of the \textit{Euclid} cluster counts .

\subsubsection{Impact of the calibration error}

In Table~\ref{tab:IOI&FOM}, we also summarise the impact of the systematic and statistical errors of our main (\rockstar) calibration and one of the auxiliary calibrations (\velociraptor) on the marginalised uncertainties in the cosmological parameters $\Omega_{\rm m}$ and $\sigma_8$. We only consider one calibration of each group of halo finders, as this test is dominated by the number of simulations used in the HMF calibration. As \velociraptor, \subfind, and \ahf\ all use the same number of simulations ---  equal to half of the set used for the \rockstar\ calibration --- they present very similar results. We compare the FOM change in the $(\Omega_{\rm m},\sigma_8)$ plane obtained by fixing the halo mass function parameters to the calibrated values with the ones obtained by marginalizing over such parameters using a multi-variate Gaussian with covariance given by the fit uncertainties. We consider again 1, 3, and 5 percent priors on the richness--mass scaling relations. For \rockstar\, the statistical uncertainty only marginally affects the cosmological inference. For \velociraptor, we observe that the only significant impact is seen for the 1 percent prior, where the FOM is over-estimated by $\sim10$ percent when the HMF parameters are left fixed. Therefore, from this test, we conclude that the residual uncertainties in the HMF parameters have a negligible impact on the corresponding cosmological constraints.
\begin{table*}
    \centering
    \caption{Summary statistics for the forecast of the impact of different halo finders and calibration errors on \textit{Euclid} cluster counts cosmological constraints on $\Omega_{\rm m}$ and $\sigma_8$. The IOI quantifies the tension in the posteriors if one uses the \rockstar\ calibration to create the synthetic data while either the \velociraptor, \subfind, or \ahf\ calibration is used for the analysis. The relative difference of the FOM assesses the attenuation of the constraining power of cluster counts if one marginalizes over the HMF parameters assuming the calibration chain as a prior. The latter statistics are presented only for the \rockstar\ and \velociraptor\ calibrations as \velociraptor, \subfind, and \ahf\ use half the simulations used for the \rockstar\ calibration and present very similar results. Errors for both statistics were estimated using bootstrap resampling.}
    \label{tab:IOI&FOM}
    \begin{tabular}{c|c|c|c|c}
    Summary statistics & richness--mass relation priors & Analysis & Synthetic catalogue & Value \\ \hline 
    \multirow{9}{*}{IOI}    & $1\,\%$ &                &           & $\hphantom{-}1.66 \pm 0.01$ \\
                            & $3\,\%$ & \velociraptor  & \rockstar & $\hphantom{-}0.77 \pm 0.01$ \\
                            & $5\,\%$ &     (Fixed)    &           & $\hphantom{-}0.65 \pm 0.01$ \\\cline{2-5}
                            & $1\,\%$ &                &           & $\hphantom{-}1.70 \pm 0.02$ \\
                            & $3\,\%$ & \subfind       & \rockstar & $\hphantom{-}0.84 \pm 0.01$ \\
                            & $5\,\%$ &     (Fixed)    &           & $\hphantom{-}0.61 \pm 0.01$ \\\cline{2-5}
                            & $1\,\%$ &                &           & $\hphantom{-}0.90 \pm 0.02$ \\
                            & $3\,\%$ & \ahf           & \rockstar & $\hphantom{-}0.61 \pm 0.01$ \\
                            & $5\,\%$ &     (Fixed)    &           & $\hphantom{-}0.47 \pm 0.00$ \\ \hline
    \multirow{6}{*}{\scalebox{1.2}{$\frac{\Delta {\rm FOM}}{{\rm FOM}}$}}
                            & $1\,\%$ &                &           & $\hphantom{-}0.04 \pm 0.05$ \\
                            & $3\,\%$ & \rockstar      & \rockstar & $\hphantom{-}0.06 \pm 0.04$ \\
                            & $5\,\%$ & (marginalised) &           & $-0.01 \pm 0.02$ \\\cline{2-5}           
                            & $1\,\%$ &                  &               & $-0.09 \pm 0.05$  \\
                            & $3\,\%$ & \velociraptor    & \velociraptor & $\hphantom{-}0.00 \pm 0.03$  \\
                            & $5\,\%$ & (marginalised)   &               & $-0.02 \pm 0.03$ \\\hline \hline
    \end{tabular}
\end{table*}

\section{\label{sec:conclusions}Conclusions}

In this paper, we carried out a detailed analysis to assess the numerical robustness of the halo mass function (HMF) predicted by \textit{N}-body simulations, and to quantify and model its deviation from universality. The variety of tests that we carried out include changing the prescription for generating initial conditions, the effect of resolution and of round-off errors in particle positions in the initial conditions, the \textit{N}-body integrator, the definition of a halo, and the halo finder. While our reference analysis were carried out assuming a vanilla $\Lambda$CDM cosmology, we also simulated the effect on the HMF of including massive neutrinos. Furthermore, in order to trace the origin of departures from universality, we also ran simulations with a purely scale-free power spectrum, both assuming Einstein--de Sitter and $\Lambda$CDM expansion histories. Finally, with the resulting high-resolution calibration of the HMF, we assess the impact of systematic effects in the cosmological parameters inference from an idealized \textit{Euclid} cluster number counts experiment. Our main conclusions can be summarised as follows.
\begin{itemize}
	\item The different gravity solvers considered in this paper agree better than 1 percent on the HMF of cluster-sized halos for particle masses smaller than $10^{10}\,\msun$. Interestingly, the \textit{N}-body integrator of \ramses, the only code based on adaptive mesh refinement among those considered here, seems to systematically predict a lower number of halos with $M\lesssim 10^{14}$ \msun\, an effect that is more apparent at lower mass resolutions;
	\item Our adopted setup for the \piccolo\ set, which includes simulations for nine different cosmological models, provides a percent-level convergence on the HMF model when compared to higher resolution simulations.
	\item Numerical artifacts, such as round-off errors, add non-Poisson fluctuations to the mass-binned distribution of halos. Choosing the mass binning accordingly mitigates the problem.
	\item The differences in the abundance of halos coming from different halo finders largely dominate all the other numerical systematic errors considered here. The final impact on cosmological constraints depends on how well the mass--observable relation is kept under control, which highlights the need of a better understanding of how halos identified in simulations are related to clusters identified in an optical/NIR photometric survey, such as those provided by \textit{Euclid}.
	\item The HMF non-universality of virial halos depends on both the background evolution and the power spectrum shape, confirming the results of~\citet{Ondaro-Mallea:2021yfv}.
	\item Our HMF model was calibrated against nine $\Lambda$CDM cosmologies evenly covering the 95 percent confidence level constraints on cosmological parameters from DES+SPT cluster counts~\citep{DES:2020cbm} and four different halo finders (\rockstar, \ahf, \subfind, \velociraptor). Our HMF calibration reproduces the abundance of virial halos more massive than $3\times10^{13}$ \msun\ with a precision of better than $1$ percent for the range of cosmological parameters studied here. However, our calibration is expected to retain its accuracy beyond this range, as the HMF modelling was validated on a more extreme set of simulations (\aetiology).
	\item Using two external sets of simulations that include massive neutrinos, we validated the model presented in~\citet{Castorina:2014}. Jointly with our calibration, the model by~\citet{Castorina:2014} reproduces the abundance of virial halos more massive than $4\times10^{13}$ \msun\ for a total neutrino mass in the range $[0.0 - 0.6]$ eV with a precision of better than 1 percent.
	\item The statistical uncertainties on the HMF calibration presented in our analysis are significantly smaller than the expected accuracy for the mass--observational relation of \textit{Euclid}. However, the difference between the HMF of the halo finders studied here is comparable to the expected accuracy for the mass--observational relation of \textit{Euclid} and, as such, could lead to a biased inference of cosmological parameters.
\end{itemize}

One of the results of our analysis is that the main source of uncertainty in the calibration of the HMF from \textit{N}-body simulations is related to the definition of halos and to the finder used to identify them. Reassuringly, the differences we found by applying four different halo finders do not compromise the accuracy of the HMF required by the \textit{Euclid} cluster survey. Still, our analysis does not provide information as to the the correspondence between a halo identified in an \textit{N}-body simulation and a cluster identified in a photometric survey, and how uncertain knowledge of this correspondence will impact on the derivation of cosmological posteriors remains an open question. Indeed, uncertainties in the relation between richness and mass enclosed within a suitably defined (and cosmology-dependent) radius, projection effects in the selection function of the cluster sample, and miscentring, are all effects that need to be controlled and convolved with the predicted HMF. While all such issues need to be addressed one by one, the analysis presented here demonstrates that the precision in the definition of a fitting function for the HMF predicted by \textit{N}-body simulations of $\Lambda(\nu)$CDM cosmologies is not a limiting factor for cluster cosmology with \textit{Euclid}. In forthcoming analyses we will verify whether a similar precision can be maintained when including uncertainties related to the astrophysics of baryons and departures from the standard $\Lambda(\nu)$CDM framework.

%
%

\begin{acknowledgements}
It is a pleasure to thank Valerio Marra for constructive comments during the production of this work, Renate Mauland-Hus and Hans Winter for the support with \ramses\ simulations, Peter Berhoozi for the support with \rockstar, Oliver Hahn for the support with \monofonic\ and \music, and Douglas Potter for the support with \pkd\ set up. Lourdes Ondaro-Mallea and Matteo Zennaro for sharing their HMF data. TC, SB, and AS are supported by the INFN INDARK PD51 grant. TC and AS are also supported by the FARE MIUR grant `ClustersXEuclid' R165SBKTMA. AS is also supported by the ERC `ClustersXCosmo' grant agreement 716762. KD acknowledges support by the Deutsche Forschungsgemeinschaft (DFG, German Research Foundation) under Germany’s Excellence Strategy - EXC-2094 - 390783311 as well as support through the COMPLEX project from the European Research Council (ERC) under the European Union’s Horizon 2020 research and innovation program grant agreement ERC-2019-AdG 882679. AR is supported by the PRIN-MIUR 2017 WSCC32 ZOOMING grant. We acknowledge the computing centre of CINECA and INAF, under the coordination of the ``Accordo Quadro (MoU) per lo svolgimento di attività congiunta di ricerca Nuove frontiere in Astrofisica: HPC e Data Exploration di nuova generazione'', for the availability of computing resources and support. We acknowledge the use of the HOTCAT computing infrastructure of the Astronomical Observatory of Trieste -- National Institute for Astrophysics (INAF, Italy) \citep[see][]{2020ASPC..527..303B,2020ASPC..527..307T}.  \AckEC
\end{acknowledgements}

%
%

\bibliography{euclid}

%

\begin{appendix}
\section{\label{sec:App-convergence}Convergence tests}

\begin{figure}
	\includegraphics[width=0.99\columnwidth]{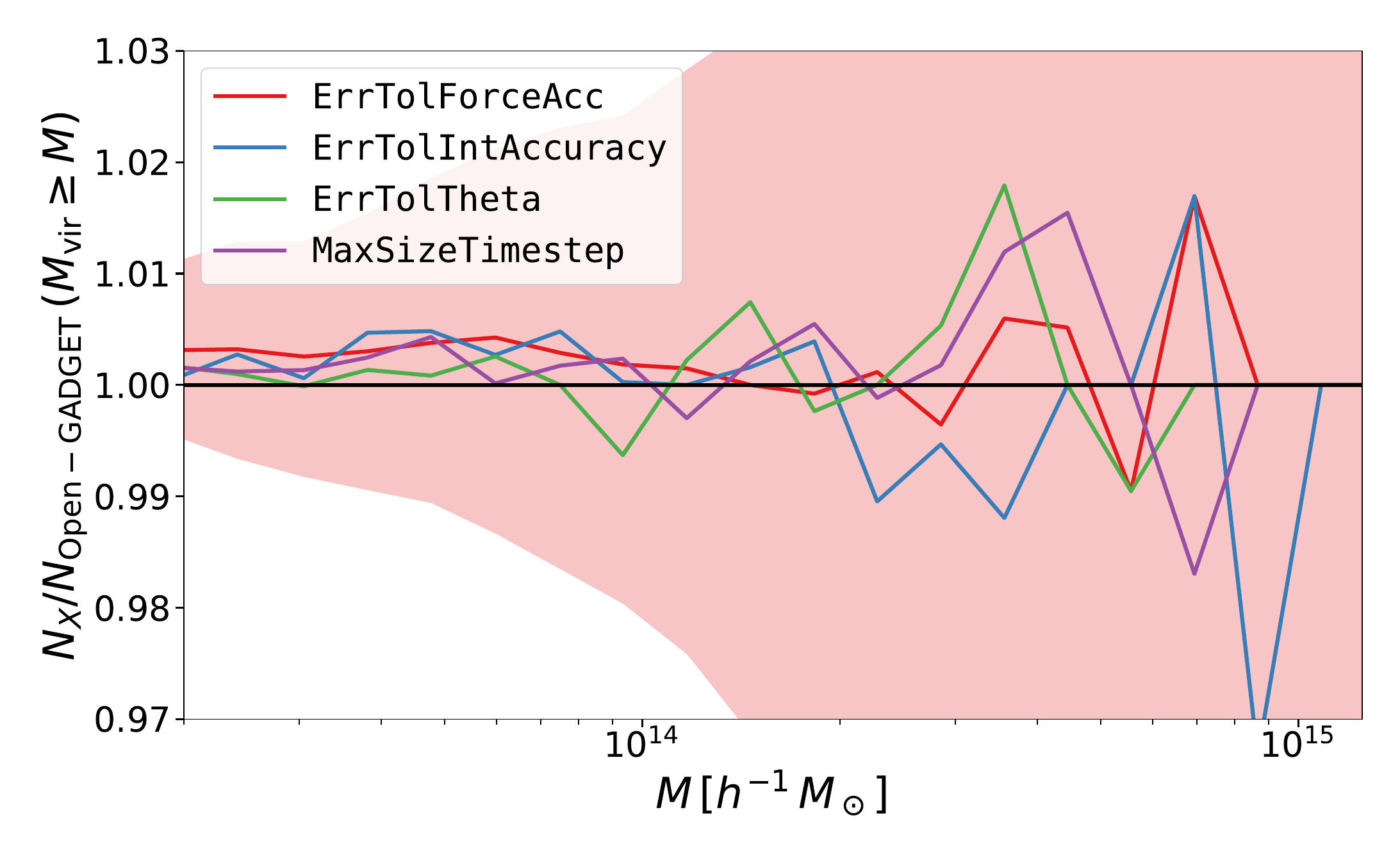}
	\caption{Sensitivity of the cumulative mass function to the accuracy parameters of \og. Each line corresponds to the ratio of the cumulative mass function obtained setting one of the accuracy parameters to half the value presented in Table~\ref{tab:og3-set} with respect to the fiducial set. For comparison, we add the $68\%$ confidence level in red for the \texttt{ErrTolForceAcc} assuming that halos in the simulations are distributed according to a Poisson distribution. }
	\label{fig:og3-convergence}
\end{figure}

In Fig.~\ref{fig:og3-convergence}, we test the sensitivity and convergence of the cumulative HMF extracted from AHF catalogues to the accuracy parameters of \og\ for one of the \tease\ simulations of $512^3$ particles displaced from an equally spaced grid according to the Zeldovich approximation at $z=99$. Each line corresponds to the ratio of the cumulative HMF obtained setting one of the accuracy parameters to half the value presented in Table~\ref{tab:og3-set}. For comparison, we also add the 68 percent confidence level in red for the \texttt{ErrTolForceAcc} assuming that the number of halos in the simulations are distributed according to a Poisson distribution. Fig.~\ref{fig:og3-convergence} shows that our parameter set provide sub-Poisson differences to higher precision sets, deviating by less than a fraction of a percent in the abundance of halos more massive than $3\times10^{13}\,\msun$. 

\begin{figure}
	\includegraphics[width=0.99\columnwidth]{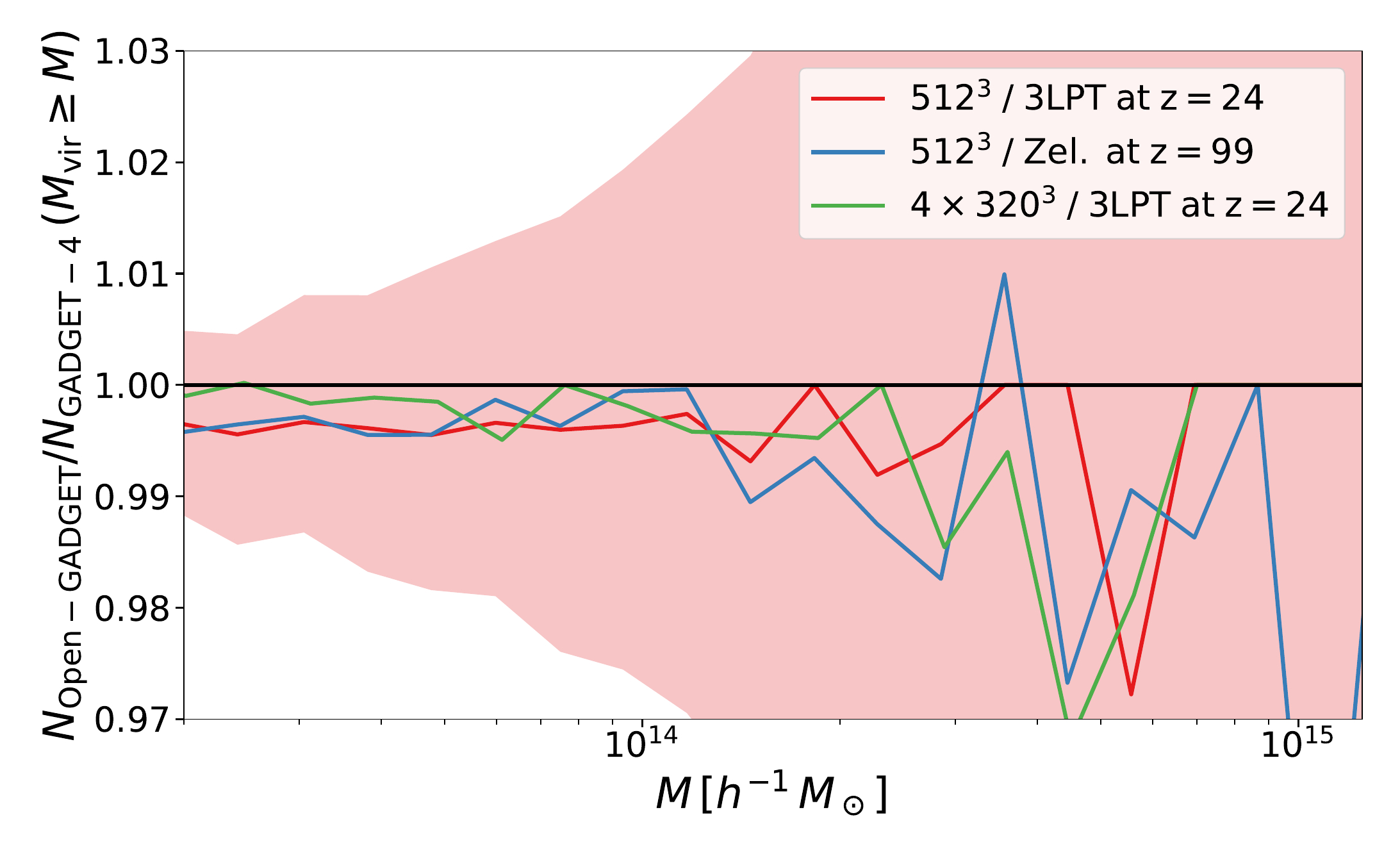}
	\caption{Comparison between the cumulative mass function extracted from \og\ and \gfour\ for three different initial conditions: $512^3$ displaced from a equally spaced grid according to 3LPT at $z=24$, the same number of particles using Zeldovich at $z=99$, and $4\times 320^3$ particles displaced from a FCC grid at $z=24$. }
	\label{fig:og3-vs-g4}
\end{figure}

In Fig.~\ref{fig:og3-vs-g4}, we present the comparison between the cumulative mass function extracted with AHF from cosmological simulations run with \og\ and \gfour. We tested the convergence of the results for three different initial conditions: $512^3$ particles displaced from an equally spaced grid according to 3LPT at $z=24$, the same number of particles using Zeldovich at $z=99$, and $4\times 320^3$ particles displaced from a face centred cubic (FCC) grid according to 3LPT at $z=24$. The box size is $500~\mpc$. In all configurations, the agreement between \og\ and \gfour\ is better than a fraction of percent for all masses considered. Due to the higher degree of planes of symmetry, the FCC configuration shows an even better agreement between the two codes, as in this configuration the usual tree algorithm delivers more accurate forces from the simulation start.

\begin{figure}
	\includegraphics[width=0.99\columnwidth]{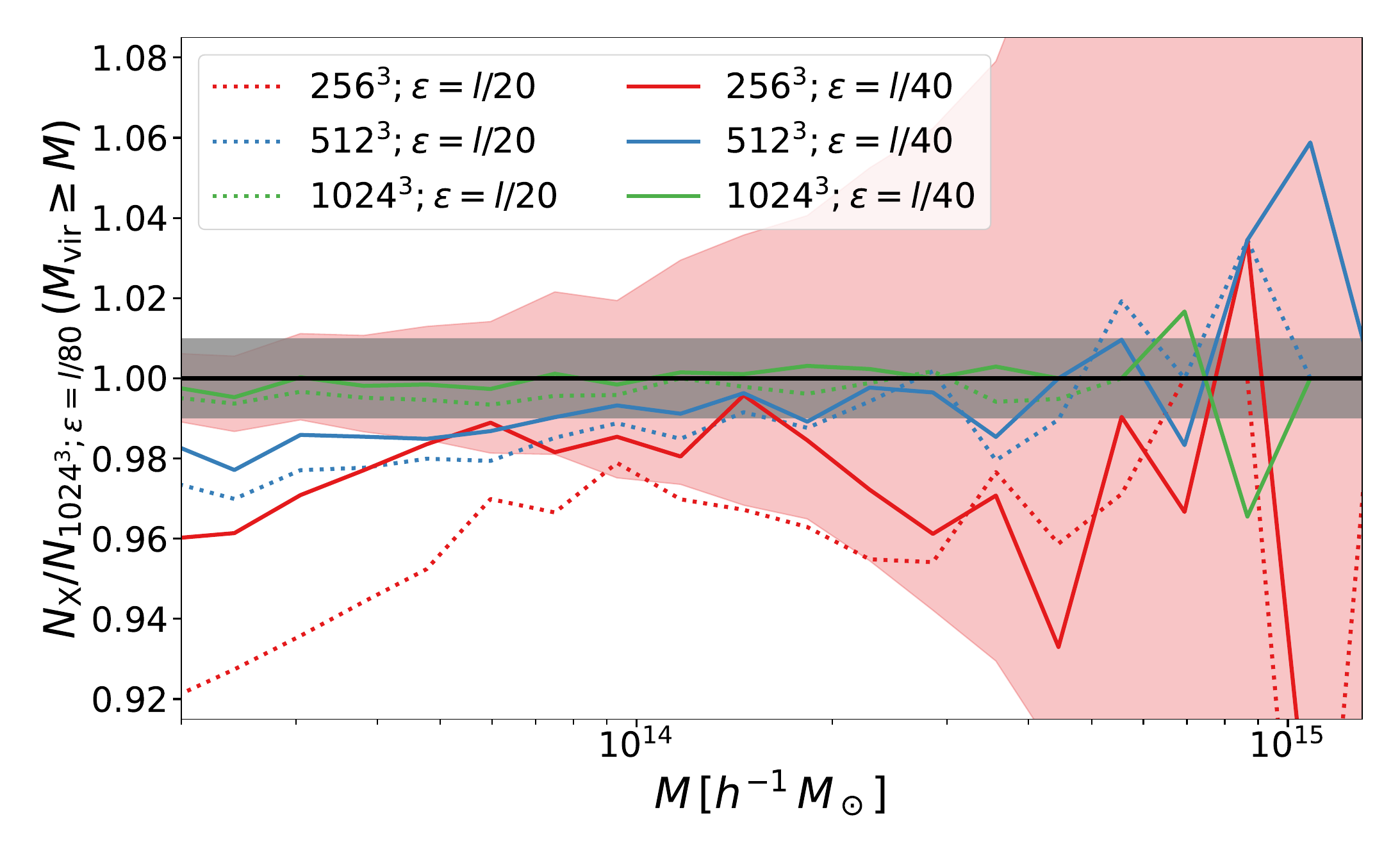}
	\caption{Convergence test for the cumulative mass function for different particle mass ($\{6.35\times 10^{11}, 7.94\times 10^{10}, 9.92\times 10^{9}\}\,\msun$, respectively, $\{256^3, 512^3, 1024^3\}$ particles in a box of $500\,\mpc$) and spatial resolution set by the gravitational softening ($\epsilon=\{l/20, l/40, l/80\}$, where $l$ is the mean interparticle distance). The filled region in grey represents the $1$ percent region around unity and the filled region in red represent the $68$ percent confidence level assuming Poisson distribution for the abundance of halos in both simulations.}
	\label{fig:convergence}
\end{figure}

In Fig.~\ref{fig:convergence}, we present a convergence test for the cumulative mass function computed from \ahf\ halo catalogue for different particle masses and spatial resolutions. We consider three mass resolutions, corresponding to the following particle masses: $\{6.35\times 10^{11}, 7.94\times 10^{10}, 9.92\times 10^{9}\}\,\msun$; respectively $\{256^3, 512^3, 1024^3\}$ particles in a box of $500\,\mpc$. For the gravitational softening we consider: $\epsilon=\{l/20, l/40, l/80\}$, where $l$ is the mean interparticle distance. For reference, the $1$ percent region is shown  with the grey-shaded area, while the $68$ percent confidence level assuming Poisson distribution for the halo distribution of the simulations with $1024^3$ particles with $\epsilon=l/40$ and $\epsilon=l/80$ is depicted in red. We note that a particle mass of few times $10^{10}\,\msun$ is enough to have a cumulative mass function that agrees to within 1 percent with the higher resolution simulation for halos more massive than $10^{14}\,\msun$. Lighter objects are more strongly affected by both the mass and spatial resolutions. For the simulation with $256^3$ particles, even the formation of very massive objects is suppressed by $\approx 4$ percent.

\section{\label{sec:App-hf}Comparison between the different halo finder calibrations}

\begin{figure}
	\includegraphics[width=0.99 \columnwidth]{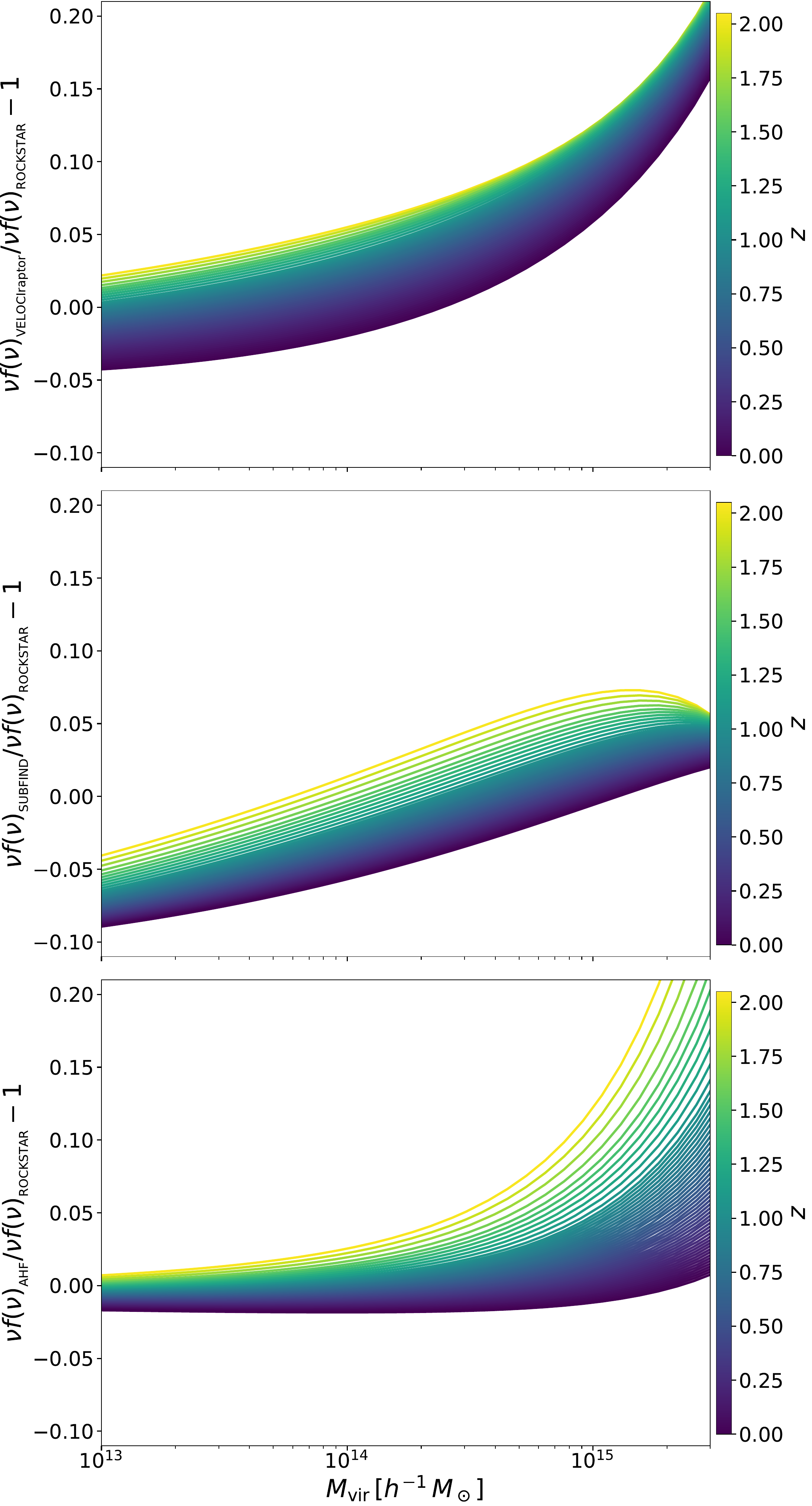}
	\caption{Ratio of the multiplicity function as a function of the halo virial mass for the different calibrations to \rockstar\ colour coded according to redshift. From \emph{top} to \emph{bottom}, we present the results for \velociraptor, \subfind, and \ahf.}
	\label{fig:comp-hf-2}
\end{figure}

The ratio of the multiplicity function as a function of the halo virial mass for the different calibrations with respect to \rockstar\ is shown in Fig.~\ref{fig:comp-hf-2} colour coded by the redshift. Fig.~\ref{fig:comp-hf-2} confirms the trends at low redshift presented in Fig.~\ref{fig:comp-hf}, but also shows their dependence with redshift. Despite the very different algorithm, \ahf\ and \rockstar\ provide a multiplicity function that agrees within $10$ percent up to $M_{\rm vir}\sim 10^{15} \msun$ for the redshift range considered. 

\rockstar's agreement with \velociraptor\ is worse than the agreement with \ahf, depending more strongly on both mass and redshifts. At low redshift, \velociraptor\ tends to suppress the number of halos with mass $M_{\rm vir}\lesssim 10^{14} \msun$ by few percents. 

\rockstar's worst agreement is with respect to \subfind, with \subfind\ suppressing the number of objects with mass $M_{\rm vir} \lesssim 10^{15} \msun$ at all redshifts.

\section{\label{sec:App-quality}Assessment of the fit-quality dependence on redshift}

\begin{figure}
	\includegraphics[width=0.99 \columnwidth]{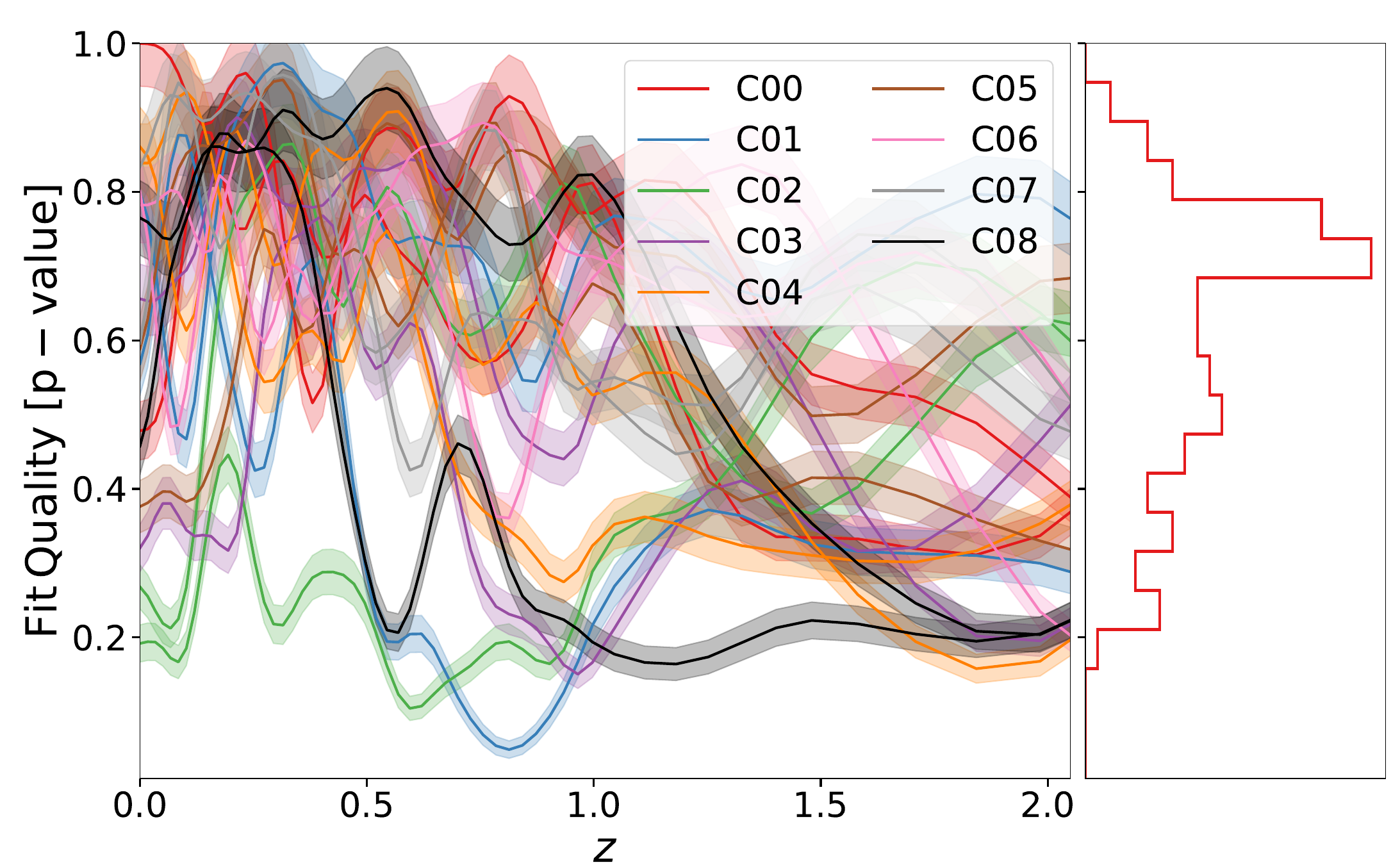}
	\caption{Dependence of \rockstar\ fit quality as a function of redshift. The fit quality (p-value) was estimated from sampling from the likelihood used for the calibration assuming the best-fit parameters. The p-value is the fraction of random catalogues that had a residual higher than the original catalogue. On the right panel, we present the histogram projection of the left panel.}
	\label{fig:cal-qual-2}
\end{figure}

In Fig.~\ref{fig:cal-qual-2}, we present the \rockstar\ fit quality dependence with redshift. The fit-quality (p-value) in Fig.~\ref{fig:cal-qual-2} was estimated from sampling $1000$ points from the likelihood used for the calibration assuming the best-fit parameters. On the right panel, we present the histogram projection of the left panel. The p-value is the fraction of random catalogues that had a residual higher than the original catalogue fit. While all the \piccolo\ snapshots were used to produce the left panel of the Fig.~\ref{fig:cal-qual-2}, only the specific snapshots used for the calibration were used to present the histogram on the right panel so as to avoid using strongly correlated snapshots.  Figure~\ref{fig:cal-qual-2} shows that our model provides robust results for the redshift range $[0, 2]$ and confirms that our choice of $\sigma_{\rm sys}$ is reasonable.
\end{appendix}

\end{document}